\documentclass[
superscriptaddress,
amsmath,
amssymb,
aps, 
prb,
twocolumn,
floatfix,
]{revtex4-2}

\usepackage{esint}

\usepackage{hyperref}
\usepackage[utf8]{inputenc}

\usepackage{verbatim}
\usepackage{amsmath}	
\usepackage{amssymb}	
\usepackage{amsfonts}
\usepackage{amsthm}			
\usepackage[capitalise]{cleveref}	
\usepackage{mathrsfs}

\usepackage{appendix}				
\usepackage{xcolor}	

\usepackage{soul}				
\sethlcolor{yellow}				

\usepackage{graphicx}			

\graphicspath{{images/}}

\allowdisplaybreaks

\renewcommand{\vec}[1]{\boldsymbol{#1}}
\let\oldhat\hat
\renewcommand{\hat}[1]{\oldhat{\boldsymbol{#1}}}

\newcommand{\bvec}[1]{\ensuremath{\mathbf{#1}}}

\newcommand{\dee}{\ensuremath{\,\mathrm{d}}}

\newcommand{\ket}[1]{\ensuremath{ \left| #1 \right> } }

\numberwithin{lemma}{section}
\numberwithin{example}{section}
\numberwithin{proposition}{section}
\numberwithin{equation}{section}
\numberwithin{theorem}{section}
\numberwithin{corollary}{section}
\numberwithin{remark}{section}
\numberwithin{definition}{section}
\numberwithin{assumption}{section}
\numberwithin{figure}{section}
\newtheorem*{definition*}{Definition}		
\newtheorem*{assumption*}{Assumption}
\newtheorem*{remark*}{Remark}
\newtheorem*{theorem*}{Theorem}
\newtheorem*{lemma*}{Lemma}
\newtheorem*{proposition*}{Proposition}
\newtheorem*{corollary*}{Corollary}
\newtheorem*{example*}{Example}
\newtheorem*{conjecture*}{Conjecture}

\usepackage{booktabs}
\usepackage{multirow}

\usepackage{xcolor}
\usepackage{braket}
\usepackage{siunitx}
\DeclareMathOperator{\tr}{tr}
\definecolor{purp}{RGB}{160, 32, 240}

\usepackage{enumitem}
\usepackage{placeins}

\newcommand{\Hbm}{H_{\text{BM}}}
\newcommand{\eps}{\varepsilon}
\newcommand{\rot}[1]{\mathfrak R\left( #1 \right)}

\renewcommand{\hat}{\oldhat}

\newcommand{\m}{\mathrm{m}}
\newcommand{\A}{\mathrm{A}}
\newcommand{\B}{\mathrm{B}}

\usepackage{graphicx}
\usepackage{dcolumn}
\usepackage{bm}

\newcolumntype{C}[1]{>{\centering\arraybackslash}m{#1}}

\begin{document}




\title{
  Relaxation effects on Hartree-Fock ground states in twisted bilayer graphene at even integer fillings
}%

\author{Tianyu Kong}%
\email{tianyuk@uchicago.edu}
\affiliation{%
  Committee on Computational and Applied Mathematics, University of Chicago, Chicago, IL 60637, USA}%

\author{Alexander B. Watson}
\email{abwatson@umn.edu }
\affiliation{%
  School of Mathematics, University of Minnesota Twin Cities, Minneapolis, Minnesota 55455, USA}%

\author{Lin Lin}%
\email{linlin@math.berkeley.edu}
\affiliation{%
  Department of Mathematics, University of California, Berkeley, CA 94720, USA}%

\author{Mitchell Luskin}%
\email{luskin@umn.edu}
\affiliation{%
  School of Mathematics, University of Minnesota Twin Cities, Minneapolis, Minnesota 55455, USA}%

\author{Kevin D. Stubbs}%
\email{Contact author: kstubbs@berkeley.edu}
\affiliation{%
  Department of Mathematics, University of California, Berkeley, CA 94720, USA}%


\date{\today}

\begin{abstract}
  A standard approach for studying magic angle twisted bilayer graphene (MATBG)'s correlated electronic phase diagram is to project the Coulomb interaction down to effective models only involving electrons in single-particle flat bands and some nearby remote bands. We provide a novel systematic derivation of a single-particle continuum model of MATBG's single-particle properties which incorporates structural relaxation while remaining in the Lagrangian frame. We project the Coulomb interactions down to electrons occupying the flat bands of this model and compute the Hartree-Fock many-body ground states at fillings $\nu = \pm 2$. 
  We find that incorporating relaxation effects drives the model into a semi-metallic phase at $- 2$ because of particle-hole asymmetry in the relaxed model's single-particle dispersion and because the flat band wavefunctions become more concentrated leading to an enhanced Hartree potential.
  Our results corroborate recent ab initio density functional theory studies which also found semi-metallic phases at $-2$. We discuss potential explanations for why such phases have not been seen in experiments.
\end{abstract}

\maketitle

\section{Introduction} \label{sec:introduction}

\subsection{Background}

Magic angle twisted bilayer graphene (MATBG) has drawn immense interest over the past few years due to the experimental observation of superconductivity and other strongly correlated behavior in this material \cite{NuckollsYazdani2024,NuckollsLeeOhEtAl2023,OhNuckollsWongEtAl2021,cao2018correlated,cao2018unconventional}.
Since the twist angle in MATBG is fairly small (about \(1.1^{\circ}\)), the number of atoms per moir\'e (the effective unit cell) is more than 10,000, making direct modeling of many-body effects a significant challenge.

Beginning with the seminal work of Bistritzer-MacDonald \cite{Bistritzer_MacDonald_2011,Watson_Kong_MacDonald_Luskin_2023}, a common approach for modeling MATBG is to define a continuum single-particle Hamiltonian which captures moir\'e-scale properties of the material while remaining theoretically and computationally tractable.
This model is then used to identify relevant degrees of freedom for reduced-order modeling of MATBG's many-body electronic properties.
These degrees of freedom generally consist of the single-particle model's nearly-flat bands at the charge neutral Fermi level together with a number of remote bands.
The many-body electronic properties of MATBG are finally modeled and predicted on the basis of effective models where electron-electron interactions are projected down to these degrees of freedom; see, e.g. \cite{BultinckKhalafLiuEtAl2020,XieMacDonald2020,2019KangVafek,LiuKhalafLeeEtAl2021,Po2018,Kwan2021,Parker2021,rr5g-3js8,Kwan02102025,SongBernevig2022,relaxedheavyfermion,x2mz-hm8y,FaulstichStubbsZhuEtAl2023,Kim2026InitioQuantumEmbedding,Sanchez2024,Sanchez2025,qlp6-hjf2,LEDWITH2021168646,Ledwith_2025,ledwith2025exoticcarriersconcentratedtopology,Schindler2022,Guo2026,HouSurWagnerEtAl2025,Bennett2024,Zhang2020,Wagner2022,Shi2022,kong2025interactingtwistedbilayergraphene}.

The predictions of these effective models can be sensitive to details of their derivations.
For example, predicted many-body ground states can change depending on the choice of effective single-particle continuum model and the choice of electron-electron interaction double-counting subtraction scheme.
The band structures of single-particle models of MATBG are known, in particular, to be sensitive to the effects of mechanical relaxation \cite{YooRelaxation2019},
where atoms move from their monolayer equilibrium positions in order to minimize their total energy within the bilayer.

The simplest way to incorporate relaxation effects into a single-particle continuum model is to directly tune the ratio of ``AA'' vs. ``AB'' hopping in the Bistritzer-MacDonald model, since relaxation tends to enlarge energetically-favorable ``AB'' stacking regions compared with ``AA'' regions; see, e.g. \cite{BultinckKhalafLiuEtAl2020}.
More systematically, relaxed atomic displacements can be computed by minimizing a continuum energy where linear elasticity is coupled to a stacking energy parametrized using DFT \cite{Carr_Massatt_Torrisi_Cazeaux_Luskin_Kaxiras_2018,Cazeaux_Luskin_Massatt_2020,Cazeaux_Clark_Engelke_Kim_Luskin_2023}.
Effective single-particle continuum models can then be derived using the resulting relaxed atomic displacements \cite{NamKoshino,Yoo_Engelke_Carr_Fang_Zhang_Cazeaux_Sung_Hovden_Tsen_Taniguchi_etal_2019,relaxedblg23,Kang_Vafek_2023,Vafek_Kang_2023}.

\subsection{Results}
 The contributions of the present work are as follows.
We first provide a novel systematic derivation of an effective single-particle continuum model for moir\'e materials such as MATBG with relaxation effects based on multiple-scale analysis.
Unlike previous approaches to relaxed models \cite{relaxedblg23,Kang_Vafek_2023,Vafek_Kang_2023,Balents2019}, our multiple-scale analysis remains entirely in the Lagrangian frame which significantly simplifies the derivation.
  For MATBG, our results closely match the results obtained by Kang and Vafek in \cite{Kang_Vafek_2023,Vafek_Kang_2023}.
  We stress that while this work focuses on MATBG, the same techniques can be used for modeling relaxation in other moir\'e materials.

After deriving the effective relaxed single-particle continuum model, or ``relaxed model'', we study the impact of relaxation on a many-body model of MATBG when restricted to the flat bands.
We compare the Hartree-Fock ground states and band structures of this model with those of the equivalent effective model parameterized by the Bistritzer-MacDonald (BM) model where relaxation is modeled by varying the ratio of ``AA'' to ``AB'' hopping.
Away from charge neutrality, we find significant \emph{qualitative} differences between the BM and relaxed model in the Hartree-Fock ground states.

Our Hartree-Fock studies are motivated by recent STM experiments which show that at filling $\nu = -2$, corresponding to the eight flat bands being one-quarter filled, the Fourier transform of the local density of states (FT-LDOS) MATBG has additional Bragg peaks in contrast to theoretical predictions \cite{Tomo2022,NuckollsLeeOhEtAl2023}.
To investigate this behavior, we first use $\nu = 0$ to benchmark out calculations and then we focus on filling factors $\nu = \pm 2$.
Away from charge neutrality, we find that relaxation corrections to the Hartree potential tend to drive the system into a \emph{semi-metallic} phase, in marked contrast to the insulating phase found when using the BM model.
These corrections can be traced to the fact that the flat band wavefunctions tend to be more concentrated in the relaxed model than those of the BM model, leading to an enhanced ``Hartree peak/dip'' \cite{Kwan02102025}.
At filling $+2$, this effect is canceled by relaxation corrections to the single-particle dispersion so that the system remains insulating.
At filling $-2$, no such cancellation occurs because the band structure of the relaxed model is strongly particle-hole asymmetric, and we find that the system is ultimately semi-metallic.
We find that these results are consistent across both the ``average'' and ``charge neutral'' double-counting subtraction schemes.  

\subsection{Comparison with literature and experiments}
Similar results, specifically semi-metallicity at filling $\nu = - 2$, have been reported both in recent \emph{ab initio} DFT studies of MATBG \cite{Kim2026InitioQuantumEmbedding} as well as other approaches based on DFT-parametrized continuum modeling \cite{HouSurWagnerEtAl2025}.
In contrast, previous works based on tuning the BM model have generally predicted $\nu = -2$ to be insulating; see, e.g., \cite{Zhang2020,Kwan2021,Wagner2022,SongBernevig2022,Shi2022} and the discussion in \cite{HouSurWagnerEtAl2025}.
Our results can be seen as providing further evidence for these phenomena and a clearer understanding of their origin.

Comparing our results to the other semi-metallic Hartree-Fock results in semimetallic\cite{HouSurWagnerEtAl2025,Kim2026InitioQuantumEmbedding}, we note that in \cite{Kim2026InitioQuantumEmbedding}, the semimetallic behavior was found to be connected to the choice of double counting subtraction.
In this work, we also find the semimetallic behavior at $\nu = -2$ is related to double counting subtraction however the mechanism we find, wavefunction concentration, cannot be easily tested in \textit{ab initio} calculations due to computational limitations.
In contrast, in \cite{HouSurWagnerEtAl2025} the authors attribute the semimetallic behavior to the inclusion of remote bands which are not considered in the present work.

Experiments have generally shown insulating behavior at filling $\nu = \pm 2$ \cite{NuckollsYazdani2024}.
It would be interesting to test whether this behavior is recovered when one of the effects neglected in the present study is incorporated into the model, such as heterostrain (strain which acts differently within each layer), which is expected to play an important role in moir\'e materials such as MATBG; see, e.g., \cite{Kwan2021,Parker2021,relaxedheavyfermion,x2mz-hm8y,rr5g-3js8,Guo2026}.

Other potentially important effects include out-of-plane layer corrugation, the effects of MATBG's remote bands, and other choices of subtraction scheme.
All of these effects could in principle be incorporated into the framework we introduce in the present work and will be the subject of future works.
Out-of-plane corrugation was taken into account in \cite{Carr2019,Carr2019a}, where it was found to increase the gap from the nearly-flat bands to remote bands.
The study \cite{HouSurWagnerEtAl2025} was built upon these models and found qualitatively similar phenomena to ours, including semi-metallicity at $\nu = -2$.
Various studies have considered the effects of remote bands; see, e.g., \cite{Sanchez2024,Sanchez2025,qlp6-hjf2,HouSurWagnerEtAl2025,Kwan2021,BultinckKhalafLiuEtAl2020}, although, to the best of our knowledge, without finding significant qualitative differences in phenomena.
For a systematic discussion of renormalization effects of remote bands, see \cite{PhysRevLett.125.257602}.
A subtraction scheme we do not consider here, the so-called ``decoupled'' subtraction scheme, has been considered in various works \cite{BultinckKhalafLiuEtAl2020,XieMacDonald2020,FaulstichStubbsZhuEtAl2023}, although again without finding significant qualitative differences in phenomena.
A more challenging potentially significant factor not included in our modeling is twist angle disorder \cite{PhysRevResearch.2.023325,Uri2020,PhysRevB.107.125413}. For more discussion, see Section \ref{sec:conclusions}.

\subsection{Structure of this work}

The structure of this work is as follows. In Section \ref{sec:cont-model-twist}, we review the single-particle BM model and sketch the derivation of our relaxed single-particle continuum model. In Section \ref{sec:inter-models-twist}, we review the general approach for modeling MATBG's many-body properties by projecting onto the single-particle models' flat bands. In Section \ref{sec:inter-model-results}, we present results for computing Hartree-Fock ground states of both the BM and relaxed models. In Section \ref{sec:conclusions}, we conclude and discuss potential extensions of this work. In Appendix \ref{sec:derivation_cont_model}, we provide a more detailed derivation of our relaxed continuum model. In Appendix \ref{sec:review-hartree-fock}, we review Hartree-Fock theory for electrons in the flat bands of MATBG. In Appendices \ref{app:subtraction} and \ref{sec:init-symm} we review the ``charge neutral'' subtraction scheme and the initial states used for our Hartree-Fock computations. In Appendix \ref{sec:wavefunction-density} we show plots of flat band wavefunctions of the relaxed model compared with those of the BM model, demonstrating that the wavefunctions of the former are more concentrated. We present results for average subtraction in the Supplementary Material \cite{supplementary}.

\section{Continuum Modeling for Twisted Bilayer Graphene} \label{sec:cont-model-twist}

In this section, we introduce the original BM model, as well as the relaxed BM model, which systematically accounts for the effect of mechanical relaxation in twisted bilayer graphene \cite{cazeaux2018energy,Carr_Massatt_Torrisi_Cazeaux_Luskin_Kaxiras_2018, Cazeaux_Clark_Engelke_Kim_Luskin_2023}. 
For details on the model derivation, please refer to the Appendix~\ref{sec:derivation_cont_model}.

\subsection{Bistritzer-MacDonald model}

A single sheet of graphene can be described by the lattice vectors $\bvec{a}_1 := a(1 / 2,\sqrt{3} / 2)^\top$, $\bvec{a}_2 := a(-1/2,\sqrt{3}/2)^\top$ with lattice constant $a = 2.46$ \r{A}.
The Bravais lattice, $\mathcal R$, and a unit cell, $\Omega$, can be expressed as \cite{2009Castro-NetoGuineaPeresNovoselovGeim}
\begin{equation}
  \begin{gathered}
    A = (\bvec{a}_1 \vert \bvec{a}_2), \quad \mathcal{R} :=  \{ \bvec{R} = A \bvec{n}: \bvec{n}\in \mathbb Z^2 \}, \\
    \Omega :=  \left\{ A \vec{\alpha} : \vec{\alpha}\in  \left[-1/2,1/2\right)^2\right\}.
  \end{gathered}
\end{equation}
In each unit cell, there are two atoms  in sub-lattices A and B, which corresponds to a relative shift in physical position $\bvec{R} + \vec{\tau}^\A$ and $\bvec{R} + \vec{\tau}^\B$ where
\begin{equation} \label{eq:relative_shift}
  \vec{\tau}^\A := (0, 0)^\top, \quad \vec{\tau}^\B : = \left(0, \frac{a}{\sqrt{3}} \right)^\top.
\end{equation}
Throughout this work, we use \(\sigma \in \{ \A, \B \}\) to denote the sublattice degree of freedom.

The monolayer Dirac points are located at
\begin{equation}
  \bvec{K} := \frac{ 4 \pi }{ 3 a } ( 1, 0 )^\top, \quad \bvec{K}' := - \bvec{K}.
\end{equation}
Near the Dirac points, the monolayer dispersion can be approximated by an effective Dirac operator \cite{2009Castro-NetoGuineaPeresNovoselovGeim}. 

TBG is obtained by stacking two identical layers with a relative interlayer twist \(\theta\) \cite{Bistritzer_MacDonald_2011,Watson_Kong_MacDonald_Luskin_2023}.
The twisted lattices for layers 1 and 2 are generated by $A_1 = \rot{-\theta / 2} A$ and $A_2 = \rot{\theta / 2} A$ respectively, where  
\begin{equation}
  \rot{\theta} = \begin{pmatrix} \cos \theta & -\sin\theta \\ \sin \theta & \cos \theta \end{pmatrix}
\end{equation}
denotes the counterclockwise rotation by $\theta$. The rotated unit cells of each layer are denoted by $\Omega_1,\Omega_2$.
We define sublattice shifts on each layer $\vec{\tau}^\sigma_1$ and $\vec{\tau}^\sigma_2$ for $\sigma \in \{\A,\B \}$ similarly. 
The Dirac points of each layer are also rotated to $\bvec{K}_1 = \rot{-\theta / 2}  \bvec{K}$, $\bvec{K}_2 = \rot{\theta / 2} \bvec{K}$. The momentum shift between Dirac points is
\begin{equation}
  \bvec{s} := \bvec{K}_2 - \bvec{K}_1 = \left(\rot{\theta/2}  - \rot{-\theta/2} \right) \bvec{K}.
\end{equation}

The moir\'e fundamental matrix between the two mismatched layers is defined as $A_\m := \left(A_1^{-1} - A_2^{-1}\right)^{-1}$, which is even valid at commensurate angles.
We denote the the lattice generated by it as
\begin{equation}
  \mathcal R_\m := \left\{ \bvec{R}_\m = A_\m \bvec{n} : \bvec{n} \in \mathbb{Z}^2 \right\}.
\end{equation}
Writing $A_\m = (\bvec{a}_{\m1} \vert \bvec{a}_{\m 2})$, the moir\'e superlattice has length $a_\m := |\bvec{a}_{\m 1}| = \frac{a}{2 \sin(\theta/2)}  \sim a\theta^{-1}$.  When $\theta$ is close to the ``magic angle'' $\approx 1.05 ^\circ$,  the resulting moir\'e superlattice vectors are roughly $100$ times the graphene lattice scale.
We can define the moir\'e reciprocal vectors from the dual relation 
\begin{equation}
  \bvec{a}_{\m i} \cdot \bvec{b}_{\m j} = 2\pi\delta_{ij}, \quad B_\m 
  =(\bvec{b}_{\m1}, \bvec{b}_{\m 2}),
\end{equation}
and the moir\'e reciprocal lattice is
\begin{equation}
  \mathcal R_\m^* := \left\{ \bvec{G}_\m = B_\m \bvec{n} : \bvec{n} \in \mathbb{Z}^2 \right\}.
\end{equation}

For conciseness of notation, we use $\bvec{R}_{\ell,\sigma}$ to represent the carbon atom located at $\bvec{R}_{\ell} + \vec \tau_\ell^\sigma$, where $\bvec{R},\sigma,\ell$ are the Bravais lattice, sublattice, and layer degrees of freedom, respectively.
The atomistic tight-binding model of TBG can be written as 
\begin{equation}\label{eq:H_tb}
  \begin{aligned}
    &H_{\text{TB}} =\\
    &\sum_{\bvec{R}, \ell, \sigma, \bvec{R}', \ell', \sigma'}  h^{\sigma\sigma'}_{\ell\ell'} \left(\bvec{R}_{\ell,\sigma} - \bvec{R}'_{\ell', \sigma'} \right)
      c_{\bvec{R}, \ell, \sigma}^\dagger c_{\bvec{R}', \ell', \sigma'}.
  \end{aligned}
\end{equation}
The creation and annihilation operators satisfy the canonical anticommutation relations
\begin{equation}
  \label{eq:car}
  \{ c^\dagger_{\bvec{R},\ell,\sigma}, c_{\bvec{R}', \ell', \sigma'} \} = \delta_{\bvec{R}, \bvec{R}'}\delta_{\ell,\ell'}\delta_{\sigma,\sigma'}.
\end{equation}
 The hopping function can be parameterized by the Slater-Koster function \cite{2009Castro-NetoGuineaPeresNovoselovGeim}, or directly from DFT calculations \cite{Fang_Kaxiras_2016}. The sublattice dependence on the hopping function $h^{\sigma\sigma'}$ accounts for the different orientation of orbitals \cite{Vafek_Kang_2023}. 

\begin{figure}[t]
 \includegraphics[width=.7\linewidth]{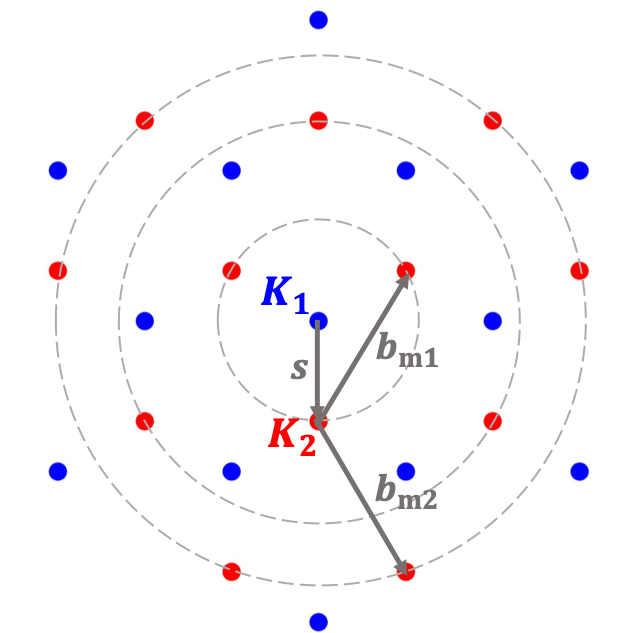}
  \caption{Dirac points $\bvec{K}_1, \bvec{K}_2$, moir\'e reciprocal lattice vectors $\bvec{b}_{\m 1}, \bvec{b}_{\m 2}$ and interlayer momentum shift $\bvec s$. Starting from $\bvec K_1$, blue points are intralayer hops and red points are interlayer hops. } 
   \label{fig:momentum_hop}
\end{figure}

 The Bistritzer-MacDonald (BM) model \cite{Bistritzer_MacDonald_2011} is the leading-order continuum approximation of the tight-binding model in the low-energy limit.  The low-energy limit at charge neutrality is concentrated at momentum states in the neighborhood of the monolayer Dirac points. 
Around the $\bvec{K}$ point (called the $\bvec{K}$ valley), the BM model consists of the monolayer Dirac operator and moir\'e-periodic interlayer tunneling
\begin{equation} \label{eq:H_BM}
  \Hbm (\bvec{r}) = 
  \begin{pmatrix} \hbar v_F \vec{\sigma} \cdot (-i \nabla_{\bvec{r}} ) &  T(\bvec{r}) \\  T^\dagger(\bvec{r}) & \hbar v_F \vec{\sigma}\cdot (-i \nabla_{\bvec{r}} + \bvec{s})	
  \end{pmatrix}.
\end{equation}
The corresponding model for the $\bvec{K}'$ valley is related by time-reversal symmetry.
The monolayer Fermi velocity satisfies $\hbar v_F = 5.75 \text{ eV}\cdot\text{\r{A}}$. The interlayer potential is
\begin{equation*}
  \label{eq:T_matrix}
  T(\bvec{r}) = \sum_{j=0}^2 T_j e^{-i \bvec{G}_j \cdot \bvec{r}}, 
  \quad T_j = 
  \begin{pmatrix}
    w_0 & w_1 e^{-i 2\pi j/3} \\ w_1 e^{i 2\pi j/3} & w_0
  \end{pmatrix},
\end{equation*}
where  $\bvec{G}_0 = 0$, $\bvec{G}_1 = \bvec{b}_{\m, 1}$ and $\bvec{G}_2 = -\bvec{b}_{\m, 2}$. After adding the interlayer momentum shift $\bvec{s}$, these moir\'e reciprocal lattice vectors form the first shell of interlayer momentum hops (see Fig.~\ref{fig:momentum_hop}). In the original BM model, where the graphene layers are assumed to be rigid, the hopping amplitudes are $w_0 = w_1 = 110 \text{ meV}$. It is common to set the ratio $\kappa = w_0/ w_1 < 1$ to model relaxation effects, where AB stacking is preferred over AA stacking configurations. Informed by calculations on the relaxed BM model to be introduced later \cite{kong2025interactingtwistedbilayergraphene}, we use $w_0 = 0.7 w_1$ in this work.

The BM model is a good first step for predicting the single-particle electronic properties of TBG at around the magic angle without making commensurate approximations.
This approximation can be rigorously justified using a multi-scale expansion of the tight-binding model (see \cite{relaxedblg23} and Appendix~\ref{sec:multiscale_derivation}). 
A higher-order continuum BM model can improve the accuracy compared to tight-binding calculations \cite{Kang_Vafek_2023,quinn2025higherordercontinuummodelstwisted}. The higher-order model includes quadratic correction to the Dirac operator in the intralayer term, and further interlayer momentum hops as well as a momentum-dependent ``non-local'' approximation in the interlayer term. 

\subsection{Tight-binding model with relaxation}

Structural relaxation is the process in which the atoms in two mismatched layers of graphene re-arrange themselves in order to minimize their total mechanical energy \cite{Yoo_Engelke_Carr_Fang_Zhang_Cazeaux_Sung_Hovden_Tsen_Taniguchi_etal_2019}. 
In this work, we model relaxation using the effective continuum model of \cite{Carr_Massatt_Torrisi_Cazeaux_Luskin_Kaxiras_2018,Cazeaux_Clark_Engelke_Kim_Luskin_2023,dai2016twisted}. In this model, the continuum relaxed displacements $\bvec{u}_{\ell}$ are assumed smooth and moir\'e-periodic: \(\bvec{u}_{\ell}(\cdot) = \bvec{u}_{\ell}(\cdot + \bvec{R}_{\m})\) for each \(\bvec{R}_{\m} \in \mathcal{R}_{\m}\). 

The displacement fields minimize the energy per moir\'e unit cell
\begin{equation}\label{eq:relax_optim}
  \begin{aligned}
    \mathcal E [\bvec{u}_{1}, \bvec{u}_{2}] = 
    \sum_{\ell=1}^2 \mathcal E_\text{intra}(\bvec{u}_{\ell}) + \mathcal E_\text{inter}(\bvec{u}_{1} -  \bvec{u}_{2}).
  \end{aligned}\end{equation}
The intralayer energy is approximated by linearizing the elastic energy of deforming the layers  
\begin{gather}
  \mathcal E_\text{intra}(\bvec{u}) :=\fint_{\Omega_\m} \frac{\lambda}{2} (\operatorname{div} \bvec{u} )^2 + \mu \vec{\eps}(\bvec{u}) : \vec{\eps}(\bvec{u}) \dee \bvec{x},
 \end{gather}
 with the infinitesimal strain tensor 
 \begin{gather}
  \vec{\eps}(\bvec{u}) := \frac{1}{2}\left(\nabla \bvec{u} + \nabla \bvec{u}^\top \right),
\end{gather} where the elastic moduli $\lambda = 7.241 \text { eV}/\text{\r{A}}^2$ and $\mu = 9.035 \text { eV}/\text{\r{A}}^2$ are fit to experiment or density functional theory (DFT) computation \cite{2009Castro-NetoGuineaPeresNovoselovGeim}.

The interlayer energy is a generalized stacking fault energy (GSFE) fit to DFT \cite{dai2016twisted} which provides the energy landscape that depends only on the relative stacking between the two layers
\begin{equation}\begin{aligned} \label{eq:thiss}
  \mathcal E_\text{inter}(\bvec{v}) :=  \frac{1}{2} \fint_{\Omega_\m}& \Phi_1({\gamma}_2\bvec{x} + \bvec{v}(\bvec{x}))   \\
  & + \Phi_2({\gamma}_1 \bvec{x} - \bvec{v} (\bvec{x})) \dee \bvec{x}.
\end{aligned}\end{equation}
where the moir\'e unit cell is
$$\Omega_\m :=  \left\{ A_\m \vec{\alpha} : \vec{\alpha}\in \left[-1/2,1/2\right)^2 \right\}.$$

Here $\Phi_1: \Omega_2 \to \mathbb R$ and $\Phi_2: \Omega_1 \to \mathbb R$ are the GSFEs in the disregistry space, and $\gamma_1, \gamma_2$ are disregistry functions that describes the local configuration (see Eq.~(\ref{eq:disregistry_def}) for their definition). The GSFE favors the AB local configuration compared to the AA configuration, so that the AB stacking region in relaxed moir\'e pattern grows, while the AA region shrinks. 

\begin{figure}[t]
\includegraphics[width=.6\linewidth]{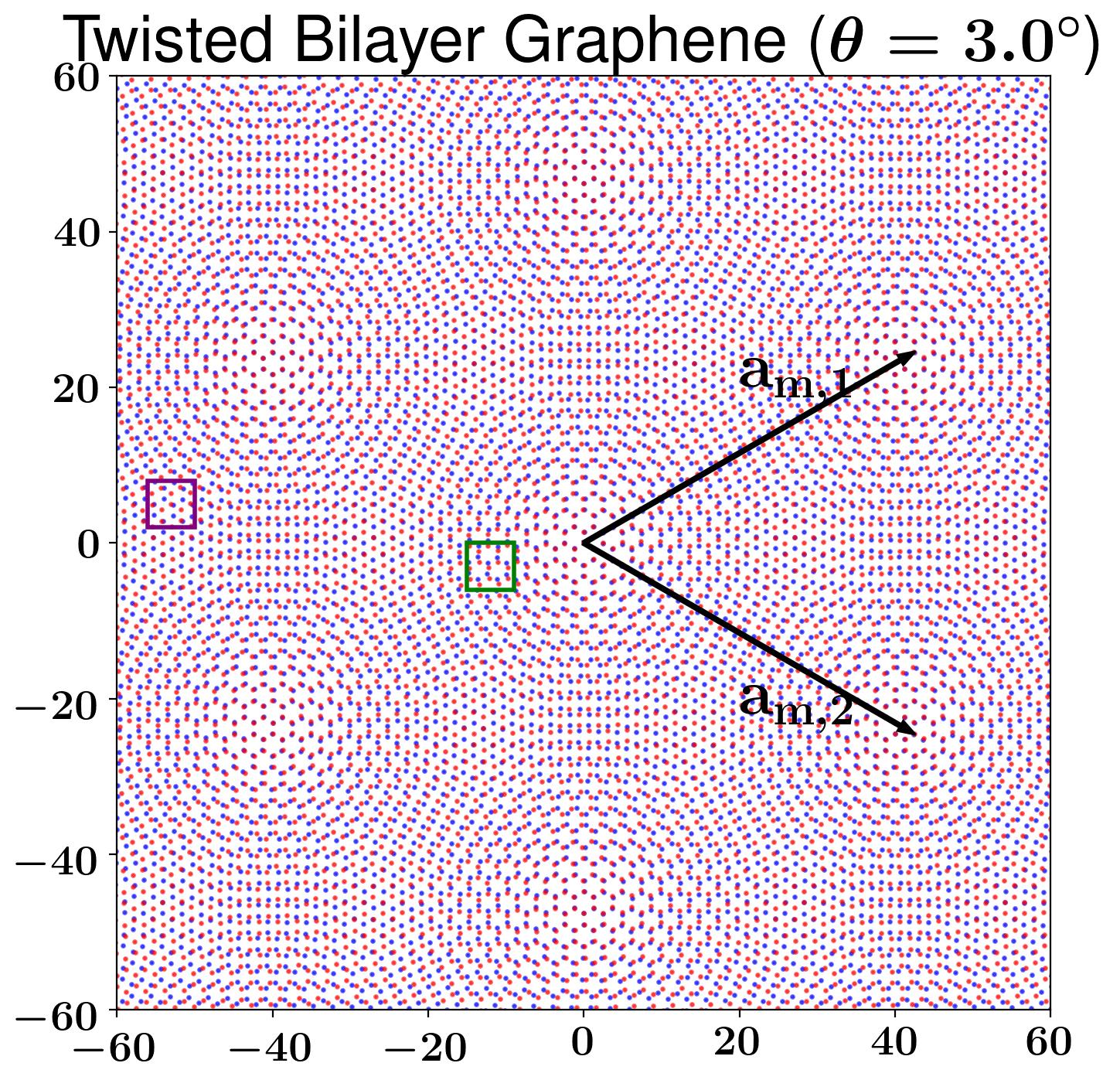}
 \includegraphics[width=.35\linewidth]{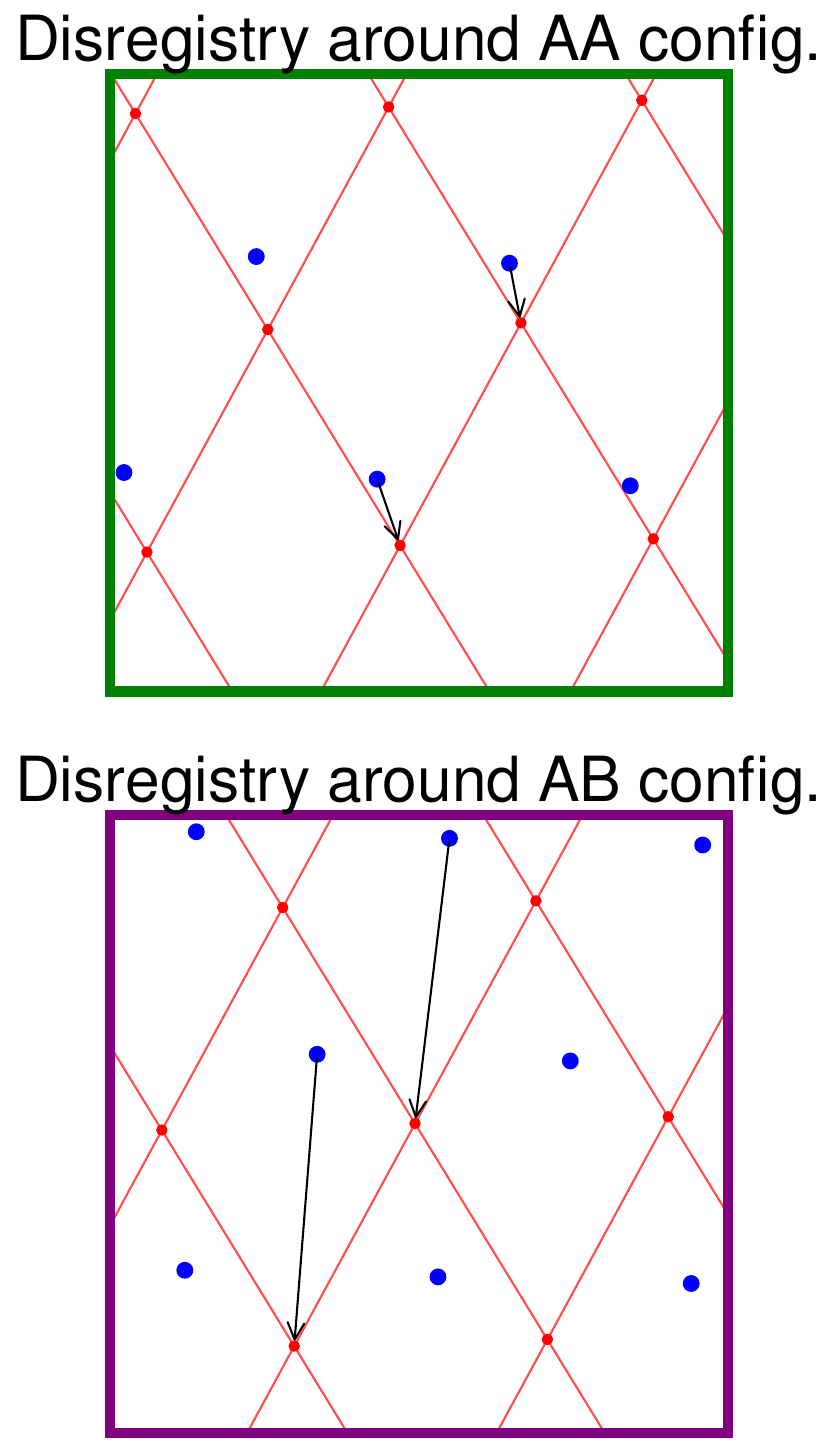}
  \caption{Atomistic configuration of twisted bilayer graphene. The disregistry, or the position of an atom measured against the unit cell of the opposite layer, are shown around both AA (green) and AB (purple) regions (for clarity only A sublattice is shown). For neighboring lattice sites, their difference in disregistry is small (approximately $\eps a$).}\label{fig:disregistry}
\end{figure}


The tight-binding model that accounts for the structural relaxation effects is 
\begin{widetext}
  \begin{equation}\label{eq:H_tb_relax}
    \begin{aligned}
      H_{\text{TB,relax}} = \sum_{\bvec{R}, \ell, \sigma, \bvec{R}', \ell', \sigma'} h^{\sigma\sigma'}_{\ell\ell'} \left( \bvec{R}_{\ell,\sigma}  + \bvec{u}_{\ell}(\bvec{R}_{\ell,\sigma})   - \bvec{R}'_{\ell',\sigma'}  - \bvec{u}_{\ell'}(\bvec{R}'_{\ell',\sigma'} ) \right) c_{\bvec{R}, \ell, \sigma}^\dagger c_{\bvec{R}', \ell', \sigma'},
\end{aligned}\end{equation}
\end{widetext}
where the hopping function $h^{\sigma \sigma'}_{\ell \ell'}$ is as in \cref{eq:H_tb} 

To pass to a continuum approximation at the moir\'e scale, it is critical to check that the relaxation model predicts that the displacements $\bvec{u}_\ell$ vary appreciably only at the moir\'e scale, i.e., slowly at the atomic scale near the magic angle. 
To see that this holds it is convenient to first make a change of variables to disregistry space displacements $\bvec{U}_1, \bvec{U}_2$ \cite{cazeaux2018energy,Carr_Massatt_Torrisi_Cazeaux_Luskin_Kaxiras_2018} such that
\begin{equation}\label{eq:u_change_of_variables}
  \bvec{u}_1(\bvec{x}) = \bvec{U}_1 ( \gamma_2  \bvec{x}), \quad \bvec{u}_2(\bvec{x}) = \bvec{U}_2 ( \gamma_1  \bvec{x}), 
\end{equation}
where
\begin{equation}\label{eq:disregistry_def}
\begin{aligned}
  \gamma_1 := I - A_1A_2^{-1}, \quad  \gamma_2 := A_2A_1^{-1} - I=\rot{\theta}\gamma_1,
\end{aligned}\end{equation}
are the disregistry functions, which map the moir\'e cell to the fundamental cells of either layer: $\gamma_1 : \Omega_\m \rightarrow \Omega_1$, $\gamma_2 : \Omega_\m \rightarrow \Omega_2$. 
The disregistry $\gamma_2$ of layer $1$ with respect to layer $2$ is $A_2A_1^{-1} - I$ rather than $I-A_2A_1^{-1}$  since it corresponds to a frame with normal pointing from layer $1$ to layer $2$ rather than from layer $2$ to layer $1.$ Note that the disregistry space displacement functions are periodic under translations by lattice vectors of the opposite layer
\begin{equation}\label{eq:u_period}
  \bvec{U}_1 (\cdot) = \bvec{U}_1(\cdot + \bvec{R}_2), \quad \bvec{U}_2 (\cdot) = \bvec{U}_2(\cdot + \bvec{R}_1).
\end{equation}

The variables $\gamma_\ell \bvec{x}$ vary appreciably only when $\bvec{x}$ varies at the moir\'e scale.
Concretely, suppose we denote the ratio between a graphene lattice and a moir\'e lattice as a dimensionless small parameter $\eps := a / a_\m = 2 \sin(\theta / 2)$, 
then we may expand the disregistry functions in orders of the small parameter $\eps$ 
\begin{equation}\label{eq:gamma_expansion}
  \begin{split}
    \gamma_1 = I - \rot{-\theta} &= -\eps \mathcal J  + \mathcal{O}(\eps^2) \\ 
    \gamma_2 = \rot{\theta} - I &= - \eps \mathcal J +  \mathcal{O}(\eps^2) 
  \end{split}
\end{equation}
where the leading order term $\mathcal J = \begin{pmatrix} 0 & -1 \\ 1 & 0 \end{pmatrix} $ is a norm 1 matrix that encodes the $\pi/2$  rotation between the moir\'e lattices and the untwisted monolayer graphene lattices.

We can conclude that the $\bvec{u}_\ell$'s vary only appreciably at the moir\'e scale as long as the $\bvec{U}_\ell$'s have derivatives independent of $\eps$. After nondimensionalization, the energy functional minimized by the $\bvec{U}_\ell$ depends only on the dimensionless parameter, \(\eta\),
\begin{equation}
  \begin{split}
    \eta & := \frac{\eps^2(\lambda + \mu)}{C_\Phi}  \quad \text{where} \\
    C_\Phi & = \max_{i \in \{ 1, 2 \}} \left(\max_{x} \Phi_{i}(x) - \min_x \Phi_{i}(x)\right)
  \end{split}
\end{equation}
where $\Phi_i$ are the GSFEs appearing in \eqref{eq:thiss}; we note that the maximum and minimum of \(\Phi_{i}\) occurs at the AA and AB stackings respectively.
For realistic parameter values near to the magic angle, the small parameter $\eps = a/a_\m \approx 0.018$, so we can conclude that $\eps^2 (\lambda + \mu) \approx 5.27 \text{ meV/\r{A}}^2$. We also have the GSFE strength $C_\Phi \approx 3.68 \text{ meV/\r{A}}^2$ \cite{Carr_Massatt_Torrisi_Cazeaux_Luskin_Kaxiras_2018,Cazeaux_Clark_Engelke_Kim_Luskin_2023}, so that the dimensionless parameter $\eta$ is estimated to be $\eta \approx 1.43$, i.e., at the magic angle, the intralayer elasticity in configuration space is approximately balanced with the interlayer stacking. In particular, we conclude that the optimal relaxation displacement in disregistry space representation satisfies  $\| \nabla \bvec{U}_\ell \| = \mathcal O(1)$. As a consequence, the real space displacements should satisfy  $\| \nabla \bvec{u}_\ell \| = \mathcal O(\eps)$, so they are slowly varying functions on the atomistic scale. For a rigorous estimate on the displacements in terms of the maximum of the \emph{derivative} of the GSFE, see \cite[Theorem 4.10] {Cazeaux_Luskin_Massatt_2020}.
For angles much smaller than the magic angle, we instead have $\eta \ll 1$, so that $\| \nabla \bvec{U}_\ell \| = \mathcal O(1 / \eps )$ and $\| \nabla \bvec{u}_\ell \| = \mathcal O(1)$; in this case, a different expansion is required to capture the electronic properties \cite{relaxedblg23,YooRelaxation2019}.

We now plug in the disregistry-based $\bvec{U}_\ell$ into Eq.~(\ref{eq:H_tb_relax}). The intralayer terms can naturally be expanded in terms of $\eps$, where the leading order term is the unrelaxed tight-binding model, and the relaxation effects are $\mathcal O(\eps)$ corrections
\begin{widetext}
  \begin{equation}\label{eq:tb_intralayer}
    \begin{aligned}
      & h^{\sigma\sigma'}_{11} \left( \bvec{R}_{1,\sigma}  + \bvec{U}_{1}(\gamma_2\bvec{R}_{1,\sigma})   - \bvec{R}'_{1,\sigma'} - \bvec{U}_{1}(\gamma_2 \bvec{R}'_{1,\sigma'} ) \right) \\
      & =   h^{\sigma\sigma'}_{11} \left(  \bvec{R}_{1,\sigma} - \bvec{R}'_{1,\sigma'}\right)  +   \nabla h^{\sigma\sigma'}_{11} \left( \bvec{R}_{1,\sigma} - \bvec{R}'_{1,\sigma'}\right)   \cdot \nabla  \bvec{U}_{1}\left(\gamma_2 \bvec{R}_{1,\sigma} \right) \cdot {\gamma_2} \left(  \bvec{R}_{1,\sigma} - \bvec{R}'_{1,\sigma'}\right) + \mathcal O(\|\gamma_2\|^2) \\
      & =   h^{\sigma\sigma'}_{11} \left( \bvec{R}_{1,\sigma} - \bvec{R}'_{1,\sigma'}\right)+ \eps \nabla h^{\sigma\sigma'}_{11} \left(  \bvec{R}_{1,\sigma} - \bvec{R}'_{1,\sigma'}\right)   \cdot \nabla  \bvec{U}_{1}\left( \tilde{\gamma}_2 \eps  \bvec{R}_{1,\sigma}  \right) \cdot {\tilde\gamma_2}  (\bvec{R}_{1,\sigma} - \bvec{R}'_{1,\sigma'})  + \mathcal O(\eps^2),
    \end{aligned}
  \end{equation}
\end{widetext}
where we define $\eps$-scaled disregistry functions
\begin{equation}\label{eq:gamma_tilde_expansion}
  \begin{split}
    \tilde{\gamma}_1 := \frac{\gamma_1}{\eps} &= - \mathcal J  + \mathcal{O}(\eps) \\ 
    \tilde{\gamma}_2 := \frac{\gamma_2}{\eps} &= - \mathcal J  + \mathcal{O}(\eps).
  \end{split}
\end{equation}
We note that we have evaluated the displacement in the relaxed intralayer hopping energies at $\bvec{U}_{1}(\gamma_2 \bvec{R}_{1,\sigma} )$ to retain the symmetry of the underlying graphene. 
We observe that the $\mathcal{O}(\eps)$ correction term has the same convolution form as the leading term, except that relaxation induces a moir\'e-periodic dependence on the slow variable $\eps  \bvec{R}_{1,\sigma}$. We could replace $\tilde{\gamma}_\ell$ by $\mathcal J$, but the difference will introduce an $\mathcal O(\eps)$ error on the moir\'e periodicity.

The interlayer terms are
\begin{equation}\label{eq:tb_interlayer}
\begin{aligned}
		h^{\sigma\sigma'}_{12} \left(   \bvec{R}_{1,\sigma}  + \bvec{U}_{1}(\gamma_2 \bvec{R}_{1,\sigma}) - \bvec{R}_{2, \sigma'} - \bvec{U}_{2}(\gamma_1\bvec{R}_{2, \sigma'} ) \right).
\end{aligned}
\end{equation}
From the definition of disregistry functions Eq.~(\ref{eq:disregistry_def}) and the periodicity conditions Eq.~(\ref{eq:u_period}), we have 
\begin{equation}
  \begin{aligned}	& \bvec{U}_1(\gamma_2\bvec{R}_{1,\sigma}) =\bvec{U}_1( \bvec{R}_{2,\sigma} -\bvec{R}_{1,\sigma} ) \\
  &=\bvec{U}_1\left(-(\bvec{R}_{1,\sigma} - \bvec{R}_{2,\sigma'}) + \vec{\tau}_2^{\sigma\sigma'} \right) ,
  \end{aligned}
\end{equation}
and 
\begin{equation}
  \begin{aligned}		
   & \bvec{U}_2(\gamma_1 \bvec{R}_{2,\sigma'} ) =\bvec{U}_2(\bvec{R}_{2,\sigma'} -\bvec{R}_{1,\sigma'})\\
    & = \bvec{U}_2\left( -(\bvec{R}_{1,\sigma} - \bvec{R}_{2,\sigma'}) + \vec{\tau}_1^{\sigma\sigma'} \right),
  \end{aligned}
\end{equation}
where the shift $\vec{\tau}_\ell^{\sigma\sigma'} = \vec{\tau}_\ell^\sigma - \vec{\tau}_\ell^{\sigma'}$ gives explicit sublattice dependence on the configuration space displacement functions. 
Note that we cannot further expand Eq.~(\ref{eq:tb_interlayer}) in terms of $\bvec{U}_\ell$ as in intralayer terms because $\bvec{U}_1(-\bvec{x} + \vec{\tau}_2^{\sigma\sigma'}) - \bvec{U}_2(- \bvec{x}+\vec{\tau}_1^{\sigma\sigma'})$ is generally not small. 

The periodicity conditions motivate defining the relaxation-dependent interlayer hopping function as
\begin{equation}\begin{aligned}
  h^{\sigma\sigma'}_{\bvec{U}}(\bvec{x}) := h_{12}^{\sigma\sigma'}(\bvec{x} + \bvec{U}_1(-\bvec{x} + \vec{\tau}_2^{\sigma\sigma'}) - \bvec{U}_2(- \bvec{x}+\vec{\tau}_1^{\sigma\sigma'}) ),
\end{aligned}\end{equation}
 so that Eq.~(\ref{eq:tb_interlayer}) can be expressed as
\begin{equation}
  h^{\sigma\sigma'}_{\bvec{U}}(\bvec{R}_{1,\sigma} - \bvec{R}_{2,\sigma'}).
\end{equation}
The argument $\bvec{R}_{1,\sigma} - \bvec{R}_{2,\sigma'}$ contains a separation between atomistic scale and moir\'e scale. To make $\eps$-dependence explicit, we denote $\widetilde{\bvec{R}}_{2,\sigma}  = A_2 A_1^{-1} \bvec{R}_{1,\sigma}$, and from the definition of disregistry function
 \begin{equation}
 \begin{aligned}
 \bvec{R}_{1,\sigma} - \bvec{R}_{2,\sigma'} = - \gamma_2  \bvec{R}_{1,\sigma} +\widetilde{\bvec{R}}_{2,\sigma} -\bvec{R}_{2,\sigma'} ,
 \end{aligned}
\end{equation}
so that we can write the interlayer hopping term as
\begin{equation}\label{eq:tb_interlayer_simplified}
  \begin{aligned}
    &h^{\sigma\sigma'}_{\bvec{U}}(- \gamma_2  \bvec{R}_{1,\sigma} +\widetilde{\bvec{R}}_{2,\sigma} -\bvec{R}_{2,\sigma'}) \\
    &= h^{\sigma\sigma'}_{\bvec{U}}(- \tilde \gamma_2 \eps \bvec{R}_{1,\sigma} +\widetilde{\bvec{R}}_{2,\sigma} -\bvec{R}_{2,\sigma'}).
  \end{aligned}
\end{equation}

Eqs.~(\ref{eq:tb_intralayer}) and (\ref{eq:tb_interlayer_simplified}) are the relaxation contributions to the tight-binding model in Lagrangian coordinates. The model has explicit multi-scale lengths with fast variables $\bvec{R}_{\ell,\sigma}$ on the atomistic scale in a convolution form, and slow variables 
$\eps \bvec{R}_{\ell,\sigma}$ varying at the moir\'e scale.

\subsection{Relaxed BM Hamiltonian}

The intralayer tunneling Eq.~(\ref{eq:tb_intralayer}) can be expanded in orders of $\eps$, the ratio between atomistic and moir\'e length. We also note that in TBG, the graphene layers are weakly interacting, and the ratio of interlayer and intralayer energy is also around $\eps$ \cite{Bistritzer_MacDonald_2011}.  So we further scale the interlayer hopping function $h_{\bvec{U}} \to \eps h_{\bvec{U}}, $ 
and expand the tight-binding Hamiltonian Eq.~(\ref{eq:H_tb}) in orders of $\eps$,
\begin{equation}\label{eq:H_ms}
	H_{\text{TB,relax}} \approx H_{\text{intra}}^{(0)} + \eps H_{\text{intra}}^{(1)} + \eps H_{\text{inter}}^{(0)} + \mathcal O(\eps^2).
\end{equation}
We will neglect the $\mathcal O(\eps^2)$ error term for the rest of the paper.

We introduce the following wave packet ansatz around the monolayer Dirac points. Let $\psi^\eps$ and $\varphi^\eps$ be continuous field operators that vary appreciably only at the moir\'e scale. Evaluating them at discrete lattice sites recovers the original operators by the projection onto low energy states.
\begin{equation}\label{eq:wavepacket_ansatz}
\begin{aligned}
		\frac{1}{|\Omega|^{\frac{1}{2}}}c_{\bvec{R},\ell,\sigma} \approx &  e^{i\bvec{K}_\ell \cdot \bvec{R}_{\ell,\sigma} } \psi_{\ell,\sigma}^\eps( \bvec{R}_{\ell,\sigma})  +  e^{i\bvec{K}_\ell' \cdot \bvec{R}_{\ell,\sigma}} \varphi^\eps_{\ell,\sigma}(\bvec{R}_{\ell,\sigma}),
	\end{aligned}
      \end{equation}
where $\psi^\eps$ and $\varphi^\eps$ satisfy the canonical anticommutation relations \begin{equation}
	\{ (\psi^\eps_{\ell,\sigma})^\dagger(\bvec{r}), \psi^\eps_{ \ell', \sigma'}(\bvec{r}') \} = \delta(\bvec{r} - \bvec{r}') \delta_{\ell,\ell'}\delta_{\sigma,\sigma'}.
\end{equation}
The moir\'e scale of the field operators $\psi^\eps, \varphi^\eps$ can be expressed explicitly by the following scaling relations on an $\eps$-independent $\psi,\varphi$
\begin{equation}\label{eq:wavepacket_scaling}
	\psi_{\ell,\sigma}^\eps (\bvec{r}) : = \eps \psi_{\ell,\sigma}( \eps \bvec{r}), \quad \varphi_{\ell,\sigma}^\eps (\bvec{r}) : = \eps \varphi_{\ell,\sigma}( \eps \bvec{r}).
\end{equation}

We then expand the annihilation operator in orders of $\eps$ so that it is also evaluated at the same point as the creation operators
\begin{equation}\begin{aligned}\label{eq:wavefunction_ms}
	\psi_{\ell,{\sigma'}}(\eps \bvec{r}')   = & \psi_{\ell,{\sigma'}}(\eps \bvec{r})   \\
	& - \eps \left(\bvec{r} - \bvec{r}'\right) ^\top \nabla \psi_{\ell,\sigma'}(\eps \bvec{r})  \\
	&  - \eps^2 \left(\bvec{r} - \bvec{r}'\right) ^\top \nabla^2 \psi_{\ell,\sigma'}(\eps \bvec{r})    \left(\bvec{r} - \bvec{r}' \right)  \\
	&  + \mathcal O(\eps^3),
\end{aligned}\end{equation}
where $\bvec{r} = \bvec{R}_{\ell,\sigma}$ and $\bvec{r}' = \bvec{R}'_{\ell,\sigma'}$ are the atomistic positions.
We then set the slow variable to be $\bvec{X} =   \eps \bvec{R}_{\ell,\sigma}$, and the expansion allows us to approximate $\psi$ by only evaluating $\psi$, $\nabla \psi$, $\nabla^2\psi$ at $\bvec{X}$. We then use a matched asymptotic expansion by plugging Eq.~(\ref{eq:wavefunction_ms}) into 
 Eq.~(\ref{eq:H_ms}), and matching all the terms with the same order of $\eps$.

 The distance between $\bvec{K}$ and $\bvec{K}'$ valleys are much larger than the moir\'e scale $\eps$, so we neglect the terms which mix different valleys (sometimes referred to as the intervalley-Hunds terms).
Expanding up to $\mathcal O(\eps^2)$, the $\bvec{K}$ valley relaxed continuum model then has the expansion  (see Appendix \ref{sec:multiscale_derivation} for details) 
\begin{equation}
	\mathsf H_{\text{eff}} \approx \mu_\mathrm{F} I + \eps \mathsf H^{(1)}_{\text{eff}} + \eps^2 \mathsf H^{(2)}_{\text{eff}}.
\end{equation}
We will show that $\mathsf H^{(1)}_{\text{eff}}$ is a BM-like term, where the relaxation effects introduce an intralayer pseudomagnetic and pseudoelectric field, and alters the interlayer potential. The $\mathsf H^{(2)}_{\text{eff}}$ term contains second-order corrections to the Dirac cones, as well as  non-local approximation (denoted nl) for the intralayer and interlayer moir\'e potentials. The non-local approximation allows for tunneling that depend on the crystal momentum $\bvec{k} - \bvec{K}_\ell$ (represented by $-i\nabla_{\bvec{r}}$ in real space), and is the main source of particle-hole asymmetry in the single-particle band structure. 

Transforming  $ \eps \mathsf H^{(1)}_{\text{eff}} + \eps^2 \mathsf H^{(2)}_{\text{eff}} $ back to the atomistic basis  $(\psi^\eps_{1,\A}, \psi^\eps_{1,\B}, \psi^\eps_{2,\A}, \psi^\eps_{2,\B})^\top$, we get
\begin{widetext}
  \begin{equation}\label{eq:H_relax}
    H_{\text{relax}} (\bvec{r}) = 
    \begin{pmatrix} 
      D_1(-i \nabla_{\bvec{r}}) + S_1(\bvec{r}) + S_1^\mathrm{nl}(\bvec{r}) \cdot (-i \nabla_{\bvec{r}} )  &	 \tilde T(\bvec{r}) + \tilde{T}^\mathrm{nl}(\bvec{r}) \cdot (-i \nabla_{\bvec{r}} )  \\  
      \left[ \tilde T(\bvec{r}) + \tilde{T}^\mathrm{nl}(\bvec{r}) \cdot (-i \nabla_{\bvec{r}} ) \right]^\dagger & D_2(-i \nabla_{\bvec{r}}) + S_2(\bvec{r}) + S_2^\mathrm{nl}(\bvec{r}) \cdot (-i \nabla_{\bvec{r}}) 
    \end{pmatrix}.
  \end{equation}
\end{widetext}
 We briefly describe each of the terms appearing in~\cref{eq:H_relax} using hopping functions $h_{\ell\ell'}$, disregistries $\gamma_\ell$ and disregistry-space displacements $\bvec{U}_\ell$  (see~\cref{sec:multiscale_derivation} for complete details). The intralayer terms include the monolayer Dirac operator with second order and linearized rotation corrections
\begin{equation}\label{eq:dirac_2nd_order}
\begin{aligned}
  D_{\ell}(\bvec{k}) :=   & \hbar v_F e^{i\sigma_3\theta_\ell} \cdot \vec{\sigma} \cdot \bvec{k}  \\
                      & + v_1 \sigma_0 \bvec{k}^2  \\
                      & + v_2  e^{-2i\sigma_3\theta_\ell}[\sigma_1 (-k_1^2 + k_2^2) + 2\sigma_2  k_1 k_2].
\end{aligned}
\end{equation}
The parameters depend only on the intralayer hopping function $h_{\ell\ell}$. Their values are  $\hbar v_F = 5.339 \text{ eV}\cdot\text{\r{A}}$, $v_1 = -0.783 \text{ eV}\cdot\text{\r{A}}^2$, and $v_2 =  -3.405 \text{ eV}\cdot\text{\r{A}}^2$. 

To simplify the expressions for intralayer and interlayer tunneling terms, we define the sublattice-shifted lattices as $\mathcal R_\ell^{\sigma\sigma'} := \mathcal R_\ell + \vec{\tau}_\ell^{\sigma\sigma'}$. The intralayer potential $S_\ell$ and $S_\ell^{\mathrm{nl}}$ terms are
 \begin{equation}
 	 \begin{aligned}
 	 &\left[S_\ell(\bvec{r})\right]^{\sigma\sigma'} \\
     & := \sum_{{\bvec{Q}}_\ell \in \mathcal R_\ell^{\sigma\sigma'}} \left[\nabla h_{\ell\ell}^{\sigma\sigma'} \left(\bvec{Q}_\ell \right)\right]^\top \nabla  \bvec{U}_{\ell}\left( \gamma_{3-\ell} \bvec{r} \right) \gamma_{3-\ell} \\
    & \qquad \bvec{Q}_\ell	 e^{-i\bvec{K}_\ell \cdot \bvec{Q}_\ell},
 \end{aligned}
\end{equation}
and 
\begin{equation}
\begin{aligned}
& \left[ S^{\mathrm{nl}}_\ell (\bvec{r}) \right]^{\sigma\sigma'} \\
& := -i \sum_{{\bvec{Q}}_\ell \in \mathcal R_\ell^{\sigma\sigma'}}   \left[\nabla h_{\ell\ell}^{\sigma\sigma'} \left(\bvec{Q}_\ell \right)\right]^\top \nabla  \bvec{U}_{\ell}\left( \gamma_{3-\ell} \bvec{r} \right) \gamma_{3-\ell} \\
     & \qquad  \left[  \bvec{Q}_\ell  \bvec{Q}_\ell^\top\right]	 e^{-i\bvec{K}_\ell \cdot  \bvec{Q}_\ell}.
 \end{aligned}
\end{equation}
They represent the local and non-local component of the pseudo-magnetic field generated by relaxation effects that break the graphene hexagonal periodicity ($\mathcal O(\eps)$ terms in Eq.~(\ref{eq:tb_intralayer})). The potentials depend on the Jacobian of $\bvec{U}_j$, as well as the gradient of the intralayer hopping function $h_{\ell\ell}$ evaluated at lattice sites. In the absence of relaxation, $\bvec{U}_j = 0$, and the intralayer potentials \(S_{\ell}\), \(S_{\ell}^{\mathrm{nl}}\) vanish.

From the periodicity of $\bvec{U}_\ell$, the intralayer potentials are periodic with respect to the moir\'e lattices. They can be decomposed into their Bloch coefficients
\begin{equation}
  S_\ell(\bvec{r}) = \sum_{\bvec{G} \in \mathcal R_\m^*}e^{-i \bvec{G} \cdot \bvec{r}} S_{\ell,\bvec{G}}, \quad S^{\mathrm{nl}}_\ell(\bvec{r}, \bvec{k}) = \sum_{\bvec{G} \in \mathcal R_\m^*} e^{-i \bvec{G} \cdot \bvec{r}} S^{\mathrm{nl}}_{\ell,\bvec{G}}.
\end{equation}

The interlayer potential $\tilde T$ and the $\bvec{k}-$dependent $\tilde T^\mathrm{nl}$ are defined as
  \begin{equation}
  	\left[ \tilde {T} (\bvec{r}) \right]^{\sigma\sigma'} := \sum_{\bvec{Q}_2 \in \mathcal R_2^{\sigma\sigma'}} h^{\sigma\sigma'}_{\bvec{U}} \left(-\gamma_2 \bvec{r} + \bvec{Q}_2  \right)  e^{-i\bvec{K}_2 \cdot \bvec{Q}_2 },
  \end{equation}
  and 
  \begin{equation}
\begin{aligned}
	  \left[ \tilde{T}^{\mathrm{nl}} (\bvec{r}) \right]^{\sigma\sigma'} := -i& \sum_{\bvec{Q}_2 \in \mathcal R_2^{\sigma\sigma'}} h^{\sigma\sigma'}_{\bvec{U}} \left(-\gamma_2 \bvec{r} + \bvec{Q}_2  \right) \\
	 & \quad \left(-\gamma_2 \bvec{r} + \bvec{Q}_2 \right)^\top e^{-i\bvec{K}_2 \cdot \bvec{Q}_2 } .
\end{aligned}
\end{equation}
 They are also periodic (up to a phase shift of $e^{-i\bvec{s} \cdot \bvec{r}}$) with respect to the moir\'e lattices, so that they can also be decomposed into their Bloch coefficients 
\begin{equation}\label{eq:T_relax}
\begin{split}
  \tilde T(\bvec{r}) & = \sum_{\bvec{G} \in \mathcal R_\m^*} e^{-i (\bvec{G} + \bvec{s}) \cdot \bvec{r}} \tilde T_{\bvec{G}}, \\
  \tilde T^{\mathrm{nl}}(\bvec{r}) & = \sum_{\bvec{G} \in \mathcal R_\m^*} e^{-i (\bvec{G} + \bvec{s}) \cdot \bvec{r}} \tilde T^{\mathrm{nl}}_{\bvec{G}}.
\end{split}
\end{equation}

From the properties of the interlayer tunneling terms $\tilde{T}$ and $\tilde{T}^{\mathrm{nl}}$, translation of the relaxed Hamiltonian Eq.~(\ref{eq:H_relax}) by a moir\'e lattice vector $\bvec{R}_\m \in \mathcal R_\m$ introduces an extra phase $e^{-i\bvec{s} \cdot \bvec{R}_\m}$. To cancel out that phase, we conjugate $H_{\mathrm{relax}}$ with a unitary operator $\mathcal U := \operatorname{diag}(1, 1, e^{-i\bvec{s} \cdot \bvec{r}}, e^{-i\bvec{s} \cdot \bvec{r}})$, so that $\mathcal U^\dagger H_{\mathrm{relax}}  \mathcal U$ is \emph{exactly} moir\'e-periodic.

The energy contributions of each term in the Relaxed BM model are listed in Tables~\ref{tab:relax_intra_params} and \ref{tab:relax_inter_params}. We keep both intralayer and interlayer Bloch coefficients to the third shell. Non-local terms are estimated by multiplying by a typical length scale of the moir\'e reciprocal momentum $ |\bvec{s}| \approx 0.033 \text{\r{A}}^{-1}$. We note that the largest intralayer energy contribution comes from the monolayer Dirac operator, which has a typical energy of $v_F |\bvec{s}| \approx 176 \text{ meV}$.  This is comparable with the interlayer energy $| \tilde T_1^{\A\B}| \approx 113.3 \text{ meV}$, so that the wave packet ansatz Eq.~(\ref{eq:wavepacket_ansatz}) accurately captures the weak coupling between layers in TBG. We also conclude that on the first shell, $| \tilde T_1^{\A\A}| / | \tilde T_1^{\A\B}| \approx 0.7$. We set this value to be $\kappa$ for the AA/AB ratio in the original BM model.

\begin{table}
  \caption{\label{tab:relax_intra_params} Intralayer energies for relaxed BM model in units of meV}
  \begin{ruledtabular}
    \begin{tabular}{lcccc}
      & \multicolumn{2}{c}{\textbf{Local}} & \multicolumn{2}{c}{\textbf{Non-local}} \\
      \cmidrule(lr){2-3} \cmidrule(lr){4-5}
      Shell & AA & AB & AA & AB \\
      \midrule
      1 & 0.1 & 17.3 & 0.0 & 0.6 \\
      2 & 0.0 & 0.3 & 0.0 & 0.0 \\
      3 & 0.0 & 1.9 & 0.0 & 0.1 \\
    \end{tabular}
  \end{ruledtabular}
\end{table}

\begin{table}
  \caption{\label{tab:relax_inter_params} Interlayer energies for relaxed BM model  in units of meV}
  \begin{ruledtabular}
    \begin{tabular}{lcccc}
      & \multicolumn{2}{c}{\textbf{Local}} & \multicolumn{2}{c}{\textbf{Non-local}} \\
      \cmidrule(lr){2-3} \cmidrule(lr){4-5}
      Shell & AA & AB & AA & AB \\
      \midrule
      1 & 78.6 & 113.3 & 7.1 & 7.6\\
      2 & 0.4 & 3.4 & 0.3  & 0.3\\
      3 & 11.4 & 11.5 & 0.6 & 0.8\\
    \end{tabular}
  \end{ruledtabular}
\end{table}

\begin{figure}[t]
 \includegraphics[width=.4\linewidth]{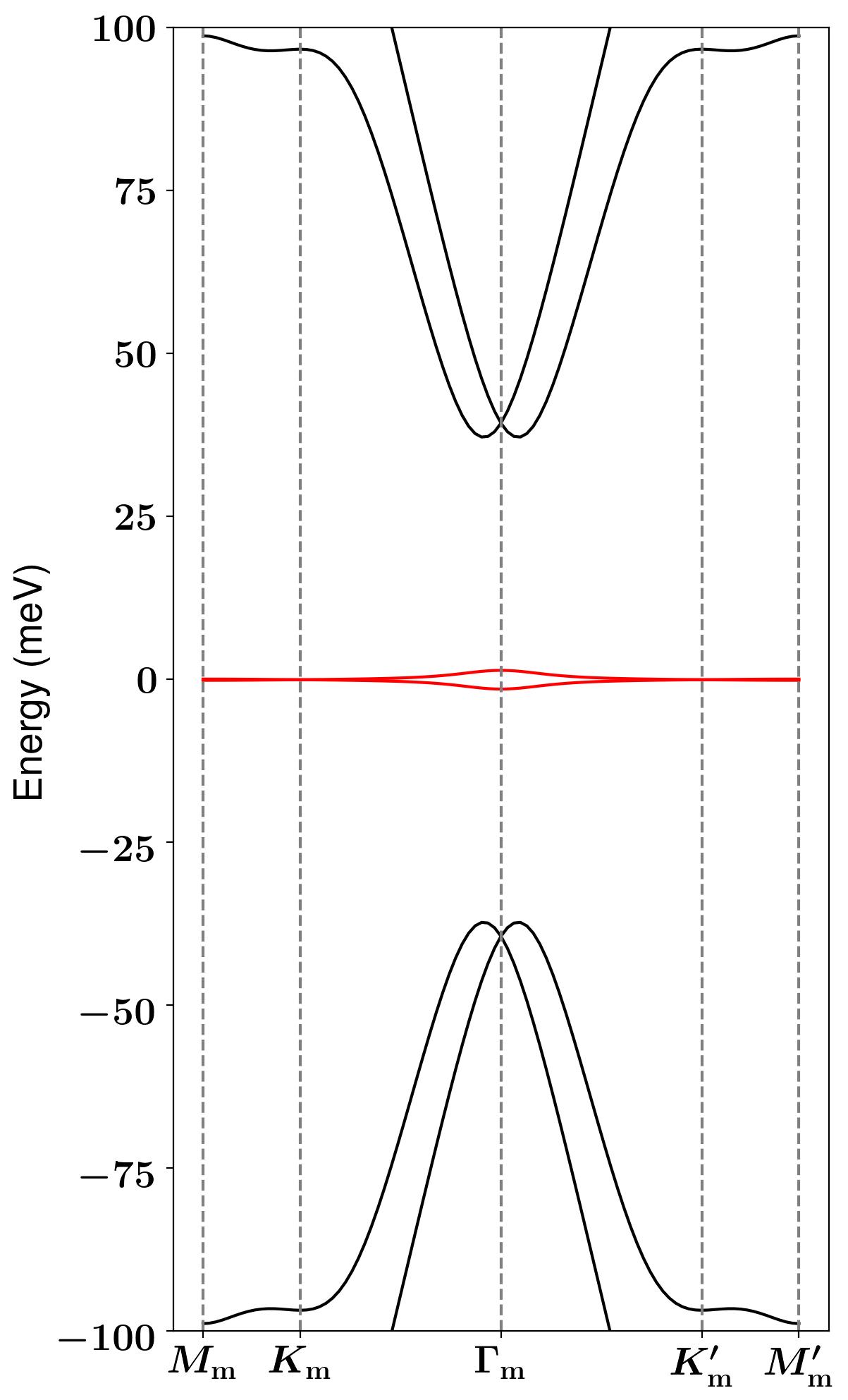}
  \includegraphics[width=.4\linewidth]{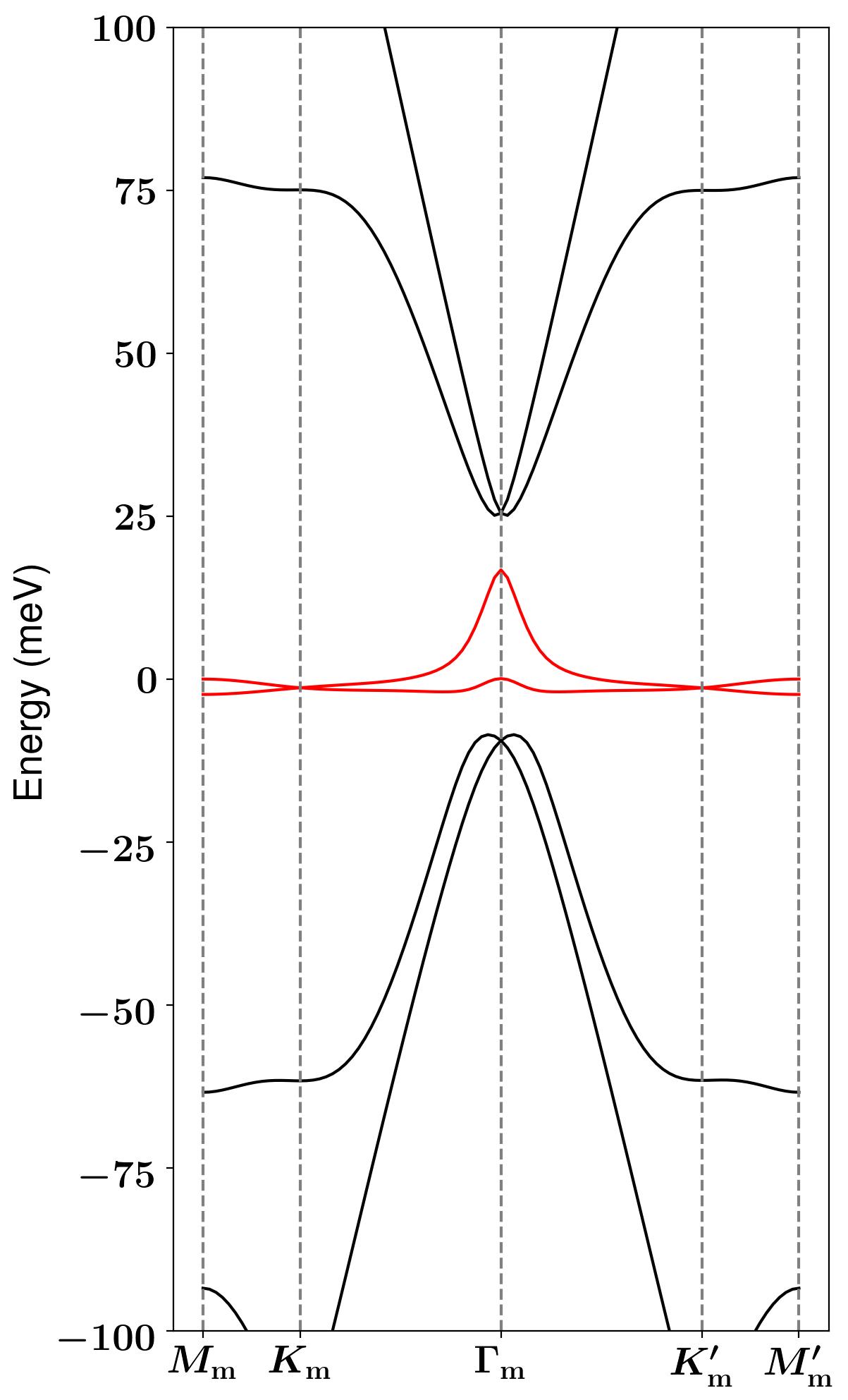}	
  \caption{Single particle band structure for BM model with $\kappa = 0.7$ (left) and relaxed model (right). The flat bands near Fermi level are highlighted in red.}
   \label{fig:non_interacting_bands}
\end{figure}

The band structure for the BM model ($\kappa=0.7$) and relaxed model along the high-symmetry lines are shown in ~\cref{fig:non_interacting_bands}. Both models have separated ``flat bands" near the Fermi level, and the interactions will be projected onto these bands. The relaxed Hamiltonian preserves \(C_{2z}\) and \(\mathcal{T}\) symmetries, but breaks the particle-hole symmetry. It also has larger single-particle dispersion ($\sim 20 \text{ meV}$) in the flat bands compared to the BM model ($\sim 5\text{ meV}$).

Finally we note that the derivation is in Lagrangian coordinates: creation operators $c^\dagger_{\bvec{R}, \ell, \sigma}$ creates an electron localized at lattice site physically located at $\bvec{R}_\ell + \vec{\tau}_\ell^\sigma + \bvec{u}_\ell(\bvec{R}_\ell + \vec{\tau}_\ell^\sigma)$, which can be described by Eulerian coordinates. We may therefore define a mapping $\bvec{Y}_\ell$ from Lagrangian to Eulerian coordinates, and a corresponding Eulerian coordinate wave function $\Psi_\ell$ by 
\begin{equation}
	\bvec{Y}_\ell (\bvec{r}) := \bvec{r} + \bvec{u}_\ell(\bvec{r}), \quad  \Psi_\ell^\eps (\bvec{r}) := \psi_\ell^\eps \left(\bvec{Y}_\ell^{-1}(\bvec{r})\right).
\end{equation}
In many-body calculations, this mapping will affect the values of form factors (see Eq.~\ref{eq:form_factor}). For realistic values of TBG at magic angle, $| \det \nabla \bvec{u}_\ell | \approx 10^{-5}$ so we neglect this factor in this study. 

\section{Interacting Models for Twisted Bilayer Graphene}
\label{sec:inter-models-twist}
In this section, we review the general framework for constructing many-body models for studying electron-electron interactions in twisted bilayer graphene.
Let \(H\) be a single particle Hamiltonian at valley \(\bvec{K}_{\m}\), such as derived in the previous section, which is periodic with respect to the moir{\'e} lattice \(\mathcal{R}_{\mathrm{m}}\).
By Bloch's theorem, there exist Bloch eigenpairs \((\lambda_{n, \bvec{k}}, \psi_{n, \bvec{k}})\) so that for all \(\bvec{R} \in \mathcal{R}_{\m}\) and \(\bvec{G} \in \mathcal{R}_{\m}^{*}\) the following holds:
\begin{equation}
  \begin{aligned}
    H \psi_{n, \bvec{k}}(\bvec{r}) & = \lambda_{n, \bvec{k}} \psi_{n, \bvec{k}}(\bvec{r}), \\[.5ex]
    \psi_{n, \bvec{k}}(\bvec{r} + \bvec{R}) & = e^{i \bvec{k} \cdot \bvec{R}} \psi_{n, \bvec{k}}(\bvec{r}), \\[.5ex]
    \psi_{n, (\bvec{k} + \bvec{G})}(\bvec{r}) & = \psi_{n, \bvec{k}}(\bvec{r}).
  \end{aligned}
\end{equation}
As \(H\) was derived by expanding near \(\bvec{K}_{\m}\), we will denote the Bloch eigenpairs \((\lambda_{n \bvec{k}}, \psi_{n \bvec{k}})\) by \((\lambda_{(n, \bvec{K}_{m}), \bvec{k}}, \psi_{(n, \bvec{K}_{\m}), \bvec{k}})\).
When \(H\) commutes with time reversal symmetry, we can determine a basis of Bloch functions at a second valley \(\bvec{K}_{\m}'=-\bvec{K}_{\m}\) by the following relations
\begin{equation}
  \begin{split}
    \psi_{(n, -\bvec{K}_{\m}), \bvec{k}}(\bvec{r}) & := \psi_{(n, \bvec{K}_{\m}), -\bvec{k}}^{*}(\bvec{r}) \\
    \lambda_{(n, -\bvec{K}_{\m}), \bvec{k}} & = \lambda_{(n, \bvec{K}_{\m}), -\bvec{k}}.
  \end{split}
\end{equation}
Hence, the following set of functions forms a complete orthogonal basis for the valleyful model of TBG:
\begin{equation}
  \Big\{ \psi_{(n, \tau), \bvec{k}}(\bvec{r}) : n \in \mathbb{Z}, \tau \in \{ \bvec{K}_{\m}, \bvec{K}_{\m}' \} \Big\}.
\end{equation}
Henceforth, we will use the Greek letter subscripts to denote a composite index of the band and valley flavors (e.g.  \(\alpha := (n, \tau)\)).

Since graphene has minimal spin-orbit coupling \cite{HuertasHernandoGuineaBrataas2006,AndreiMacDonald2020}, we can define a spinful model for TBG by simply adding a spin flavor to the model.
Given the valleyful Bloch functions \(\psi_{\alpha \bvec{k}}\), we define creation operators \(f_{(\alpha, s) \bvec{k}}^{\dagger}\) which creates the state \(\psi_{\alpha \bvec{k}}\) in spin \(s\) (and similarly for the annihilation operator \(f_{(\alpha, s), \bvec{k}}\)).
This family of creation and annihilation operators satisfy the canonical anticommutation relations
\begin{equation}
  \begin{aligned}
    \{ f_{(\alpha, s) \bvec{k}}^{\dagger}, f_{(\beta, s') \bvec{k}'}^{\dagger}  \} = \{ f_{(\alpha, s) \bvec{k}}, f_{(\beta, s') \bvec{k}'} \} = 0 \\[1ex]
    \{ f_{(\alpha, s) \bvec{k}}^{\dagger}, f_{(\beta, s') \bvec{k}'}  \} = \delta_{\alpha \beta} \delta_{s s'} \delta_{\bvec{k} - \bvec{k}' \in \mathcal{R}_{\m}^{*}}.
  \end{aligned}
\end{equation}
Next, we fix a discrete momentum grid \(\mathcal{K} \subseteq \Gamma_{\m}^{*}\) with \(N_{\mathcal{K}}\) points and a set of active bands \(\mathcal{N} \subseteq \mathbb{Z}\).
The corresponding interacting model projected to the bands \(\mathcal{N}\) for TBG is written as
\begin{equation}
  \label{eq:h-tbg}
  H_{\mathrm{TBG}} = H_{0} - H_{\mathrm{sub}} + H_{I} 
\end{equation}
where \(H_{0}\) is the projection of the single particle Hamiltonian \(H\) to the bands \(\mathcal{N}\), \(H_{\mathrm{sub}}\) is the double counting subtraction, and \(H_{I}\) is the interacting term.
Both \(H_{0}\) and \(H_{\mathrm{sub}}\) are single particle operators which can be written:
\begin{align}
    H_{0}
    & := \sum_{\bvec{k} \in \mathcal{K}}^{} \sum_{s \in \{ \uparrow, \downarrow \}}^{} \sum_{\alpha}^{} \lambda_{\alpha \bvec{k}} f_{(\alpha, s),\bvec{k}}^{\dagger} f_{(\alpha, s), \bvec{k}}, \\
    H_{\mathrm{sub}}
    & := \sum_{\bvec{k} \in \mathcal{K}}^{} \sum_{s \in \{ \uparrow, \downarrow \}}^{} \sum_{\alpha, \beta}^{} [ H_{\mathrm{sub}}(\bvec{k}) ]_{\alpha, \beta} f_{(\alpha, s) \bvec{k}}^{\dagger} f_{(\beta, s) \bvec{k}}.
\end{align}
Here, the sums over \(\alpha, \beta = (n, \tau)\) run over both the set of active bands \(n \in \mathcal{N}\) and valley \(\tau \in \{ \bvec{K}_{\m}, -\bvec{K}_{\m} \}\).
The double counting subtraction term \(H_{\mathrm{sub}}\) is used to correct for the fact that the original tight binding model for \(H\) implicitly includes part of the electron-electron interactions.
We will discuss the double counting subtraction Hamiltonian in more detail in \cref{sec:double-count-subtr}, however, for the moment we will let \(H_{\mathrm{sub}}(\bvec{k})\) be an arbitrary \(\bvec{k}\)-resolved, single particle Hamiltonian.

The interacting term, \(H_{I}\), can be written in terms of the form factor \(\Lambda_{\bvec{k}}(\bvec{q}')\) which is an \(\# \mathcal{N} \times \# \mathcal{N}\) matrix whose entries can be calculated as follows:
\begin{equation}\label{eq:form_factor}
  [\Lambda_{\bvec{k}}(\bvec{q}')]_{\alpha,\beta} := \int_{\Omega_{\m}}^{} e^{-i \bvec{q}' \cdot \bvec{r}} \psi_{\alpha \bvec{k}}^{*}(\bvec{r}) \psi_{\beta (\bvec{k} + \bvec{q}')}(\bvec{r}) \dee \bvec{r}.
\end{equation}
Due to the large distance between the two valleys, we take \([\Lambda_{\bvec{k}}(\bvec{q}')]_{(n, \tau),(n', \tau')} = 0\) unless \(\tau = \tau'\).

Given the form factor, we define the many-body operator \(\rho(\bvec{q}')\)
\begin{equation}
  \rho(\bvec{q}') := \sum_{\bvec{k} \in \mathcal{K}}^{} \sum_{s \in \{ \uparrow, \downarrow \}}^{}  \sum_{\alpha, \beta}^{} [\Lambda_{\bvec{k}}(\bvec{q}')]_{\alpha,\beta}  f_{(\alpha, s) \bvec{k}}^{\dagger}  f_{(\beta, s), (\bvec{k} + \bvec{q}')},
\end{equation}
so that \(H_{I}\) is written
\begin{equation}
  H_{I} = \frac{1}{2 N_{\mathcal{K}} | \Omega_{\m}|} \sum_{\bvec{q}' \in \mathcal{K} + \mathcal{R}_{\mathrm{m}}^{*}}^{} V(\bvec{q}')  : \rho(-\bvec{q}') \rho(\bvec{q}') :
\end{equation}
where \(: \cdot :\) denotes normal ordering and \(V(\bvec{q}')\) denotes the (screened) Coulomb interaction in momentum space; see \cref{eq:1}.

\subsection{Hartree-Fock Theory for TBG}
We briefly recall some of the important facts of Hartree-Fock theory as they relate to our later discussions and analysis.
For a given set of momentum, \(\mathcal{K}\), a translation-invariant Hartree-Fock state, \(\ket{\Psi}\), is completely described by its one-body reduced density matrix (1-RDM) \(P\) where
\begin{equation}
  \begin{split}
    P & := \bigoplus_{\bvec{k} \in \mathcal{K}} P(\bvec{k}) ,\\
    [P(\bvec{k})]_{(\alpha, s), (\beta, s')} & := \braket{\Psi | f_{(\beta, s') \bvec{k}}^{\dagger} f_{(\alpha, s) \bvec{k}} | \Psi}.
  \end{split}
\end{equation}
In this case, the 1-RDM is an orthogonal projection and the total number of electrons in state \(\ket{\Psi}\) equals \(\tr(P)\).
We can write the energy of any Hartree-Fock state as
\begin{equation}
  \label{eq:hf-energy}
  \begin{split}
    & \braket{\Psi | H_{\mathrm{TBG}} | \Psi } \\
    &  = \tr{ \Big( (H_{0} - H_{\mathrm{sub}}) P  \Big)} + \frac{1}{2} \tr{\Big( (J[P] - K[P]) P  \Big)}
  \end{split}
\end{equation}
where \(J[P]\) and \(K[P]\) are single body operators referred to as the ``Hartree'' and ``exchange'' potentials respectively; see~\cref{sec:review-hartree-fock} for explicit formulas in terms of the form factor and \(V(\mathbf{q}')\).

Given a target number of electrons, \(N_{e}\), solving the Hartree-Fock equations for \(H_{\mathrm{TBG}}\) involves finding a 1-RDM \(P\) which minimizes the Hartree-Fock energy (\cref{eq:hf-energy}) subject to the particle number constraint \(\tr(P) = N_{e}\).
The first variation of this objective is given by the matrix
\begin{equation}
  \label{eq:fock-matrix}
  F[P] := H_{0} - H_{\mathrm{sub}} + J[P] - K[P] 
\end{equation}
which is referred to as the Fock matrix.
For translation invariant 1-RDMs, the Fock matrix can also be decomposed as a direct sum \(\bigoplus_{k \in \mathcal{K}} F[P](\bvec{k})\).
The eigenvalues of the map \(\bvec{k} \mapsto F[P](\bvec{k})\) then defines the Hartree-Fock band structure for \(P\).

\subsection{Double Counting Subtraction}
\label{sec:double-count-subtr}
The double counting subtraction term, \(H_{\mathrm{sub}}\), plays a critical role in the properties of of the many-body ground states of TBG \cite{FaulstichStubbsZhuEtAl2023,Kim2026InitioQuantumEmbedding,VafekKang2020}.
A common choice of double counting subtraction in the TBG is to use a ``mean-field'' subtraction based on Hartree-Fock theory:
\begin{equation}
  H_{\mathrm{sub}} = J[P_{\mathrm{sub}}] - K[P_{\mathrm{sub}}].
\end{equation}
Here, \(J[P_{\mathrm{sub}}]\) and \(K[P_{\mathrm{sub}}]\) are the Hartree and exchange potentials induced by the 1-RDM \(P_{\mathrm{sub}}\).
The 1-RDM \(P_{\mathrm{sub}}\) is typically taken to be a fixed reference density at half-filling.

In this work, we generally choose \(P_{\mathrm{sub}}\) to project onto the lower energy half of the single-particle flat bands, with Fermi-Dirac smoothing (for details, see Appendix \ref{app:subtraction}). We refer to this as ``charge neutral'' subtraction.
Our main results do not change significantly when we replace this subtraction with the ``average subtraction'' scheme where \(P_{\mathrm{sub}} = \frac{1}{2} I\); see \cite{supplementary}.

\subsection{Exact Ground States at the Chiral Limit}
Previous theoretical work on the chiral limit of TBG has found that at integer fillings, the set of many-body ground states are massively degenerate \cite{BultinckKhalafLiuEtAl2020,BernevigSongRegnaultEtAl2021,StubbsRagoneMacDonaldEtAl2025}.
In more physically realistic models of TBG, such as the Bistritzer-MacDonald or heavy fermion model, this degeneracy is broken, however many of the previously degenerate states remain close in energy and can be distinguished by symmetry.
To fully explore the different symmetry sectors under perturbation, we initialize our calculations in the standard symmetric ansatzes: valley polarized (VP), valley Hall (VH), quantum Hall (QH), Kramers intervalley coherent (KIVC), and time-reversal symmetric intervalley coherent (TIVC).
We additionally initialize inter-orbital inter-valley coherent (IOVC) identified in recent work as a competing ground state when relaxation and remote bands are included \cite{HouSurWagnerEtAl2025};
see \cref{sec:init-symm} for a more complete discussion.

\section{Numerical Results for the Relaxed Interacting Model}
\label{sec:inter-model-results}

To understand the impact of relaxation on the corresponding interacting model, we compare Hartree-Fock results between the Bistritzer-MacDonald model with \(\kappa = 0.7\) and the relaxed model outlined in the previous sections and three different filling factors \(\nu = 0\) (charge neutrality), \(\nu = -2\) (hole doped), and \(\nu = + 2\) (particle doped).
To isolate the effects which are purely due to relaxation, we focus on \(H_{\mathrm{TBG}}\) projected to the flat bands leaving the impact of remote bands to future work.
We perform all our calculations on a \(12 \times 12\) Monkhorst-Pack grid centered at the moir{\'e} gamma point \(\vec{\Gamma}_{\m}\).
Additionally, we use the double-gate screened Coulomb potential
\begin{equation} \label{eq:1}
V(\bvec{q}') := \frac{2 \pi}{\epsilon} \frac{\tanh{( |\bvec{q}'| d / 2)}}{|\bvec{q}'|}
\end{equation}
with relative permittivity \(\epsilon = 10.79\) and gate distance \(d = 30\, \mathrm{nm}\).

\begin{figure}[t]
  \includegraphics[width=.8\linewidth]{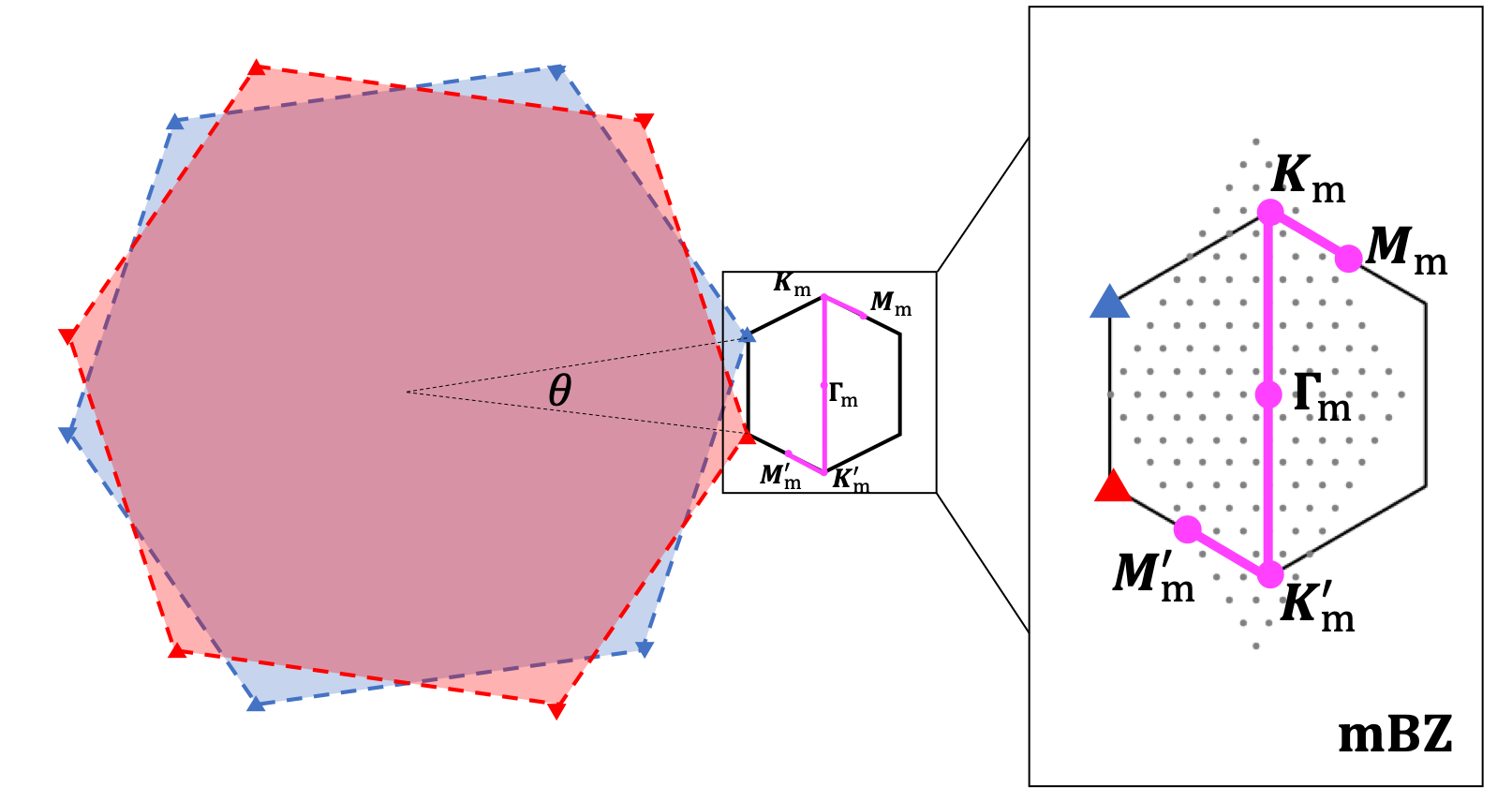}
  \caption{Plot of the monolayer Brillouin zones (red and blue), moir{\'e} Brillouin zone and the high symmetry line, and the uniform grid $\mathcal K$ for interacting calculations.}
  \label{fig:bz-high-symmetry}
\end{figure}

In our numerical tests, we pay particular attention to two related classes of many-body states: those which fully occupy one of the two spins and those which leave one of the spins fully empty.
Away from charge neutrality, these states typically have lower energies compared to other states. 

In addition to reporting the energy orderings at the even integer fillings, we plot the Hartree-Fock band structure.
For our plots of the Hartree-Fock band structure, we smoothly interpolate the Hartree-Fock band as described in \cref{sec:single-shot}.

\subsection{Hartree-Fock Results}
\begin{figure*}[t]
  \centering
  \includegraphics[width=.9\linewidth]{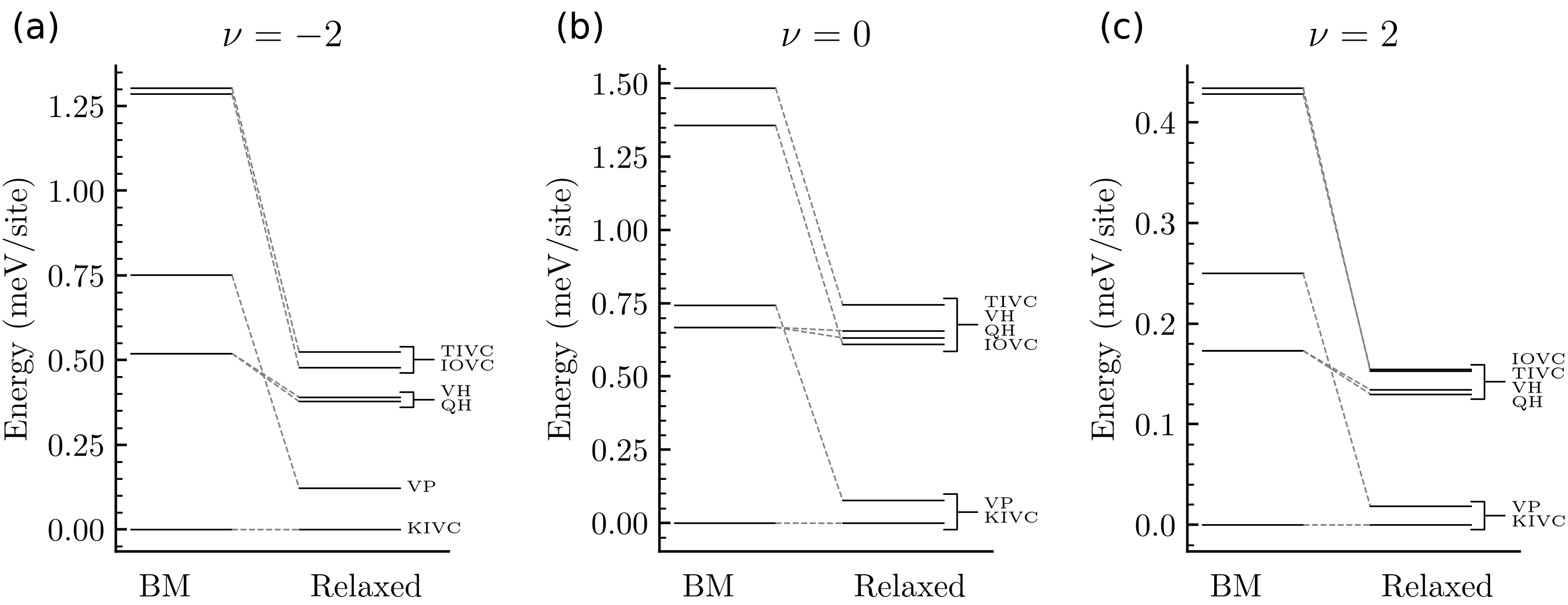}
  \caption{
    Energy orderings of self-consistent Hartree-Fock solutions starting from different initializations in the Bistritzer-MacDonald (BM) and Relaxed interacting models at filling factor \(\nu = -2\) (a), \(\nu = 0\) (b), \(\nu = + 2\) (c).
    Energies are shifted so that \(E_{KIVC} = 0\).
    All calculations are performed on a \(12 \times 12\) Monkhorst-Pack grid with charge neutral subtraction.
  }
  \label{fig:hf-results}
\end{figure*}

\begin{figure*}[t]
  \centering
  \includegraphics[width=.8\linewidth]{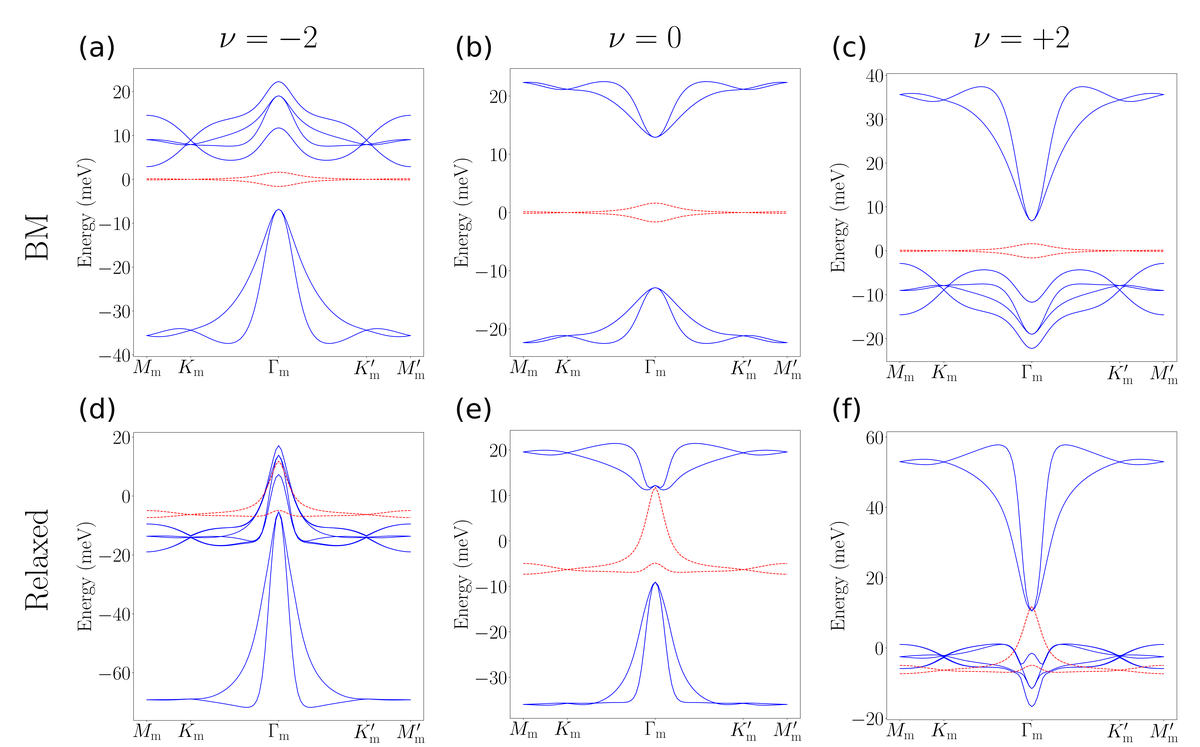}
  \caption{
    Single particle bands (\textcolor{red}{red, dashed}) and Hartree-Fock bands (\textcolor{blue}{blue, solid}) for the Bistritzer-MacDonald (BM) (a,b,c) and Relaxed (d,e,f) interacting models at filling factors \(\nu = -2\) (a,d), \(\nu = 0\) (b,e), \(\nu = +2\) (c,f) starting from KIVC initialization.
    All calculations are performed on a \(12 \times 12\) Monkhorst-Pack grid with charge neutral subtraction. 
    Note that while the Hartree-Fock minimizer for the BM model is an insulator at both \(\nu = +2\) and \(\nu = -2\), the Hartree-Fock minimizer for the Relaxed model is only an insulator at \(\nu = +2\) and semimetallic at \(\nu = - 2\).
    All bands are shifted by a reference energy \(E_{\mathrm{ref}}\) which is chosen so that Bistritzer-MacDonald model at \(\nu = 0\) has \(\tr(F[P] - E_{\mathrm{ref}}) = 0\).}
  \label{fig:kivc-bands}
\end{figure*}

We begin by reporting the relative energy orderings of self-consistent Hartree-Fock calculations with different initializations between the interacting Bistritzer-MacDonald and the relaxed models in~\cref{fig:hf-results}.
Consistent with existing results for the interacting Bistritzer-MacDonald model, at \(\nu \in \{ -2, 0, +2 \}\) we find the lowest energy Hartree-Fock state is an insulating KIVC state \cite{KwanWagnerBultinck2021,Kwan2021,Wagner2022,SongBernevig2022,zhang2020correlated}.
For the relaxed model, while the KIVC state is the lowest energy state at these three fillings it is not always an insulator; in particular at \(\nu \in \{ 0, +2 \}\) this state is insulating but a semimetal for \(\nu = -2\) (see~\cref{fig:kivc-bands}).
These Hartree-Fock results raise two important questions: (i) Why is there particle hole asymmetry between \(\nu = -2\) and \(\nu = +2\)?, (ii) Why is the ground state metallic at \(\nu = -2\)?
As we will show in the next section, the particle hole asymmetry can be largely attributed to the particle hole asymmetry in the single particle dispersion.
The metallic behavior, however, has a more interesting source: wavefunction concentration. 

\vspace{.2in} 

\subsection{Particle Hole Asymmetry, Metallic Behavior, and Wavefunction Squeezing}
To address the two questions raised by our Hartree-Fock experiments, we begin by recalling the expression for the Fock matrix.
Since the double counting subtraction Hamiltonian takes the form \(H_{\mathrm{sub}} = J[P_{\mathrm{sub}}] - K[P_{\mathrm{sub}}]\) for a 1-RDM \(P\), we can simplify \cref{eq:fock-matrix} to:
\begin{equation}
  \label{eq:fock-matrix-combined}
  F[P] = H_{0} + J[P - P_{\mathrm{sub}}] - K[P - P_{\mathrm{sub}}].
\end{equation}

\begin{figure*}[t]
  \centering
  \includegraphics[width=.8\linewidth]{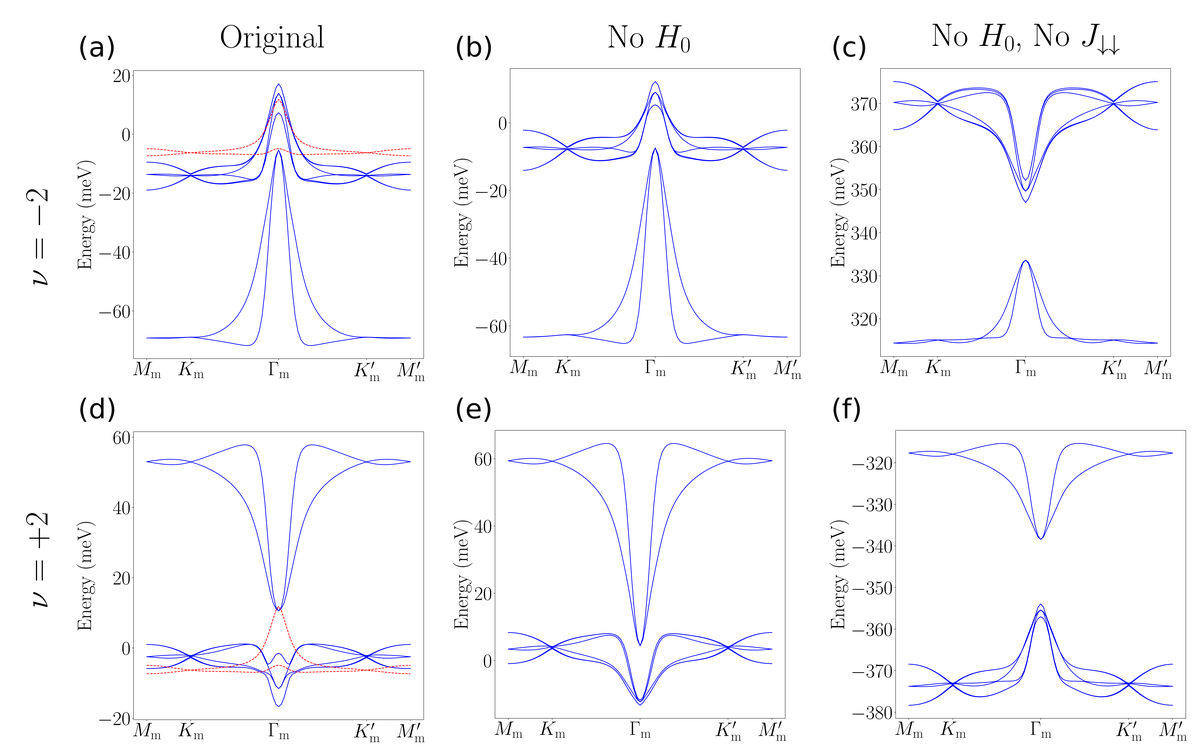}
  \caption{
    Plots of the single particle bands (\textcolor{red}{red, dashed}) and Hartree-Fock bands (\textcolor{blue}{blue, solid}) for the Relaxed interacting model at filling factors \(\nu = -2\) (a) and \(\nu=+2\) (d) starting from KIVC initalization on a \(12 \times 12\) grid with charge neutral subtraction.
    Note that in (d) the size of the band gap at \(\Gamma_{\m}\) is of comparable size to the single particle dispersion.
    Plots (b,e) are the same as (a,d) except \(H_{0}\) in \cref{eq:fock-decomposition} was set to zero.
    Plots (c,f) are the same as (a,d) except both \(H_{0}\) and \(J[P_{\downarrow\downarrow} - P_{\mathrm{sub},\downarrow\downarrow}]\) in \cref{eq:fock-decomposition} were set to zero.
  }
  \label{fig:dispersion-comparison}
\end{figure*}

Compared to the Bistritzer-MacDonald model, the single particle bands in the relaxed model are much larger in magnitude and have significant particle hole asymmetry so a natural candidate for the particle hole asymmetry between \(\nu = \pm 2\) is this single particle dispersion.
In \cref{fig:dispersion-comparison}(a,b), we plot the eigenvalues of the Fock matrix for KIVC in the relaxed model with and without the single particle dispersion (i.e., set \(H_{0} \equiv 0\)) at \(\nu = -2\).
Comparing to the results at \(\nu = +2\) (\cref{fig:dispersion-comparison}(d,e)) we see that setting \(H_{0} \equiv 0\) restores particle hole symmetry but metallic features.
Therefore, the semimetallic behavior cannot be completely due to the single particle dispersion; to explain this effect, we must take a closer look at properties of the Fock matrix.

Since for TBG, the single particle Hamiltonian \(H_{0}\) has no spin orbit coupling, we can separate the Hartree as \(J[P - P_{\mathrm{sub}}] = J[P_{\uparrow\uparrow} - P_{\uparrow\uparrow,\mathrm{sub}}] + J[P_{\downarrow\downarrow} - P_{\downarrow\downarrow, \mathrm{sub}}]\) and the full Fock matrix can be written
\begin{widetext}
  \begin{equation} \label{eq:fock-decomposition}
    \begin{split}
      F[P]
      & = H_{0} + J[P_{\uparrow\uparrow} - P_{\uparrow\uparrow,\mathrm{sub}}] - K[P - P_{\mathrm{sub}}] + J[P_{\downarrow\downarrow} - P_{\downarrow\downarrow,\mathrm{sub}}] \\
      & = H_{0} + J[P_{\uparrow\uparrow} - P_{\uparrow\uparrow,\mathrm{sub}}] - K[P - P_{\mathrm{sub}}]
        +
        \begin{cases}
          J[0 - P_{\downarrow\downarrow,\mathrm{sub}}] & \nu = -2, \\
          J[P_{\downarrow\downarrow} - P_{\downarrow\downarrow,\mathrm{sub}}] & \nu = 0, \\
          J[I - P_{\downarrow\downarrow,\mathrm{sub}}] & \nu = +2, \\
        \end{cases}
    \end{split}
  \end{equation}
\end{widetext}

where in the second line we have used the assumption that, for $\nu = \pm 2$, \(P\) either leaves down spin fully empty or fully filled.
We now provide an explanation for why this term is larger for the relaxed model and for how this causes the semimetallic behavior.

The fact that the down spin Hartree term is larger in the relaxed model when $\nu = \pm 2$ can be traced back to the fact that the flat band Bloch wavefunctions of the relaxed model tend to be more concentrated than those of the BM model. To see that the wavefunctions becoming more concentrated tends to increase the strength of the Hartree potential, recall that the $\bvec{k}$-resolved Hartree energy is given by
\begin{equation}
\int_{\mathbb{R}^{2}}^{}
  \int_{\mathbb{R}^{2}}^{} \varrho_{\bvec{k}}[P](\bvec{r}) V(\bvec{r} - \bvec{r}')  \varrho_{\bvec{k}}[P](\bvec{r}') \dee \bvec{r} \dee \bvec{r}',
\end{equation}
where the electronic density is given by
\begin{equation}
  \label{eq:real-density-k}
  \varrho_{\bvec{k}}[P](\bvec{r}) = \sum_{m,n}^{} \overline{\psi_{n \bvec{k}}(\bvec{r})} [P(\bvec{k})]_{mn} \psi_{m \bvec{k}}(\bvec{r}).
\end{equation}
Now suppose the Bloch functions become more concentrated; mathematically this corresponds to performing the substitution \(\psi_{n \bvec{k}}(\bvec{r}) \mapsto c \psi_{n \bvec{k} }(c \bvec{r})\) for some \(c > 1\) (the factor of \(c\) is included to ensure the squeezed Bloch functions remain normalized).
Substituting these new Bloch functions into the Hartree energy, we see that after changing variables
\begin{equation}
  \begin{split}
    & c^{4}\int_{\mathbb{R}^{2}}^{} \int_{\mathbb{R}^{2}}^{} \varrho_{\bvec{k}}[P](c \bvec{r}) V(\bvec{r} - \bvec{r}')  \varrho_{\bvec{k}}[P](c \bvec{r}') \dee \bvec{r} \dee \bvec{r}' \\
    & \hspace{1em} =\int_{\mathbb{R}^{2}}^{} \int_{\mathbb{R}^{2}}^{} \varrho_{\bvec{k}}[P](\bvec{r}) V(c^{-1} (\bvec{r} - \bvec{r}'))  \varrho_{\bvec{k}}[P](\bvec{r}') \dee \bvec{r} \dee \bvec{r}' .\\
  \end{split}
\end{equation}
Since $V(\bvec{r}) \sim |\bvec{r}|^{-1}$ as $|\bvec{r}| \rightarrow \infty$, this suggests the Hartree energy at $\bvec{k}$ will increase by a roughly a factor of $c$.

\begin{figure}[t]
  \centering
  \includegraphics[width=.9\linewidth]{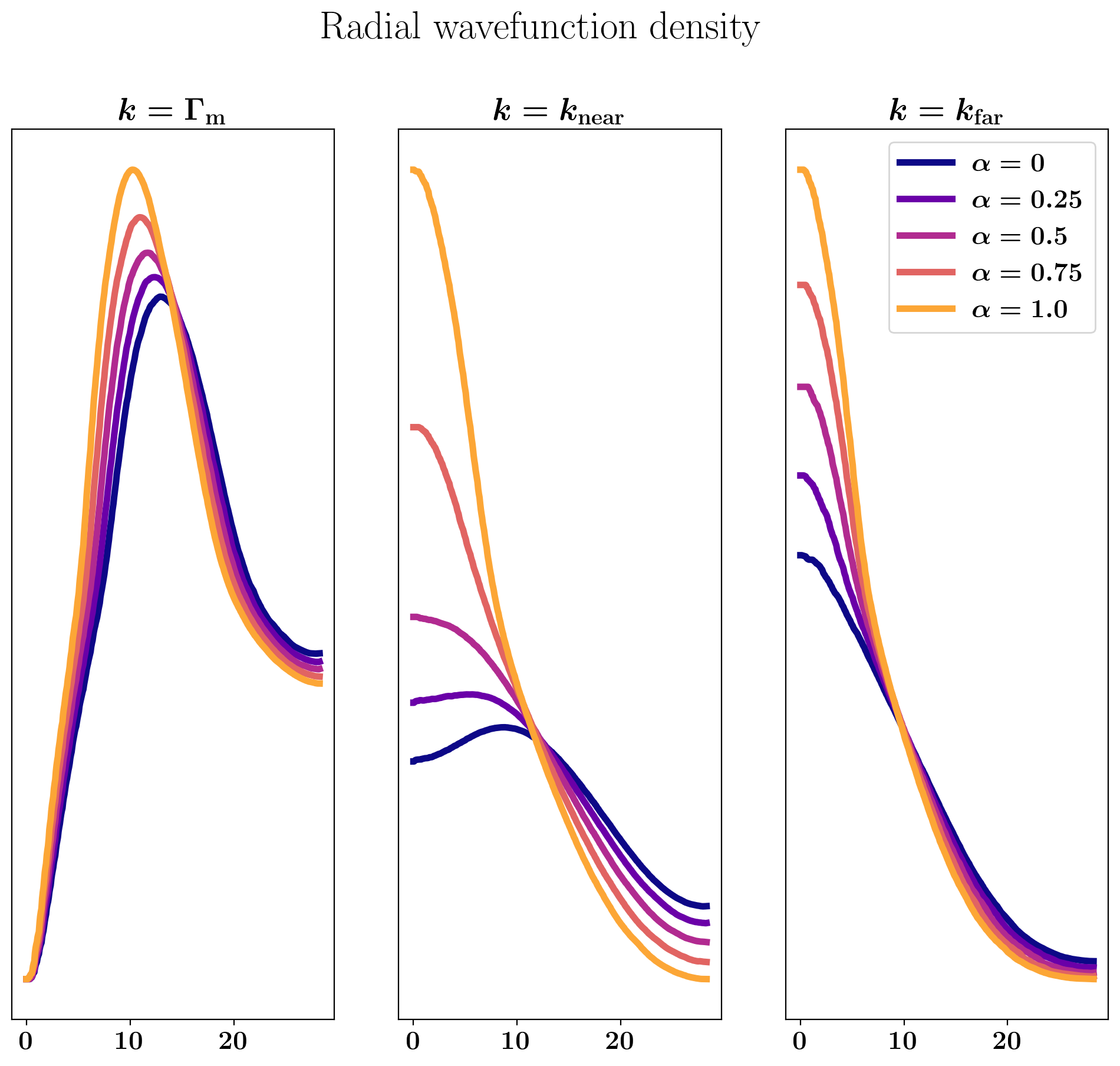}
  
  \caption{Plots of the radial density \(\int_0^{2\pi} \varrho_{\bvec{k}}\left[P_{\downarrow\downarrow,\mathrm{sub}}\right](r,\theta) \dee \theta \) for the Hamiltonian \( (1 - \alpha) H_\mathrm{BM} + \alpha H_{\mathrm{relax}}\) on a single unit cell. We find that the density is more concentrated in the relaxed model at a representative \(\bvec{k} \neq \bvec{\Gamma}_{\m}\) (here we take $\bvec{k}_{\mathrm{near}} = \bvec{\Gamma} - \tfrac{1}{12} (\bvec{b}_{\text{m},1} + \bvec{b}_{\text{m},2})$ and  $\bvec{k}_{\text{far}}  = \bvec{\Gamma}_{\m} - \tfrac{5}{12} (\bvec{b}_{\m,1} + \bvec{b}_{\m,2})$, and the result is similar at other $\bvec{k}$) compared with \(\bvec{k} = \vec{\Gamma}_{\m}\). Note that the wavefunction at $\bvec{k} = \bvec{\Gamma}_{\m}$ must have a node to transform correctly under $C_{3z}$.}
  \label{fig:radial_density}
\end{figure}

To study the concentration of the wavefunctions in the relaxed BM model, we define the angular average of the electron density by:
\begin{equation}
  D_{\bvec{k}}(r) := \frac{1}{2 \pi} \int_0^{2\pi} \varrho_{\bvec{k}}\left[P_{\downarrow\downarrow,\mathrm{sub}}\right](r,\theta) \dee \theta.
\end{equation}
We then consider the family of Hamiltonians which interpolates between the BM and relaxed models
\begin{equation}
  H(\alpha) := (1 - \alpha) H_{\mathrm{BM}} + \alpha H_{\mathrm{relax}},
\end{equation}
and plot $D_{\bvec{k}}(r)$ for different values of \(\alpha \in [0, 1]\) and $\bvec{k}$ in the Brillouin zone.
We find that there are three different regimes which depend on how close \(\bvec{k}\) is to \(\bvec{\Gamma}_{\m}\).
To illustrate these regimes, in~\cref{fig:radial_density} we show plots for three representative points: \(\bvec{\Gamma}_{\m}\), \(\bvec{k}_{\text{near}}\), and \(\bvec{k}_{\text{far}}\) where
\begin{equation}
  \begin{split}
    \bvec{k}_{\text{near}} & = \bvec{\Gamma}_{\m} - \tfrac{1}{12} (\bvec{b}_{\m,1} + \bvec{b}_{\m,2}) \\
    \bvec{k}_{\text{far}} & = \bvec{\Gamma}_{\m} - \tfrac{5}{12} (\bvec{b}_{\m,1} + \bvec{b}_{\m,2}).
  \end{split}
\end{equation}
See~\cref{sec:wavefunction-density} for full plots of \(\varrho_{\bvec{k}}[\textstyle{\frac{1}{2}} P_{\downarrow\downarrow}]\).
Observe that the wavefunction concentrates for \(\bvec{k}_{\mathrm{near}}\) and \(\bvec{k}_{\mathrm{far}}\) but less so for \(\bvec{k} = \bvec{\Gamma}_{\m}\).
Near \(\bvec{\Gamma}_{\m}\), the wavefunction has a symmetry-protected node at the origin (the wavefunction transforms under $C_{3z}$ with eigenvalue $\neq 1$).

Following the argument given above, this implies that the Hartree potential near \(\bvec{k}_{\mathrm{near}}\) and \(\bvec{k}_{\mathrm{far}}\) will be significantly larger as compared to the one from the Bistritzer-MacDonald model.

We can now understand the semi-metallic behavior by the following argument. 
At $\nu = -2$, the down spin Hartree potential $J[0-P_{\downarrow\downarrow,\mathrm{sub}}] = - J[P_{\downarrow\downarrow,\mathrm{sub}}]$ is \emph{negative}, and larger for the relaxed model than the BM model for $\bvec{k} \neq \bvec\Gamma_{\m}$. As a result, the Hartree-Fock bands of the relaxed model are shifted \emph{downwards} compared to those of the BM model, everywhere except at $\bvec{k} = \bvec\Gamma_{\m}$, where they are roughly the same, resulting in the bands overlapping near to $\bvec\Gamma_{\m}$.
At $\nu = +2$, the same Hartree potential is $J[I - P_{\downarrow\downarrow,\mathrm{sub}}]$, which is positive, so that the bands are shifted \emph{upwards} instead.

\section{Conclusions}
\label{sec:conclusions}
In this work, we have introduced a systematic approach for deriving an effective single particle model in MATBG with relaxation while retaining all symmetries and remaining in the Lagrangian frame.
While we focus on MATBG in this work, our approach could be extended to other bilayer moir\'e materials.

After deriving an effective single particle model with systematic relaxation, we construct a many-body model for MATBG and compare the results to the interacting Bistritzer-MacDonald model which accounts for relaxation by modifying the interlayer hopping coefficients.
By performing Hartree-Fock calculations at even integer fillings, \(\nu \in \{ -2, 0 ,+2 \}\), we find that the relaxed model deviates from the Bistritzer-MacDonald model in two qualitatively different ways: (i) particle hole asymmetry between \(\nu = \pm 2\), and (ii) semimetallic behavior at \(\nu = -2\).
We find that while the particle hole asymmetry in the relaxed interacting model is largely due to the single particle dispersion, the semimetallic behavior has a different origin.
Specifically, due to lattice relaxation, we find that the wavefunction density near \(\bvec{\Gamma}_{\m}\) (but not equal to \(\bvec{\Gamma}_{\m}\)) becomes more concentrated.
Away from charge neutrality, this concentration leads to an increase in the Hartree term of the double counting subtraction away from \(\bvec{G}_{\m}\) which is large enough to close the Hartree-Fock gap leading to semimetallic behavior.
Furthermore, the impact of wavefunction concentration on the Hartree term of the double counting subtraction appears to be robust; that is, it is not very sensitive to the underlying double-counting subtraction choice (see \cite{supplementary}).

Our finding of semi-metallic behavior at \(\nu = -2\) is consistent with recent \emph{ab initio} and DFT-parametrized continuum studies \cite{Kim2026InitioQuantumEmbedding,HouSurWagnerEtAl2025}, in contrast to the insulating behavior typically predicted by models based on tuning the Bistritzer-MacDonald model \cite{Zhang2020,Kwan2021,Wagner2022,SongBernevig2022,Shi2022}.
This is at odds with experiments, which generally report insulating behavior at \(\nu = \pm 2\) \cite{NuckollsYazdani2024}.
Resolving the differences between our current model and experiments will require including physical effects which our present model omits.

A few promising candidates for resolving these differences and fit well within the framework introduced in this work include: out-of-plane layer corrugation, which enlarges the gap between flat and remote bands \cite{Carr2019,Carr2019a} and produces semi-metallicity at \(\nu=-2\) in the related study \cite{HouSurWagnerEtAl2025}; the inclusion of remote bands \cite{Sanchez2024,Sanchez2025,qlp6-hjf2,HouSurWagnerEtAl2025,Kwan2021,BultinckKhalafLiuEtAl2020,PhysRevLett.125.257602}; and the choice of more physically realistic subtraction schemes \cite{BultinckKhalafLiuEtAl2020,XieMacDonald2020,FaulstichStubbsZhuEtAl2023,PhysRevLett.125.257602}.
Another important factor not included in this work is heterostrain, which has been shown to have a significant impact on lowest energy states in MATBG and experimental observations \cite{Kwan2021,Parker2021,relaxedheavyfermion,x2mz-hm8y,rr5g-3js8,Guo2026,NuckollsLeeOhEtAl2023}.
Finally, twist angle disorder \cite{PhysRevResearch.2.023325,Uri2020} is a further effect which is believed to have an important impact on the experimentally observed properties of MATBG, though incorporating within our framework poses a significant technical challenge.
We leave a systematic study of these effects, particularly heterostrain, to future work.







 \begin{acknowledgments}
   TK's, ML's, LL's and KDS's  research was partially supported by Simons Targeted Grant Award No. 896630. ML's research was also partially supported in part by NSF Award No. 2422469, and  AW's research was supported in part by grant NSF DMS-2406981. Part of this research was performed while the authors were visiting the Institute for Mathematical and Statistical Innovation (IMSI), which is supported by the National Science Foundation (Grant No. DMS-2425650). The authors are grateful for useful conversations with Dan Parker, Tomohiro Soejima, Patrick Ledwith, and Oskar Vafek.
\end{acknowledgments}

\bibliography{bibliography,bibliography0}

\appendix

\crefalias{section}{appendix}


\section{Derivation of continuum model for relaxed TBG from tight-binding}\label{sec:derivation_cont_model}

\subsection{Effective Hamiltonian}\label{sec:multiscale_derivation}

We present a brief derivation of the effective Hamiltonian using matched asymptotic expansions. For more detail and proof, please refer to \cite{Kong_relaxed}. 
We first recall that the tight-binding Hamiltonian Eq.~(\ref{eq:H_ms}) can be expressed in orders of $\eps$,
\begin{equation}
	H_{\text{TB}} \approx H_{\text{intra}}^{(0)} + \eps H_{\text{intra}}^{(1)} + \eps H_{\text{inter}}
\end{equation}
For conciseness, we will focus on the terms on layer 1 only, as the terms on layer two can be derived by symmetry.
The $\mathcal O(1)$ term $H_{\text{intra}}^{(0)}$ is the \textit{unrelaxed} monolayer graphene Hamiltonian
\begin{equation}\label{eq:H_intra_0}
\begin{aligned}
		H_{\text{intra}}^{(0)} = \sum_{\bvec{R}, \sigma, \bvec{R}', \sigma'} h^{\sigma\sigma'}_{11} \left(\bvec{R}_{1,\sigma} - \bvec{R}'_{1,\sigma'}  \right) c_{\bvec{R}, 1, \sigma}^\dagger c_{\bvec{R}', 1, \sigma'}.
\end{aligned}
\end{equation}

The higher-order corrections contain the relaxation effects on the intralayer $H_{\text{intra}}^{(1)}$, defined as 
\begin{equation}\begin{aligned}\label{eq:H_intra_1}
  H_{\text{intra}}^{(1)} & = \sum_{\bvec{R}, \sigma, \bvec{R}', \sigma'} \nabla h^{\sigma\sigma'}_{11} \left( \bvec{R}_{1,\sigma} - \bvec{R}'_{1,\sigma'} \right)^\top \\
  & \qquad \nabla  \bvec{U}_{1}\left(-\tilde\gamma_{2} \eps \bvec{R}_{1,\sigma}  \right) \tilde\gamma_2  (  \bvec{R}_{1,\sigma} - \bvec{R}'_{1,\sigma'}) \\
  & \qquad   c_{\bvec{R}, 1, \sigma}^\dagger c_{\bvec{R}', 1, \sigma'}.
\end{aligned}\end{equation}

  The higher-order term also contains the interlayer tunneling with relaxation effects. In terms of the relaxed interlayer hopping function
\begin{equation}\begin{aligned}
	h^{\sigma\sigma'}_{\bvec{U}}(\bvec{x}) = h^{\sigma\sigma'}_{12}(\bvec{x} + \bvec{U}_1(\bvec{x} - \vec{\tau}_2^{\sigma\sigma'}) - \bvec{U}_2(- \bvec{x}+\vec{\tau}_1^{\sigma\sigma'}) ),
\end{aligned}\end{equation}
we have
\begin{equation}\label{eq:H_inter}
 \begin{aligned}
 	 H_{\text{inter}}  = \sum_{\bvec{R}, \sigma, \bvec{R}', \sigma'} 
 	 h^{\sigma\sigma'}_{\bvec{U}} \left(\bvec{R}_{1,\sigma} - \bvec{R}'_{2,\sigma'}    \right)  c_{\bvec{R}, 1, \sigma}^\dagger c_{\bvec{R}', 2, \sigma'}.
 \end{aligned}
\end{equation}

Our strategy is to approximate the tight-binding problem by an effective continuum Hamiltonian. We derive the multiscale Hamiltonian by plugging in the wave packet ansatz Eq.~(\ref{eq:wavefunction_ms}) to Eq.~(\ref{eq:H_ms}), and matching all the terms with the same order of $\eps$. The effective continuum Hamiltonian would have the expansion
\begin{equation}
	\mathsf H_{\text{eff}} = \mathsf H^{(0)}_{\text{eff}} + \eps \mathsf  H^{(1)}_{\text{eff}} + \eps^2 \mathsf  H^{(2)}_{\text{eff}} + \dots.
\end{equation}
\subsubsection{Identities involving Bloch transform}
Defining the Bloch transform on tight-binding wave functions 
\begin{equation}
	\widetilde{\varphi}_{\ell, \sigma}(\bvec{k}) = \sum_{\bvec{R}} e^{-i\bvec{k} \cdot \bvec{R}_{\ell,\sigma} } \varphi_{\bvec{R}, \ell, \sigma},
\end{equation}
the mono-layer Bloch Hamiltonian is
\begin{equation}
	(\widetilde{H_{\ell \ell} \varphi})_{\ell,\sigma}(\bvec{k}) = \sum_{\sigma'} \widetilde {h_{\ell\ell}^{\sigma\sigma'}}(\bvec{k}) \widetilde \varphi_{\ell,\sigma'}(\bvec{k}).
\end{equation}
Denoting the sublattice-shifted lattices as $\mathcal R_\ell^{\sigma\sigma'} := \mathcal R_\ell + \vec{\tau}_\ell^{\sigma\sigma'}$, we have that
\begin{equation}
	\widetilde {h_{\ell\ell}^{\sigma\sigma'}}(\bvec{k}) = \sum_{\bvec{Q}_\ell \in \mathcal R_\ell^{\sigma\sigma'}} h_{\ell\ell}^{\sigma\sigma'} \left(\bvec{Q}_\ell \right) e^{-i\bvec{k} \cdot \bvec{Q}_\ell }.
\end{equation}
The gradient and Hessian of $\widetilde {h_{\ell\ell}^{\sigma\sigma'}}(\bvec{k})$ with respect to momentum $\bvec{k}$ are therefore \begin{equation}
    \begin{split}
        \nabla_{\bvec{k}} \widetilde {h_{\ell\ell}^{\sigma\sigma'}}(\bvec{k}) =  -i \sum_{\bvec{Q}_\ell \in \mathcal R_\ell^{\sigma\sigma'}} h_{\ell\ell}^{\sigma\sigma'} \left(\bvec{Q}_\ell \right) e^{-i\bvec{k} \cdot \bvec{Q}_\ell }\bvec{Q}_\ell,
    \end{split}
\end{equation} 
and 
\begin{equation}
    \begin{split}
        \nabla_{\bvec{k}}^2 \widetilde {h_{\ell\ell}^{\sigma\sigma'}}(\bvec{k}) = - \sum_{\bvec{Q}_\ell \in \mathcal R_\ell^{\sigma\sigma'}} h_{\ell\ell}^{\sigma\sigma'} \left(\bvec{Q}_\ell \right) e^{-i\bvec{k} \cdot \bvec{Q}_\ell}   \bvec{Q}_\ell \bvec{Q}_\ell^\top.
    \end{split}
\end{equation}
From \cite{quinn2025higherordercontinuummodelstwisted}, and recalling that $\theta_1 = -\theta/2$, $\theta_2 = \theta/2$, we have the following identities at Dirac points
 \begin{equation}\label{eq:hk_0}
 \begin{gathered}
       \widetilde{h^{\A\A}_{\ell\ell}} (\bvec{K}_\ell) = \widetilde{h^{\B\B}_{\ell\ell}} (\bvec{K}_\ell)  = \mu_F, \\  
       \widetilde{h^{\A\B}_{\ell\ell}} (\bvec{K}_\ell) =  \widetilde{h^{\B\A}_{\ell\ell}} (\bvec{K}_\ell) = 0,
 \end{gathered}
 \end{equation}
 where $\mu_F$ is the Fermi level of neutral monolayer graphene. The gradients are
 \begin{equation}\label{eq:hk_1}
 \begin{gathered}
 	       \nabla_{\bvec{k}} \widetilde{h^{\A\A}_{\ell\ell}} (\bvec{K}_\ell)  = \nabla_{\bvec{k}} \widetilde{h^{\B\B}_{\ell\ell}} (\bvec{K}_\ell)   = \bvec{0}, \\   \nabla_{\bvec{k}} \widetilde{h^{\A\B}_{\ell\ell}} (\bvec{K}_\ell)  = v_F e^{i\theta_\ell}   (1,-i)^\top, \\
 	       \nabla_{\bvec{k}} \widetilde{h^{\B\A}_{\ell\ell}} (\bvec{K}_\ell)  = v_F e^{-i\theta_\ell}   (1,i)^\top.
 \end{gathered}
 \end{equation}
 where $v_F$ is the Fermi velocity.
 And lastly, the Hessians are 
 \begin{equation}\label{eq:hk_2}
 \begin{gathered}
        \nabla^2_{\bvec{k}} \widetilde{h^{\A\A}_{\ell\ell}} (\bvec{K}_\ell) = \nabla^2_{\bvec{k}} \widetilde{h^{\B\B}_{\ell\ell}} (\bvec{K}_\ell) = v_1 I, \\  
        \nabla^2_{\bvec{k}} \widetilde{h^{\A\B}_{\ell\ell}} (\bvec{K}_\ell) = v_2 e^{-i2\theta_\ell} \begin{pmatrix}
            1 & i\\
            i & -1
        \end{pmatrix}, \\
        \nabla^2_{\bvec{k}} \widetilde{h^{\B\A}_{\ell\ell}} (\bvec{K}_\ell) = v_2 e^{i2\theta_\ell} \begin{pmatrix}
            1 & -i\\
            -i & -1
        \end{pmatrix},
  \end{gathered}
 \end{equation}

\subsubsection{$\mathcal O(1)$ terms}

The overall $\mathcal O(1)$ term  $\mathsf H^{(0)}_{\text{eff}}$ is obtained by plugging in the $\mathcal O(1)$ expansion of $\psi$ into Eq.~(\ref{eq:H_intra_0}) 
\begin{widetext}
\begin{equation}
\begin{aligned}
	 \mathsf H^{(0)}_\mathrm{eff} & = \eps^2 \sum_{\bvec{R}, \sigma, \bvec{R}', \sigma'} h_{11}^{\sigma\sigma'} \left(\bvec{R}_{1,\sigma} - \bvec{R}'_{1,\sigma'} \right) e^{-i \bvec{K}_1 \cdot  \bvec{R}_{1,\sigma} } \psi^\dagger_{1,\sigma}(\bvec{X})  e^{i \bvec{K}_1 \cdot \bvec{R}'_{1,\sigma'}} \psi_{1,\sigma'}(\bvec{X}) \\
	 & =  \eps^2 \sum_{\bvec{R}} \sum_{\bvec R'} \sum_{\sigma, \sigma'}  h_{11}^{\sigma\sigma'} \left(\bvec{R}_{1,\sigma} - \bvec{R}'_{1,\sigma'} \right) e^{-i \bvec{K}_1 \cdot \left(\bvec{R}_{1,\sigma} - \bvec{R}'_{1,\sigma'} \right)} \psi^\dagger_{1,\sigma}(\bvec{X}) \psi_{1,\sigma'}(\bvec{X}) \\
	 & \approx  \int_{\mathbb R^2} \sum_{\sigma, \sigma'} \left[\sum_{ \bvec{Q}_1 \in \mathcal R_1^{\sigma\sigma'}} h_{11}^{\sigma\sigma'} \left(\bvec{Q}_1  \right) e^{-i\bvec{K}_1 \cdot \bvec{Q}_1 } \right]  \psi^\dagger_{1,\sigma}(\bvec{X})  \psi_{1, \sigma'}(\bvec{X}) \dee \bvec{X}.
\end{aligned}
\end{equation}
\end{widetext}
Here we introduce a change of variable ${\bvec{Q}}_1 = \bvec{R}_{1,\sigma} - \bvec{R}'_{1,\sigma'} \in \mathcal R_1^{\sigma\sigma'}$, and approximate the discrete summation by the integration $ \eps^2 \sum_{\bvec{R}} \psi (\eps \bvec{R}) \approx \int \psi(\eps \bvec{R}) \dee (\eps \bvec{R})$.

 From Eq.~\ref{eq:hk_0}, the $\mathcal O(1)$ term is a constant shift in the Fermi level
\begin{equation}
	\mathsf H^{(0)}_{\mathrm{eff}} =   \mu_\mathrm{F}  \int_{\mathbb R^2} \sum_{\sigma, \ell }  \psi^\dagger_{\ell,\sigma}(\bvec{X})  \psi_{\ell, \sigma}(\bvec{X}) \dee \bvec{X} = \mu_\mathrm{F} I.
\end{equation}

\subsubsection{$\mathcal O(\eps)$ terms}

The $\mathcal O(\eps)$ term $\mathsf H^{(1)}_{\text{eff}}$ is the sum of three terms. The first term is obtained by plugging in the $\mathcal O(\eps)$ expansion of $\psi$ into Eq.~(\ref{eq:H_intra_0})
\begin{equation}\begin{aligned}
& \int_{\mathbb R^2}  \sum_{\sigma, \sigma'}\sum_{ \bvec{Q}_1 \in \mathcal R_1^{\sigma\sigma'}} h_{11}^{\sigma\sigma'} \left(\bvec Q_1 \right) e^{-i\bvec{K}_1 \cdot \bvec Q_1} (-\bvec Q_1)^\top \\
 & \qquad \psi^\dagger_{1,\sigma}(\bvec{X})  \nabla_{\bvec{X}} \psi_{1, \sigma'}(\bvec{X}) \dee \bvec{X} \\[1.5ex]
= & \int_{\mathbb R^2}  \sum_{\sigma, \sigma'} \psi^\dagger_{1,\sigma}(\bvec{X})  \left[\nabla_{\bvec{k}} \widetilde {h_{11}^{\sigma\sigma'}}(\bvec{K}_1)  \right]^\top (-i \nabla_{\bvec{X}})\psi_{1, \sigma'}(\bvec{X}) \dee \bvec{X}.
\end{aligned}\end{equation} 
By using the identities in Eq.~\ref{eq:hk_1}, we recover the twisted Dirac operator (the first term in Eq.~\ref{eq:dirac_2nd_order}).
 
 The second term contains Eq.~(\ref{eq:H_intra_1}) and the $\mathcal O(1)$ expansion of $\psi$ 
 \begin{equation}\begin{aligned}
 	\int_{\mathbb R^2}& \sum_{\sigma, \sigma'}\sum_{\bvec{Q}_1 \in \mathcal R_1^{\sigma\sigma'}} \nabla h_{11}^{\sigma\sigma'} \left(\bvec Q_1 \right)^\top \nabla  \bvec{U}_{1}\left(\tilde\gamma_2 \bvec{X} \right)  \tilde\gamma_2  \\
  &    \bvec Q_1 	 e^{-i\bvec{K}_1 \cdot \bvec Q_1}   \psi^\dagger_{1,\sigma}(\bvec{X})  \psi_{1, \sigma'}(\bvec{X}) \dee \bvec{X}.
 \end{aligned}\end{equation}
 We can then define the moir\'e-periodic potential from intralayer relaxation effects 
 \begin{equation}
 	 \begin{aligned}
 	 \left [\mathcal S_1(\bvec{X}) \right]^{\sigma\sigma'} := \sum_{\bvec{Q}_1 \in \mathcal R_1^{\sigma\sigma'}} & \left[\nabla h_{11}^{\sigma\sigma'} \left(\bvec{Q}_1\right)\right]^\top \nabla  \bvec{U}_{1}\left( \tilde\gamma_2 \bvec{X} \right)\tilde\gamma_2 \\
     & \qquad \bvec{Q}_1	 e^{-i\bvec{K}_1 \cdot \bvec{Q}_1}.
 \end{aligned}
\end{equation}

 The third term contains Eq.~(\ref{eq:H_inter}) and the $\mathcal O(1)$ expansion of $\psi$ 
  \begin{equation}	
  \begin{aligned}
  	\int_{\mathbb R^2}  \sum_{\sigma,\sigma'} \sum_{\bvec{Q}_2 \in \mathcal R_2^{\sigma\sigma'}} & h^{\sigma\sigma'}_{\bvec{U}} \left(-\tilde\gamma_2 \bvec{X} + \bvec{Q}_2 \right)  e^{-i\bvec{K}_2 \cdot \bvec Q_2} \\
  	 & \psi^\dagger_{1,\sigma}(\bvec{X})  \psi_{2, \sigma'}(\bvec{X}) \dee \bvec{X}.
  \end{aligned}
  \end{equation}
 Here we have used the fact that the rotation of layers give $\bvec{K}_1 \cdot \bvec{R}_{1,\sigma}  = \bvec{K}_2 \cdot \widetilde{\bvec{R}}_{2,\sigma}$, where $\widetilde{\bvec{R}}_{2,\sigma}  = A_2 A_1^{-1}  \bvec{R}_{1,\sigma}$. 
 We then introduce a change of variables $\bvec{Q}_2 = \widetilde{\bvec{R}}_{2,\sigma}  - \bvec{R}_{2,\sigma'}'$ to get the desired result. 
  The interlayer hopping potential $ \tilde {\mathcal T}^{\sigma\sigma'}$ is then
  \begin{equation}
  	\left [ \tilde {\mathcal T}(\bvec{X})\right]^{\sigma\sigma'} := \sum_{\bvec{Q}_2 \in \mathcal R_2^{\sigma\sigma'}} h^{\sigma\sigma'}_{\bvec{U}} \left(-\tilde\gamma_2 \bvec{X} + \bvec{Q}_2 \right)  e^{-i\bvec{K}_2 \cdot \bvec{Q}_2}.
  \end{equation}

The potentials $\mathcal S_\ell$ and $\tilde {\mathcal T}$ have scaled-moir\'e periodicity. The  \textit{exact} periodicity can be recovered by recalling $\bvec{X} = \eps(\bvec{R}_1 + \vec{\tau}_1^\sigma)$ and rescaling the potentials as
\begin{equation}
\begin{gathered}
	S_1(\bvec{r}) := \eps \mathcal S_1(\eps \bvec{r}), \quad S_2(\bvec{r}) := \eps \mathcal S_2(\eps \bvec{r}), \\
	\tilde T(\bvec{r}) := \eps \tilde{\mathcal T}(\eps  \bvec{r}).
\end{gathered}
\end{equation}
The disregistry maps a moir\'e lattice vector to an atomistic lattice vector, i.e. $\gamma_2 (\bvec{x} + \bvec{R}_\m)= \gamma_2 \bvec{x} + \bvec{R}_2$, and the disregistry space displacement functions satisfy $\bvec{U}_1(\bvec{x} + \bvec{R}_2) = \bvec{U}_1(\bvec{x})$, we may conclude that 
\begin{equation}\label{eq:S_moire_period}
\begin{split}
	S_\ell(\bvec{r}  + \bvec{R}_\m) &=  S_\ell(\bvec{r} ), \\
	 \tilde {T} (\bvec{r} + \bvec{R}_\m ) &= \tilde {T} (\bvec{r} ) e^{i(\bvec{K}_1 - \bvec{K}_2) \cdot \bvec{R}_\m}.
\end{split}
\end{equation}

Therefore we can conclude that the leading order term the relaxed Hamiltonian is
 \begin{widetext}
 	 \begin{equation}
 	 \begin{aligned}
  	\eps \mathsf H^{(1)}_{\text{eff}} &= \eps \int_{\mathbb R^2}   \sum_{\sigma,\sigma'}  \psi^\dagger_{1,\sigma}(\bvec{X})  \left \{ \left[\nabla_{\bvec{k}} \widetilde {h_{11}^{\sigma\sigma'}}(\bvec{K}_1)^\top (-i\nabla_{\bvec{X}}) + \left[\mathcal S_{1}(\bvec{X}) \right]^{\sigma\sigma'}\right]  \psi_{1,\sigma'}(\bvec{X})  +  \left[\tilde {\mathcal T}(\bvec{X}) \right]^{\sigma\sigma'} \psi_{2, \sigma'}(\bvec{X}) \right \} \dee \bvec{X}  + (1 \leftrightarrow 2),\\
  	 &=  \int_{\mathbb R^2}   \sum_{\sigma,\sigma'}  (\psi^\eps_{1,\sigma})^\dagger (\bvec{r}) \left \{ \left[\nabla_{\bvec{k}} \widetilde {h_{11}^{\sigma\sigma'}}(\bvec{K}_1)^\top (-i\nabla_{\bvec{r}}) +  \left[S_{1}(\bvec{r})\right]^{\sigma\sigma'}\right]  \psi^\eps_{1,\sigma'}(\bvec{r})  +  \left[\tilde {T}(\bvec{r})\right]^{\sigma\sigma'} \psi^\eps_{2, \sigma'}(\bvec{r}) \right \} \dee \bvec{r}  + (1 \leftrightarrow 2).
  \end{aligned}
 \end{equation}
 \end{widetext}

\subsubsection{$\mathcal O(\eps^2)$ terms}

The $\mathcal O(\eps^2)$ higher-order corrections can be derived similarly. It also consists of three terms, in which the derivation is analogous to $\mathcal O(\eps)$ terms. 

First, we have the quadratic correction to the Dirac cone, which is obtained by plugging second-order derivatives of $\psi$ with order $\mathcal O(\eps^2)$ into Eq.~(\ref{eq:H_intra_0}) 
\begin{equation}\label{eq:H_cont_1}
\begin{aligned}
  & \frac{1}{2} \int_{\mathbb R^2}  \sum_{\sigma, \sigma'}\psi^\dagger_{1,\sigma}(\bvec{X})\sum_{ \bvec{Q}_1 \in \mathcal R_1^{\sigma\sigma'}}  h^{\sigma\sigma'}_{11} \left(\bvec{Q}_1 \right) e^{-i\bvec{K}_1 \cdot \bvec{Q}_1} \bvec Q_1^\top  \\
 & \qquad  \nabla^2 \psi_{1, \sigma'}(\bvec{X})  \bvec Q_1 \dee \bvec{X}. \\[1.5ex]
 & = \frac{1}{2} \int_{\mathbb R^2}  \sum_{\sigma, \sigma'} \psi^\dagger_{1,\sigma}(\bvec{X}) \tr \left[\nabla^2_{\bvec{k}} \widetilde {h_{11}^{\sigma\sigma'}}(\bvec{K}_1)    \nabla^2 \psi_{1, \sigma'}(\bvec{X})  \right]  \dee \bvec{X}.
\end{aligned}\end{equation}

Then we have the non-local approximation of intralayer relaxation potential, which is obtained from Eq.~(\ref{eq:H_intra_1}) and the $\mathcal O(\eps)$ expansion of $\psi$. 
 \begin{equation}\begin{aligned}
 	\int_{\mathbb R^2}& \sum_{\sigma, \sigma'} \sum_{ \bvec{Q}_1 \in \mathcal R_1^{\sigma\sigma'}} -i\nabla h_{11}^{\sigma\sigma'} \left(\bvec Q_1\right)^\top \nabla  \bvec{U}_{1}\left(\tilde\gamma_2 \bvec{X} \right)  \tilde\gamma_2 \\
  &   \bvec Q_1 \bvec Q_1^\top e^{-i\bvec{K}_1 \cdot \bvec Q_1}   \psi^\dagger_{1,\sigma}(\bvec{X}) (-i \nabla )\psi_{1, \sigma'}(\bvec{X}) \dee \bvec{X},
 \end{aligned}\end{equation}
 and so we can extract the coefficients for the non-local terms as  
 \begin{equation}
 \begin{split}
 & \left[\mathcal S_1^\mathrm{nl}(\bvec{X})	\right]^{\sigma\sigma'}\\
 & :=  -i \sum_{\bvec{Q}_1 \in \mathcal R_1^{\sigma\sigma'}}  \left[\nabla h_{11}^{\sigma\sigma'} (\bvec{Q}_1)\right]^\top \nabla  \bvec{U}_{1}\left(\tilde\gamma_2 \bvec{X} \right)  \tilde\gamma_2 \\
  &  \qquad  \qquad  \left[ \bvec{Q}_1 \bvec{Q}_1^\top\right] e^{-i\bvec{K}_1 \cdot \bvec{Q}_1}.
 \end{split}
\end{equation}

Lastly, we have the non-local approximation of interlayer hopping potential, by plugging in the $\mathcal O(\eps)$ expansion of $\psi$ with first derivatives into Eq.~(\ref{eq:H_inter})
\begin{equation}
  \begin{aligned}
  	\int_{\mathbb R^2} &  \sum_{\sigma,\sigma'} \sum_{\bvec{Q}_2\in\mathcal R_2^{\sigma\sigma'}}  -i h^{\sigma\sigma'}_{\bvec{U}} \left(-\tilde\gamma_2 \bvec{X} + \bvec{Q}_2 \right)  e^{-i\bvec{K}_2 \cdot \bvec{Q}_2}  \\
  	 &  \left(-\tilde\gamma_2 \bvec{X} + \bvec{Q}_2 \right)^\top  \psi^\dagger_{1,\sigma}(\bvec{X}) (-i \nabla)\psi_{2, \sigma'}(\bvec{X}) \dee \bvec{X}.
  \end{aligned}
\end{equation}
Similarly the coefficients for the non-local terms are 
\begin{equation}
\begin{aligned}
	 & \left[ \tilde{\mathcal T}^{\mathrm{nl}} (\bvec{X}) \right]^{\sigma\sigma'} \\
	 & := \sum_{\bvec{Q}_2 \in \mathcal R_2^{\sigma\sigma'}} -i h^{\sigma\sigma'}_{\bvec{U}} \left(-\tilde\gamma_2 \bvec{X} + \bvec{Q}_2 \right)\left(-\tilde\gamma_2 \bvec{X} + \bvec{Q}_2 \right)^\top e^{-i\bvec{K}_2 \cdot \bvec{Q}_2} .
\end{aligned}
\end{equation}

These coefficients satisfy a similar scaled moir\'e periodicity. We can rescale them by
\begin{equation}
	S^{\mathrm{nl}}_\ell(\bvec{r}) := \eps \mathcal S^{\mathrm{nl}}_\ell(\eps \bvec{r}), \quad  {\mathcal T}^{\mathrm{nl}} (\bvec{r}) := \eps \tilde{\mathcal T}^{\mathrm{nl}}(\eps \bvec{r}),
\end{equation}
such that we have the same periodicity conditions as Eq.~(\ref{eq:S_moire_period}) for the non-local approximations
\begin{equation}
\begin{split}
	S^{\mathrm{nl}}_\ell(\bvec{r}  + \bvec{R}_\m) &=  S^{\mathrm{nl}}_\ell(\bvec{r}), \\
	 \tilde {T}^{\mathrm{nl}} (\bvec{r} + \bvec{R}_\m ) &= \tilde {T}^{\mathrm{nl}} (\bvec{r} ) e^{i(\bvec{K}_1 - \bvec{K}_2) \cdot \bvec{R}_\m}.
\end{split}
\end{equation}

Therefore the higher-order corrections to the relaxed TBG continuum Hamiltonian are 
\begin{widetext}
 \begin{equation}\label{eq:H_cont_2}
 	 \begin{aligned}
  	&\eps^2 \mathsf H^{(2)}_{\text{eff}} \\
  	& = \eps^2 \int_{\mathbb R^2}  \sum_{\sigma,\sigma'}  \psi^\dagger_{1,\sigma}(\bvec{X})  \left [\frac{1}{2} \tr \left(\nabla^2_{\bvec{k}} \widetilde {h_{11}^{\sigma\sigma'}}(\bvec{K}_1)  \nabla^2_{\bvec{X}}  \right)+\left[\mathcal S_1^\mathrm{nl}(\bvec{X})	\right]^{\sigma\sigma'}  (-i\nabla_{\bvec{X}})\right]  \psi_{1,\sigma'}(\bvec{X})   \\
  	& \qquad\qquad  +  \psi^\dagger_{1,\sigma}(\bvec{X}) \left[ \tilde{\mathcal T}^{\mathrm{nl}}(\bvec{X}) \right]^{\sigma\sigma'}  (-i\nabla_{\bvec{X}})  \psi_{2, \sigma'}(\bvec{X})  \dee \bvec{X}  + (1 \leftrightarrow 2) \\
  	& = \int_{\mathbb R^2}  \sum_{\sigma,\sigma'}  (\psi^\eps_{1,\sigma})^\dagger (\bvec{r})  \left\{ \left [\frac{1}{2} \tr \left(\nabla^2_{\bvec{k}} \widetilde {h_{11}^{\sigma\sigma'}}(\bvec{K}_1)  \nabla^2_{\bvec{r}}  \right)+\left[ S_1^\mathrm{nl}	(\bvec{r})\right]^{\sigma\sigma'}  (-i\nabla_{\bvec{r}})\right]  \psi^\eps_{1,\sigma'}(\bvec{r}) + 
  	 \left[ \tilde{ T}^{\mathrm{nl}}(\bvec{r}) \right]^{\sigma\sigma'}  (-i\nabla_{\bvec{r}})  \psi^\eps_{2, \sigma'}(\bvec{r})\right\}  \dee \bvec{r}  \\
  	 & \qquad+ (1 \leftrightarrow 2).
  \end{aligned}
\end{equation}	
 \end{widetext}
 Together with Eq.~(\ref{eq:H_cont_1}), we recover the atomistic scale continuum Hamiltonian Eq.~(\ref{eq:H_relax}).

\section{A Review of Hartree-Fock Theory for Twisted Bilayer Graphene}
\label{sec:review-hartree-fock}
Given a (translation-invariant) Hartree-Fock Slater determinant \(\ket{\Psi}\), we define the 1-RDM \(P(\bvec{k})\) by \([P(\bvec{k})]_{nm} := \braket{\Psi | f_{m \bvec{k}}^{\dagger} f_{n \bvec{k}} | \Psi }\).
Using Wick's theorem together with translation invariance, we can write the two-body term in \(H_{I}\) as a difference of two terms corresponding to the symmetric and anti-symmetric terms in the Wick's contractions.
Specifically,
\begin{equation}
  \label{eq:wicks-1}
  \braket{\Psi | f_{m \bvec{k}}^{\dagger} f_{m' \bvec{k}'}^{\dagger} f_{n' (\bvec{k}' - \bvec{q})} f_{n (\bvec{k} + \bvec{q})} | \Psi } = (*) - (**) 
\end{equation}
where
\begin{equation}
  \label{eq:wicks-2}
  \begin{split}
  (*) &=  \delta_{\bvec{q} \in \mathcal{R}_{\m}^{*}} \,[P(\bvec{k})]_{n m}\,[P(\bvec{k}')]_{n' m'} \\[.5ex]
   (**) &= \delta_{\bvec{k}' - \bvec{k} - \bvec{q} \in \mathcal{R}_{\m}^{*}}\,[P(\bvec{k})]_{n' m}\,[P(\bvec{k}+\bvec{q})]_{n m'}.
  \end{split}
\end{equation}
A standard calculation using~\cref{eq:wicks-1,eq:wicks-2} shows that \(\braket{\Psi | H_{I} |  \Psi} = J[P] - K[P]\) where
\begin{widetext}
  \begin{align}
      J[P]
    & = \frac{1}{N_{\mathcal{K}} |\Omega_{\mathrm{m}}|} \sum_{\bvec{k}, \bvec{k}' \in \mathcal{K}}^{}  \sum_{\bvec{G} \in \mathcal R_{\mathrm{m}}^{*}}^{} \sum_{\alpha, \beta}^{}   V(\bvec{G}) \tr\big(\Lambda_{\bvec{k}'}(-\bvec{G}) P(\bvec{k}') \big)  [\Lambda_{\bvec{k}}(\bvec{G})]_{\alpha \beta} f_{\alpha \bvec{k}}^{\dagger}   f_{\beta \bvec{k}} \label{eq:hartree-full} \\
    K[P]
    & = \frac{1}{N_{\mathcal{K}} |\Omega_{\mathrm{m}}|} \sum_{\bvec{k}, \bvec{k}' \in \mathcal{K}}^{}  \sum_{\bvec{G} \in \mathcal{R}_{\mathrm{m}}^{*}}^{} \sum_{\alpha, \beta, \gamma, \gamma'}^{} V(\bvec{k}' - \bvec{k} + \bvec{G})  [\Lambda_{\bvec{k}}(\bvec{k}' - \bvec{k} + \bvec{G})]_{\alpha \gamma} [P(\bvec{k}')]_{\gamma \gamma'} [\Lambda_{\bvec{k}'}(\bvec{k} - \bvec{k}' + \bvec{G})]_{\gamma' \beta} f_{\alpha \bvec{k}}^{\dagger} f_{\beta \bvec{k}} \label{eq:exchange-full} 
  \end{align}
\end{widetext}

\subsection{Interpolated Hartree-Fock Bands}
\label{sec:single-shot}
To generate the Hartree-Fock band structure along the high symmetry line \(\bvec{M}_{\m} \to \bvec{K}_{\m} \to \vec{\Gamma}_{\m} \to \bvec{K}_{\m}' \to \bvec{M}_{\m}' \) we begin by performing on self-consistent Hartree-Fock calculation on a Monkhorst pack grid \(\mathcal{K}\) to obtain a converged Hartree-Fock one-body reduced density matrix \(P_{*}\).
To interpolate these bands along a high symmetry line \(\mathcal{K}_{\mathrm{HS}}\), we compute the modified Fock matrix
\begin{equation}
  \begin{split}
    F_{\mathrm{interp}}[P_{*}]
    & = H_{0} + J_{\mathrm{interp}}[P_{*} - P_{\mathrm{sub}}] \\
    & \hspace{4em} - K_{\mathrm{interp}}[P_{*} - P_{\mathrm{sub}}]  .
  \end{split}
\end{equation}
Here, \(H_{0}\) is the single particle bands evaluated along \(\mathcal{K}_{\mathrm{HS}}\) and \(J_{\mathrm{interp}}\) and \(J_{\mathrm{interp}}\) are as in~\cref{eq:hartree-full,eq:exchange-full} except the sum over \(\bvec{k}\) is taken over \(\mathcal{K}_{\mathrm{HS}}\) instead of \(\mathcal{K}\).
We note that this calculate requires evaluating the form factor \([\Lambda_{\bvec{k}}(\bvec{k}' - \bvec{k} + \bvec{G})]_{mn} = \braket{u_{m \bvec{k}}, u_{n (\bvec{k}' + \bvec{G})}}\) which was not calculated in the original self-consistent iteration.
In particular, this involves computing the overlap between single particle states at momentum \(\bvec{k} \in \mathcal{K}_{\mathrm{HS}}\) and momentum \(\bvec{k}' \in \mathcal{K}\).

\section{Charge Neutral Subtraction} \label{app:subtraction}
Given a single particle Hamiltonian, \(H(\bvec{k})\), we define the charge neutral subtraction through Fermi-Dirac smearing which depends on two parameters: the inverse temperature, \(\beta\), and the chemical potential \(\mu\).
Given these parameters, the subtraction Hamiltonian is:
\begin{equation}
  P_{\mathrm{sub}}(\bvec{k}) = \left( 1 + \exp\Big\{{} \beta (H(\bvec{k}) - \mu) \Big\} \right)^{-1}.
\end{equation}
For the experiments in the main text, we take \(\beta = 10^{-5}\) and set \(\mu\) so that \(P\) is half-filled (i.e. \(\sum_{\bvec{k}}^{} \tr{(P(\bvec{k}))} = 4 N_{\mathcal{K}}\)).

\section{Initialization and Symmetries}
\label{sec:init-symm}
Following \cite{BultinckKhalafLiuEtAl2020}, we consider the five symmetric candidate states: valley polarized (VP), valley Hall (VH), quantum Hall (QH), Kramers intervalley coherent (KIVC), and time-reversal symmetric intervalley coherent (TIVC), as well as the intervalley coherent (IOVC) found in \cite{HouSurWagnerEtAl2025}.
Each of these states are specified by their 1-RDM which is \(\bvec{k}\) independent and can be written as \(P = \frac{1}{2} ( I + Q)\) where
\begin{equation}
  \renewcommand{\arraystretch}{1.3}
  \begin{array}{lll}
    Q_{\text{VP}} =  \tau_{z} \sigma_{0}  & Q_{\text{VH}} =  \tau_{0} \sigma_{z} & Q_{\text{QH}} =  \tau_{z} \sigma_{z} \\
    Q_{\text{KIVC}} =  \tau_{y} \sigma_{y}  & Q_{\text{TIVC}} = \tau_{x} \sigma_{x} & Q_{\text{IOVC}} = \tau_{x} \sigma_{z}.
  \end{array}
\end{equation}


\section{Plot of Wavefunction Density}
\label{sec:wavefunction-density}
\begin{figure*}[t]
  \centering
  \begin{tabular}{cccccc}
    & \(\alpha = 0.00\) & \(\alpha = 0.25\) & \(\alpha = 0.50\) & \(\alpha = 0.75\) & \(\alpha = 1.00\) \\
    \raisebox{10ex}{\rotatebox[origin=c]{90}{\(\bvec{k} = \bvec{\Gamma}_{\m}\)}}
    & \includegraphics[width=.18\linewidth]{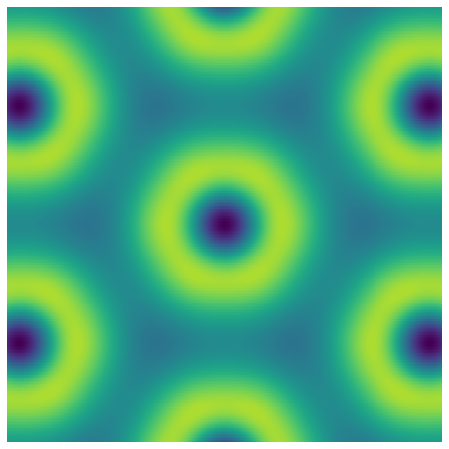} 
    & \includegraphics[width=.18\linewidth]{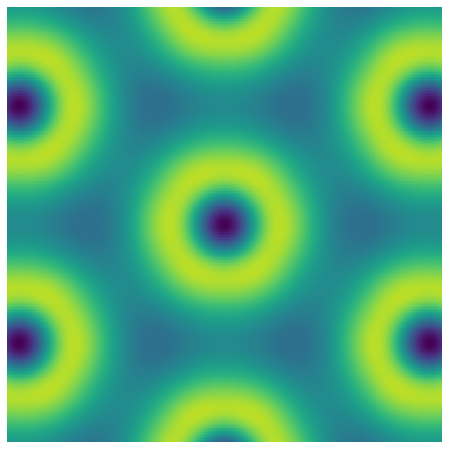} 
    & \includegraphics[width=.18\linewidth]{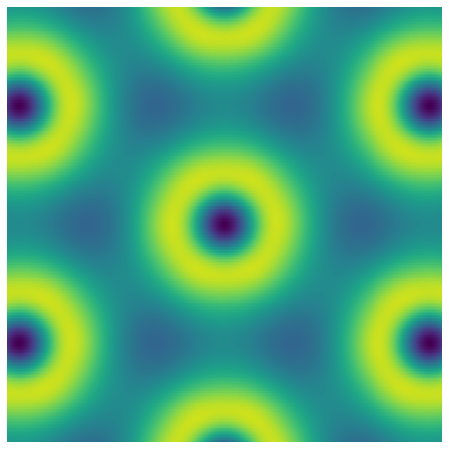} 
    & \includegraphics[width=.18\linewidth]{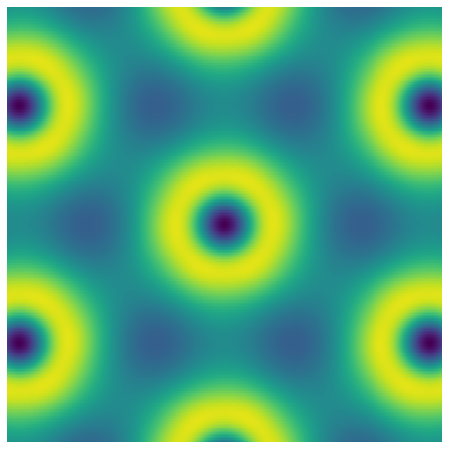} 
    & \includegraphics[width=.18\linewidth]{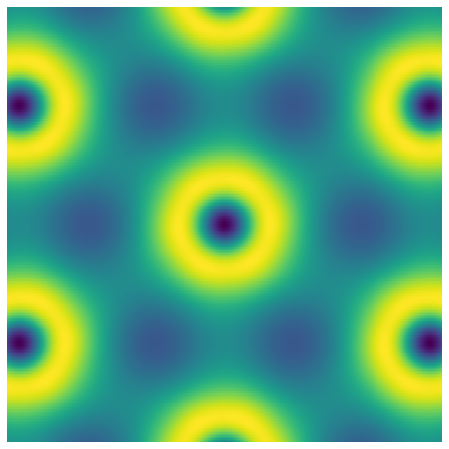} 
    \\
    \raisebox{10ex}{\rotatebox[origin=c]{90}{\(\bvec{k} = \bvec{k}_{\mathrm{near}}\)}}
    & \includegraphics[width=.18\linewidth]{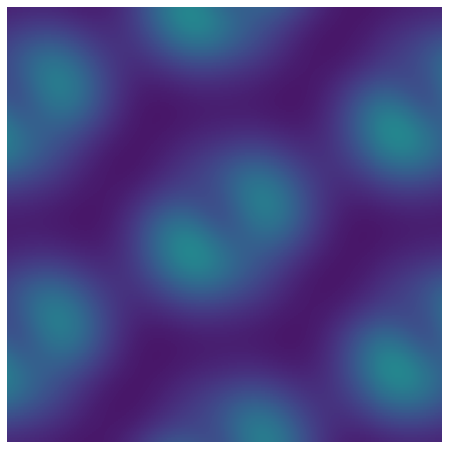} 
    & \includegraphics[width=.18\linewidth]{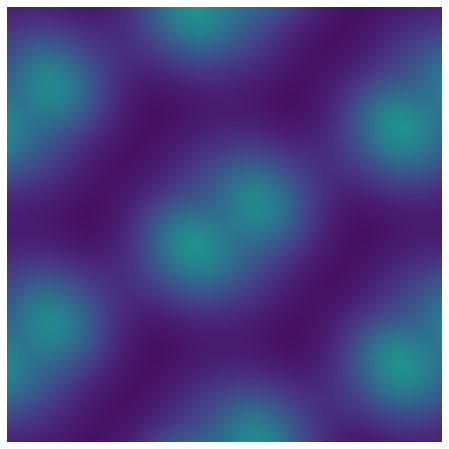} 
    & \includegraphics[width=.18\linewidth]{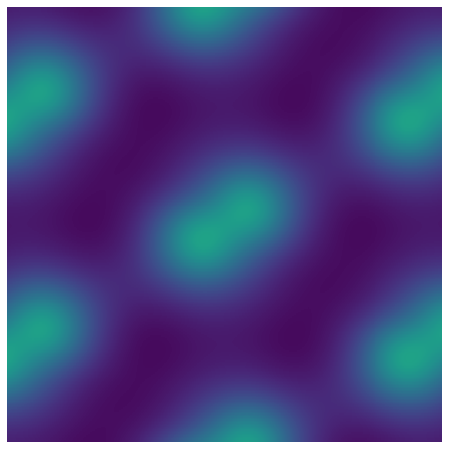} 
    & \includegraphics[width=.18\linewidth]{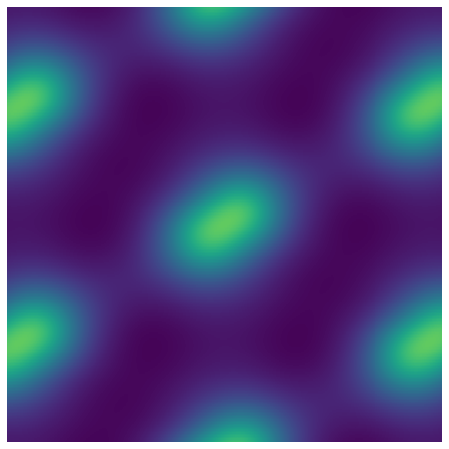} 
    & \includegraphics[width=.18\linewidth]{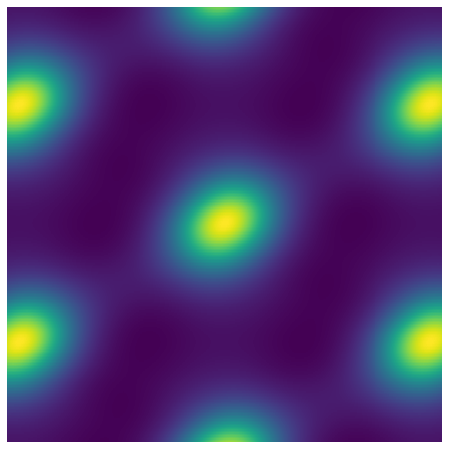} 
        \\
    \raisebox{10ex}{\rotatebox[origin=c]{90}{\(\bvec{k} = \bvec{k}_{\mathrm{far}}\)}}
    & \includegraphics[width=.18\linewidth]{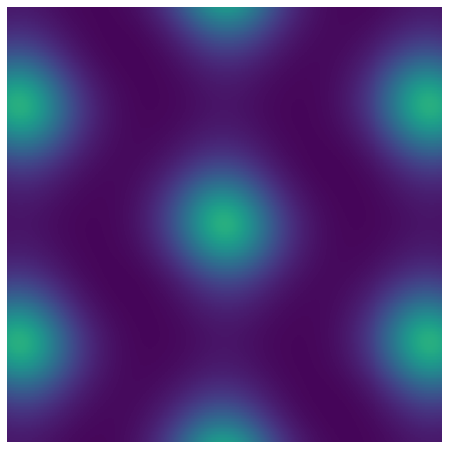} 
    & \includegraphics[width=.18\linewidth]{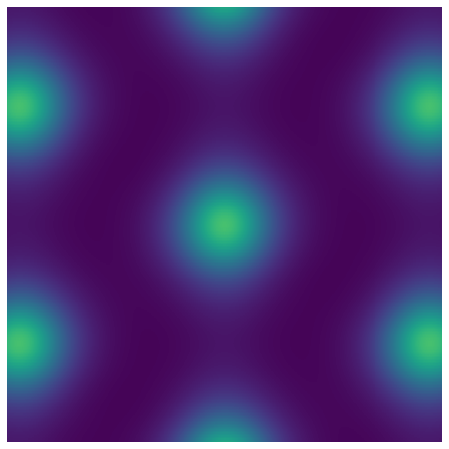} 
    & \includegraphics[width=.18\linewidth]{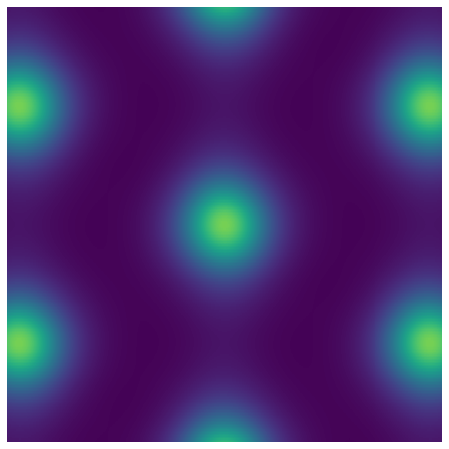} 
    & \includegraphics[width=.18\linewidth]{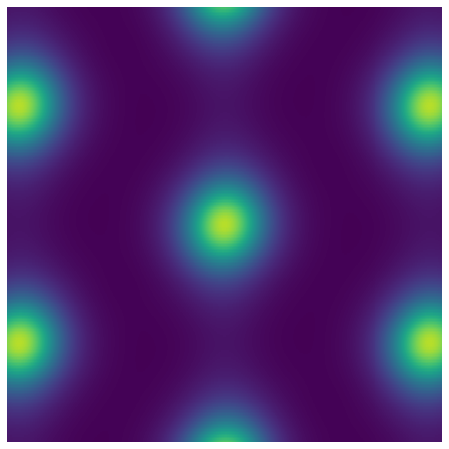} 
    & \includegraphics[width=.18\linewidth]{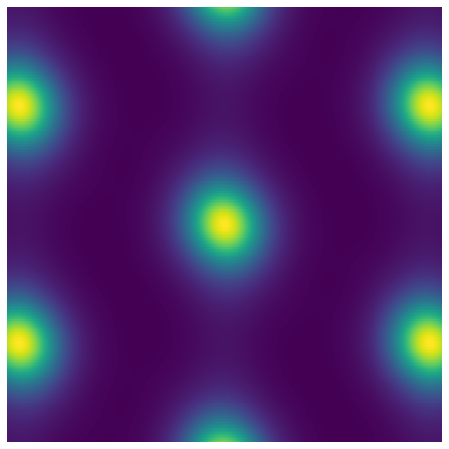} 
     \end{tabular}
     \caption{Plots of the real space density \(\varrho_{\bvec{k}}\left[P_{\downarrow\downarrow,\mathrm{sub}}\right](\bvec r)\) for the Hamiltonian \( (1-\alpha) H_\mathrm{BM} + \alpha H_{\mathrm{relax}}\). Note that the density is more concentrated at \(\bvec{k} \neq \vec{\Gamma}_{\m}\) (recall that we take $\bvec{k}_{\mathrm{near}} = \bvec{\Gamma} - \tfrac{1}{12} (\bvec{b}_{\text{m},1} + \bvec{b}_{\text{m},2})$ and  $\bvec{k}_{\text{far}}  = \bvec{\Gamma}_{\m} - \tfrac{5}{12} (\bvec{b}_{\m,1} + \bvec{b}_{\m,2})$) as compared to \(\bvec{k} = \vec{\Gamma}_{\m}\) for the relaxed model.
     }
  \label{eq:wavefunction-density}
\end{figure*}

In this section, we plot the \(\bvec{k}\)-resolved real space density, \(\varrho_{\bvec{k}}[P_{\downarrow\downarrow,\mathrm{sub}}](\bvec{r})\) as described in~\cref{eq:real-density-k} for the Hamiltonian
\begin{equation}
  H(\alpha) := (1 - \alpha) H_{\mathrm{BM}} + \alpha H_{\mathrm{relax}}
\end{equation}
as we vary \(\alpha \in [0, 1]\) (see~\cref{eq:wavefunction-density}).
We note that the density plot is very similar to the results reported in recent DFT studies \cite[Fig. 2]{ZhuBennettLarsonEtAl2026}.

\end{document}


\title{Supplemental Material for ``Relaxation effects on Hartree-Fock ground states in twisted bilayer graphene at even integer fillings''}

\author{Tianyu Kong}%
\email{tianyuk@uchicago.edu}
\affiliation{%
  Committee on Computational and Applied Mathematics, University of Chicago, Chicago, IL 60637, USA}%

\author{Alexander B. Watson}
\email{abwatson@umn.edu }
\affiliation{%
  School of Mathematics, University of Minnesota Twin Cities, Minneapolis, Minnesota 55455, USA}%
  
\author{Lin Lin}%
\email{linlin@math.berkeley.edu}
\affiliation{%
  Department of Mathematics, University of California, Berkeley, CA 94720, USA}%

\author{Mitchell Luskin}%
\email{luskin@umn.edu}
\affiliation{%
  School of Mathematics, University of Minnesota Twin Cities, Minneapolis, Minnesota 55455, USA}%

\author{Kevin D. Stubbs}%
\email{Contact author: kstubbs@berkeley.edu}
\affiliation{%
  Department of Mathematics, University of California, Berkeley, CA 94720, USA}%

\maketitle

We provide the analogous figures to those in Section IV in the main text using the average subtraction scheme instead of the charge neutral scheme.
For understanding our plots, we recall our decomposition of the Fock matrix:
\begin{equation} \label{eq:fock-decomposition}
\begin{split}
  F[P]
  & = H_{0} + J[P_{\uparrow\uparrow} - P_{\uparrow\uparrow,\mathrm{sub}}] - K[P - P_{\mathrm{sub}}] + J[P_{\downarrow\downarrow} - P_{\downarrow\downarrow,\mathrm{sub}}] \\
  & = H_{0} + J[P_{\uparrow\uparrow} - P_{\uparrow\uparrow,\mathrm{sub}}] - K[P - P_{\mathrm{sub}}]
    +
    \begin{cases}
      J[0 - P_{\downarrow\downarrow,\mathrm{sub}}] & \nu = -2, \\
      J[P_{\downarrow\downarrow} - P_{\downarrow\downarrow,\mathrm{sub}}] & \nu = 0, \\
      J[I - P_{\downarrow\downarrow,\mathrm{sub}}] & \nu = +2, \\
    \end{cases}
\end{split}
\end{equation}
After performing self-consistent Hartree-Fock calculations, we find both the KIVC and VP states are close in energy so we report the band structures for both.




\begin{figure*}[t]
  \centering
  \includegraphics[width=.9\linewidth]{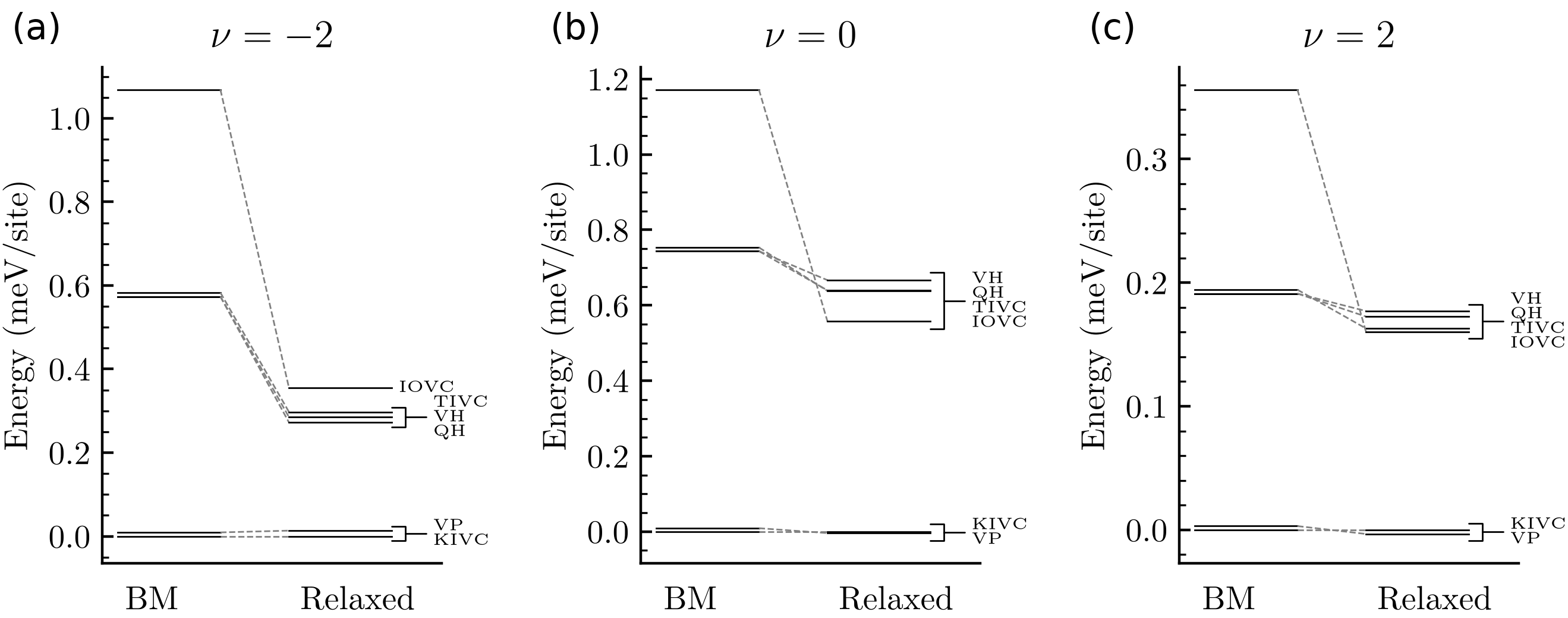}
  \caption{
    Energy orderings of self-consistent Hartree-Fock solutions starting from different initializations in the Bistritzer-MacDonald (BM) and Relaxed interacting models at filling factor \(\nu = -2\) (a), \(\nu = 0\) (b), \(\nu = + 2\) (c).
    Energies are shifted so that \(E_{KIVC} = 0\).
    All calculations are performed on a \(12 \times 12\) Monkhorst-Pack grid with average subtraction.
  }
  \label{fig:hf-results-avg}
\end{figure*}


\begin{figure*}[t]
  \centering
  \includegraphics[width=.8\linewidth]{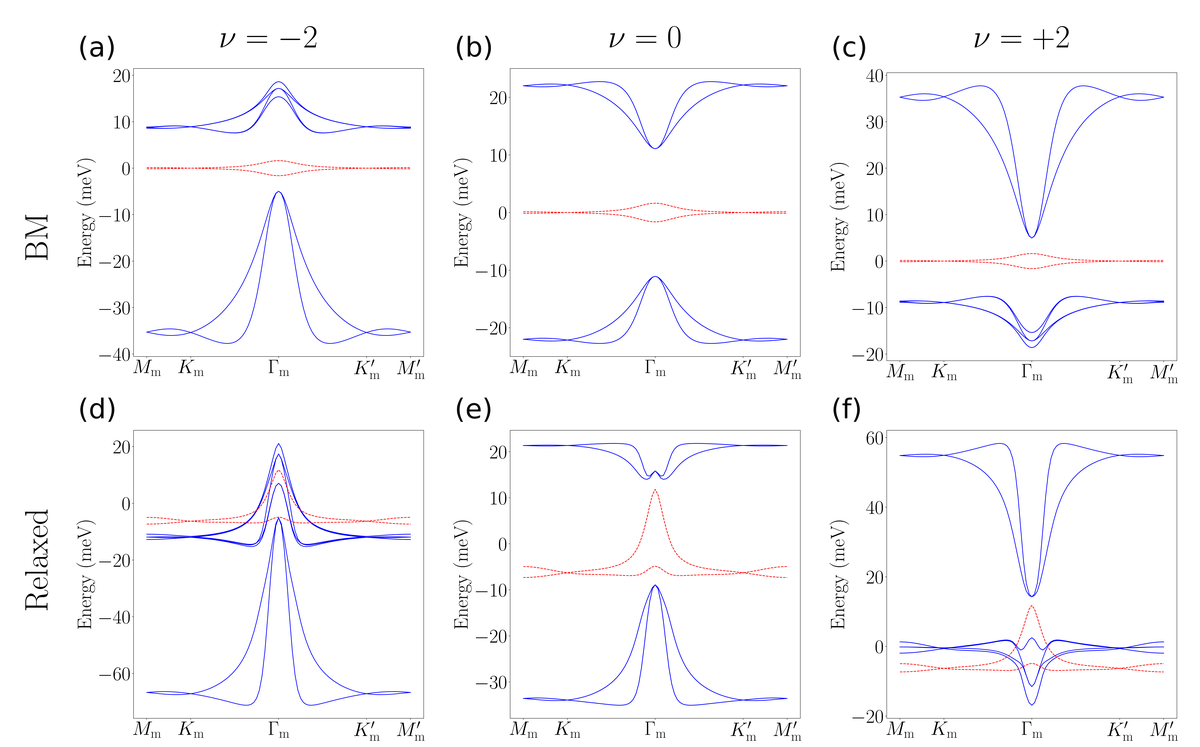}
  \caption{
    Single particle bands (\textcolor{red}{red, dashed}) and Hartree-Fock bands (\textcolor{blue}{blue, solid}) for the Bistritzer-MacDonald (BM) (a,b,c) and Relaxed (d,e,f) interacting models at filling factors \(\nu = -2\) (a,d), \(\nu = 0\) (b,e), \(\nu = +2\) (c,f) starting from KIVC initialization.
    All calculations are performed on a \(12 \times 12\) Monkhorst-Pack grid with average subtraction.
    }
\end{figure*}

\begin{figure*}[t]
  \centering
  \includegraphics[width=.8\linewidth]{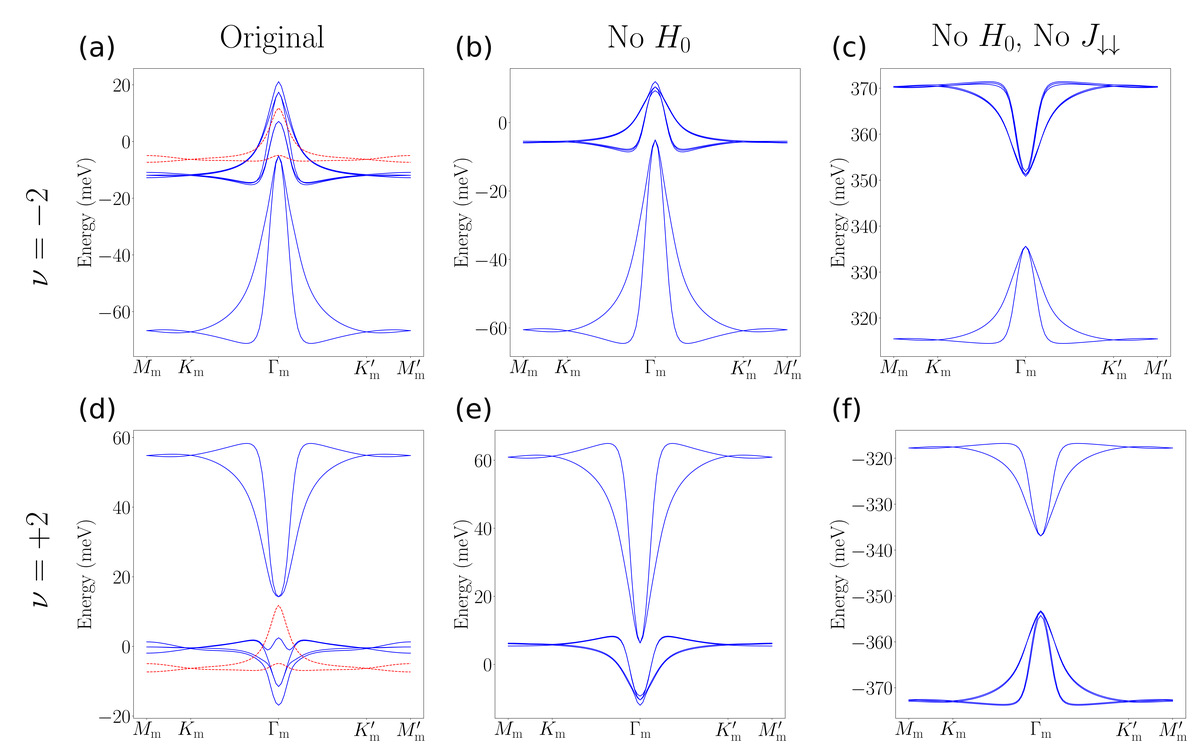}
  \caption{
    Plots of the single particle bands (\textcolor{red}{red, dashed}) and Hartree-Fock bands (\textcolor{blue}{blue, solid}) for the Relaxed interacting model at filling factors \(\nu = -2\) (a) and \(\nu=+2\) (d) starting from KIVC initalization on a \(12 \times 12\) grid with average subtraction.
    Note that in (d) the size of the band gap at \(\Gamma_{\m}\) is of comparable size to the single particle dispersion.
    Plots (b,e) are the same as (a,d) except \(H_{0}\) in \cref{eq:fock-decomposition} was set to zero.
    Plots (c,f) are the same as (a,d) except both \(H_{0}\) and \(J[P_{\downarrow\downarrow} - P_{\mathrm{sub},\downarrow\downarrow}]\) in \cref{eq:fock-decomposition} were set to zero.
  }
  \label{fig:dispersion-comparison}
\end{figure*}


\begin{figure*}[t]
  \centering
  \includegraphics[width=.8\linewidth]{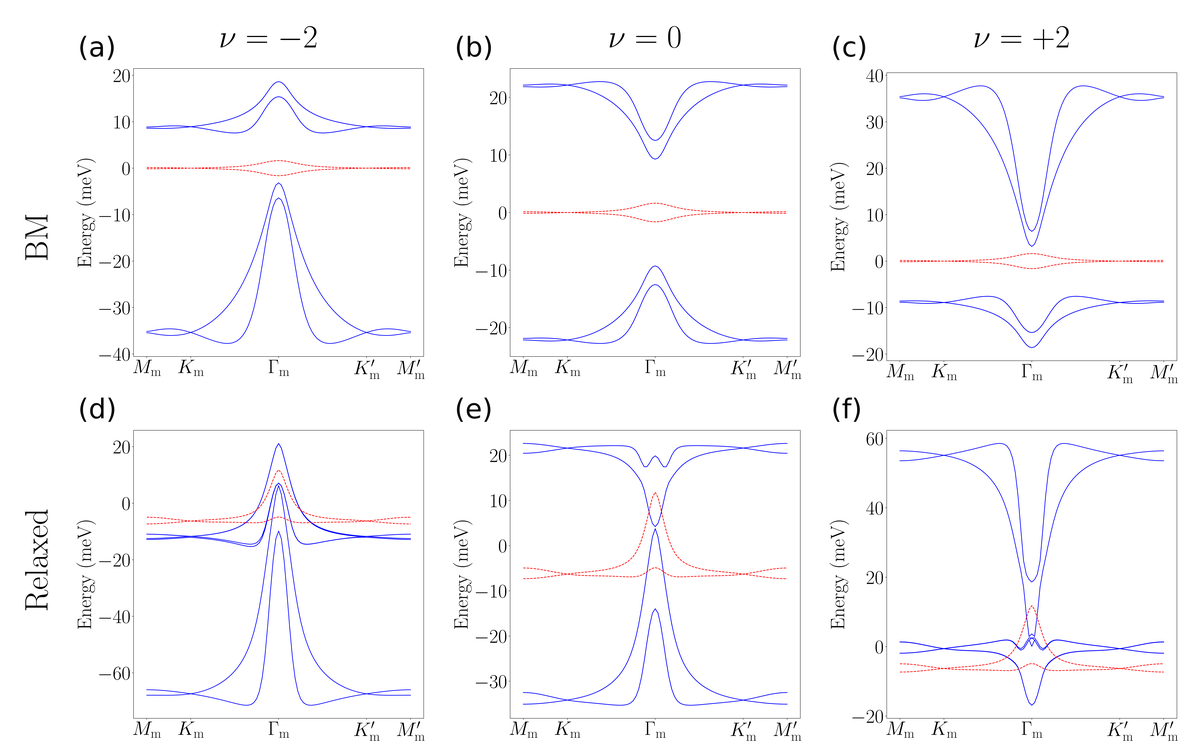}
  \caption{
    Single particle bands (\textcolor{red}{red, dashed}) and Hartree-Fock bands (\textcolor{blue}{blue, solid}) for the Bistritzer-MacDonald (BM) (a,b,c) and Relaxed (d,e,f) interacting models at filling factors \(\nu = -2\) (a,d), \(\nu = 0\) (b,e), \(\nu = +2\) (c,f) starting from VP initialization.
    All calculations are performed on a \(12 \times 12\) Monkhorst-Pack grid with charge neutral subtraction.
    }
  \label{fig:vp-bands-avg}
\end{figure*}

\begin{figure*}[t]
  \centering
  \includegraphics[width=.8\linewidth]{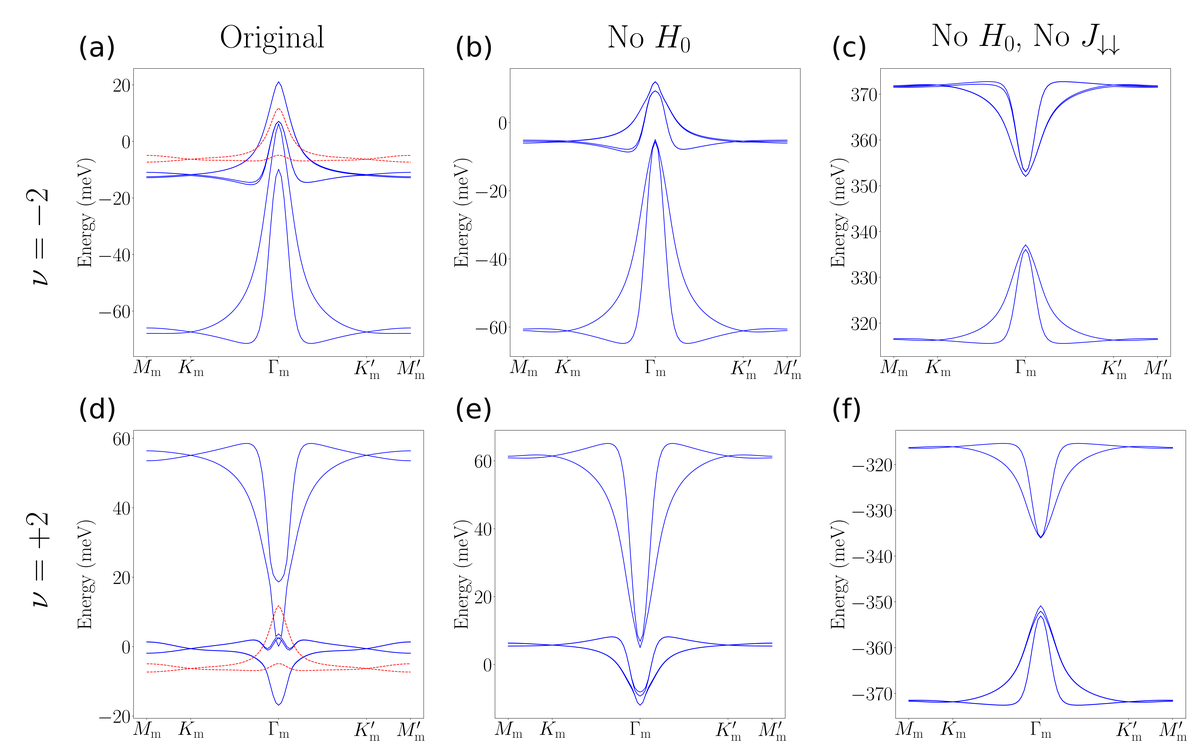}
  \caption{
    Plots of the single particle bands (\textcolor{red}{red, dashed}) and Hartree-Fock bands (\textcolor{blue}{blue, solid}) for the Relaxed interacting model at filling factors \(\nu = -2\) (a) and \(\nu=+2\) (d) starting from KIVC initalization on a \(12 \times 12\) grid with average subtraction.
    Note that in (d) the size of the band gap at \(\Gamma_{\m}\) is of comparable size to the single particle dispersion.
    Plots (b,e) are the same as (a,d) except \(H_{0}\) in \cref{eq:fock-decomposition} was set to zero.
    Plots (c,f) are the same as (a,d) except both \(H_{0}\) and \(J[P_{\downarrow\downarrow} - P_{\mathrm{sub},\downarrow\downarrow}]\) in \cref{eq:fock-decomposition} were set to zero.
  }
  \label{fig:dispersion-comparison}
\end{figure*}


\begin{figure*}[t]
  \centering
  \begin{tabular}{cccccc}
    & \(\alpha = 0.00\) & \(\alpha = 0.25\) & \(\alpha = 0.50\) & \(\alpha = 0.75\) & \(\alpha = 1.00\) \\
    \rotatebox{90}{\(\bvec{k} = \vec\Gamma\)} 
    & \includegraphics[width=.18\linewidth]{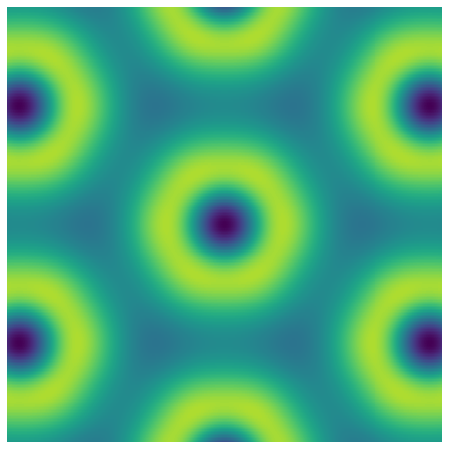} 
    & \includegraphics[width=.18\linewidth]{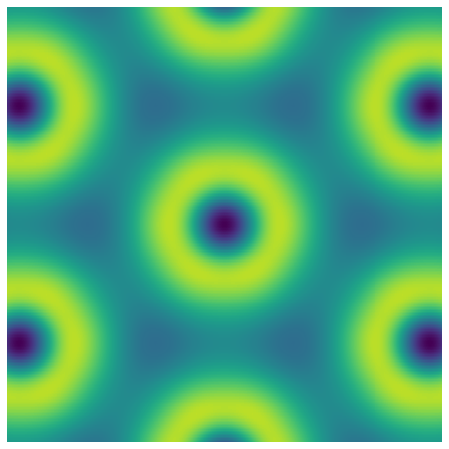} 
    & \includegraphics[width=.18\linewidth]{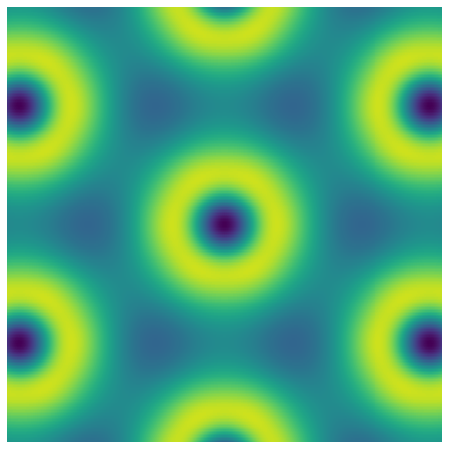} 
    & \includegraphics[width=.18\linewidth]{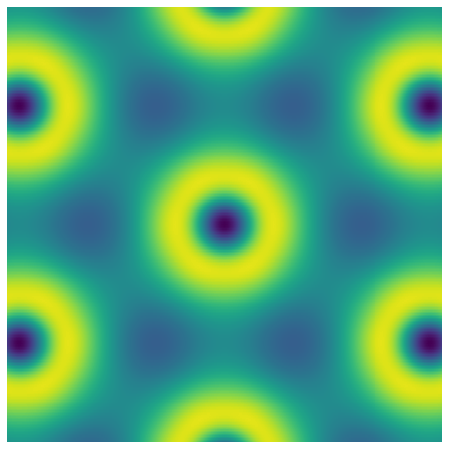} 
    & \includegraphics[width=.18\linewidth]{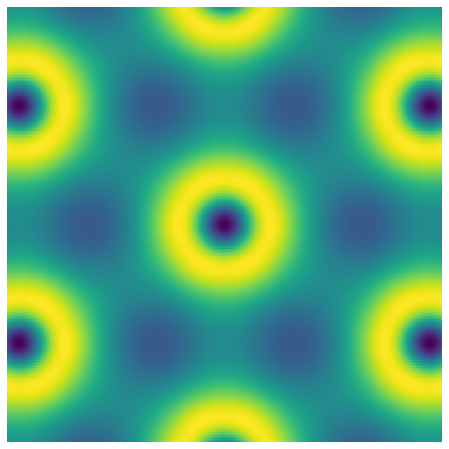} 
    \\    \rotatebox{90}{\(\bvec{k} = \bvec{k}_{\mathrm{near}}\)} 
    & \includegraphics[width=.18\linewidth]{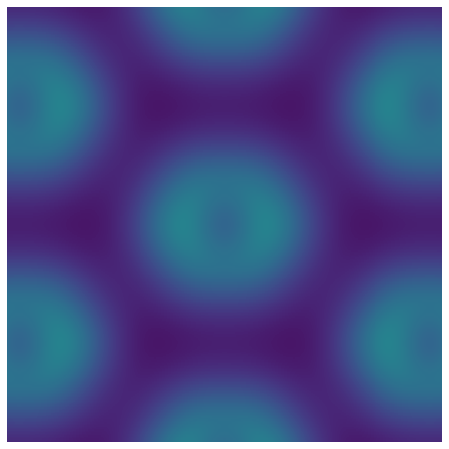} 
    & \includegraphics[width=.18\linewidth]{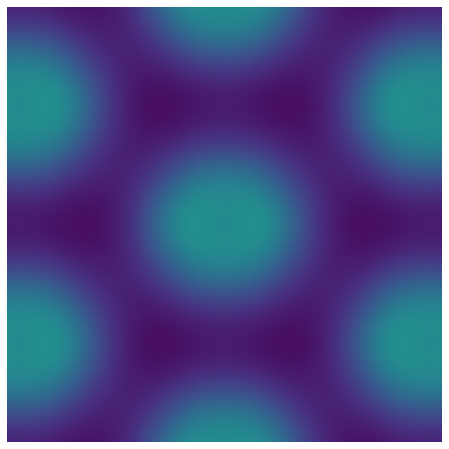} 
    & \includegraphics[width=.18\linewidth]{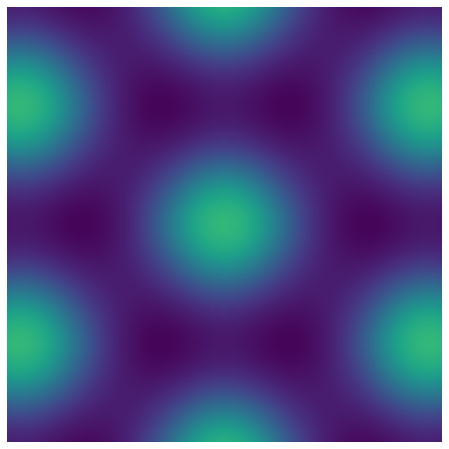} 
    & \includegraphics[width=.18\linewidth]{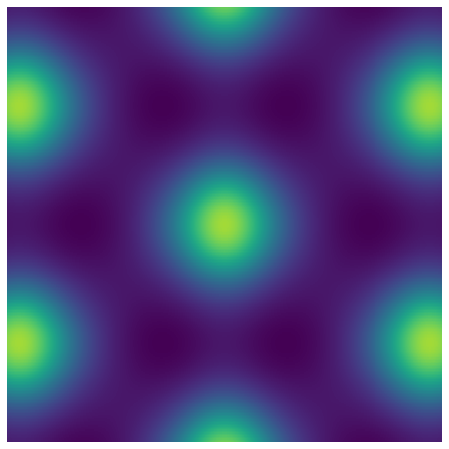} 
    & \includegraphics[width=.18\linewidth]{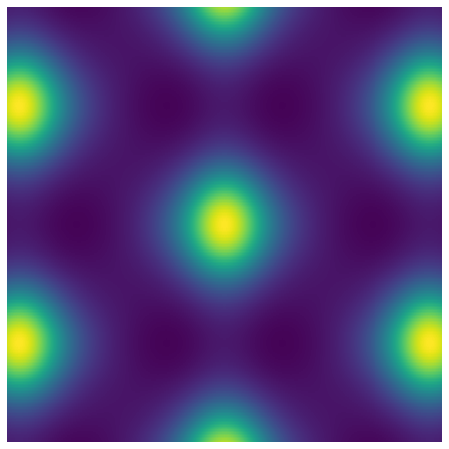} 
        \\    
      \rotatebox{90}{\(\bvec{k} = \bvec{k}_{\mathrm{far}}\)} 
    & \includegraphics[width=.18\linewidth]{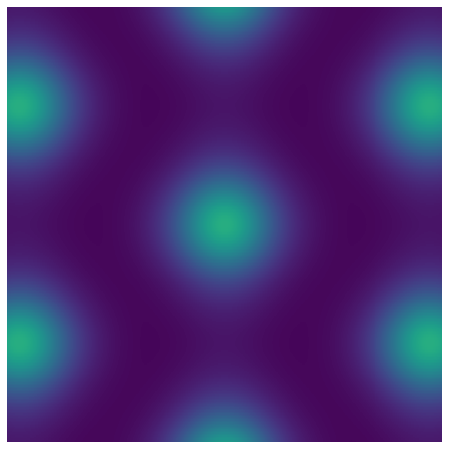} 
    & \includegraphics[width=.18\linewidth]{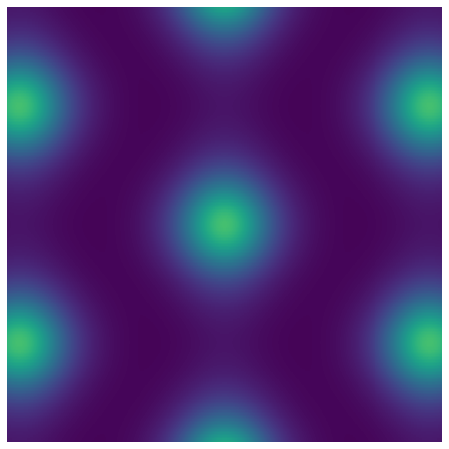} 
    & \includegraphics[width=.18\linewidth]{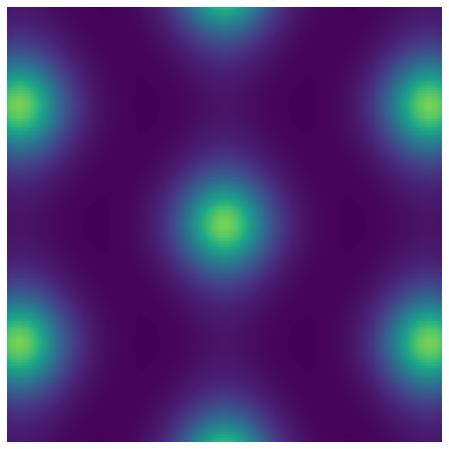} 
    & \includegraphics[width=.18\linewidth]{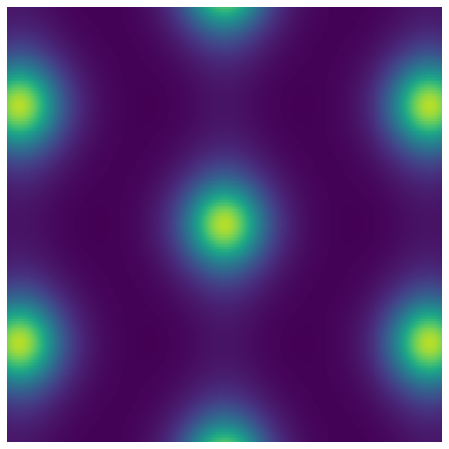} 
    & \includegraphics[width=.18\linewidth]{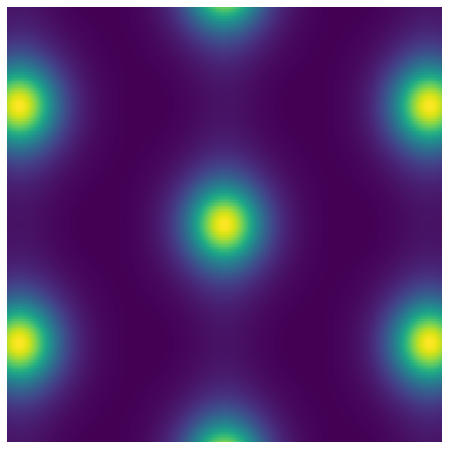} 
     \end{tabular}
       \caption{Plots of the real space density \(\varrho_{\bvec{k}}\left[\frac{1}{2} I\right](\bvec r)\) (the density matrix for average subtraction) for the Hamiltonian \( (1-\alpha) H_\mathrm{BM} + \alpha H_{\mathrm{relax}}\). Note that the density is more concentrated at \(\bvec{k} \neq \vec{\Gamma}_{\m}\) (recall that we take $\bvec{k}_{\mathrm{near}} = \bvec{\Gamma} - \tfrac{1}{12} (\bvec{b}_{\text{m},1} + \bvec{b}_{\text{m},2})$ and  $\bvec{k}_{\text{far}}  = \bvec{\Gamma}_{\m} - \tfrac{5}{12} (\bvec{b}_{\m,1} + \bvec{b}_{\m,2})$) as compared to \(\bvec{k} = \vec{\Gamma}_{\m}\) for the relaxed model.}
  \label{eq:wavefunction-density}
\end{figure*}

\begin{figure}[t]
  \centering
  \includegraphics[width=.9\linewidth]{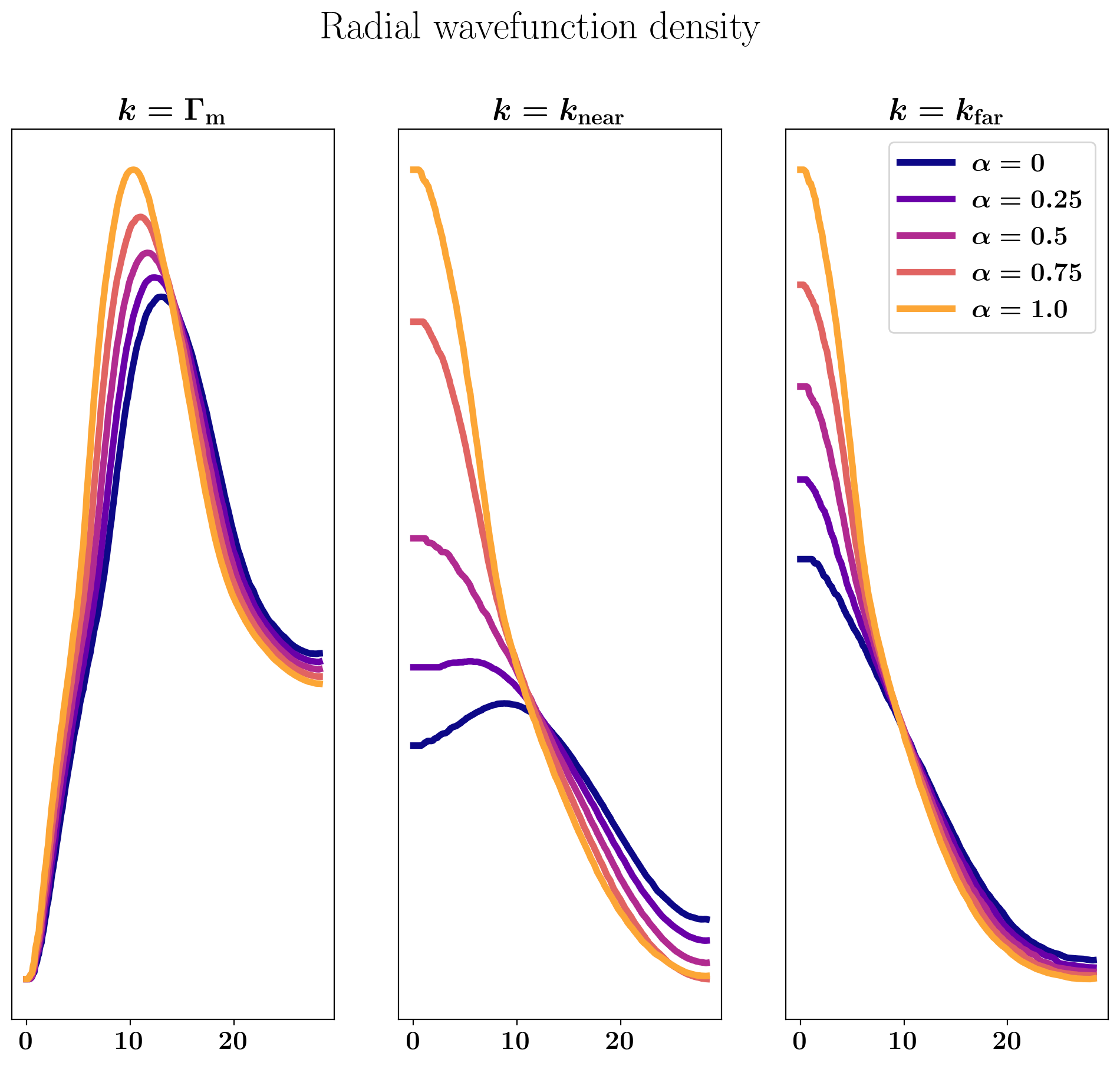}
  \caption{Plots of the radial density \(\int_0^{2\pi} \varrho_{\bvec{k}}\left[\frac{1}{2}I\right](r,\theta) \dee \theta \) for the Hamiltonian \( (1 - \alpha) H_\mathrm{BM} + \alpha H_{\mathrm{relax}}\) on a single unit cell. We find that the density is more concentrated in the relaxed model at a representative \(\bvec{k} \neq \bvec{\Gamma}_{\m}\) (here we take $\bvec{k}_{\mathrm{near}} = \bvec{\Gamma} - \tfrac{1}{12} (\bvec{b}_{\text{m},1} + \bvec{b}_{\text{m},2})$ and  $\bvec{k}_{\text{far}}  = \bvec{\Gamma}_{\m} - \tfrac{5}{12} (\bvec{b}_{\m,1} + \bvec{b}_{\m,2})$, and the result is similar at other $\bvec{k}$) compared with \(\bvec{k} = \vec{\Gamma}_{\m}\). Note that the wavefunction at $\bvec{k} = \bvec{\Gamma}_{\m}$ must have a node to transform correctly under $C_{3z}$.  }
  \label{fig:radial_density}
\end{figure}